\nonstopmode
\documentstyle[twocolumn,eqsecnum,aps,prd,amsfonts,floats]{revtex}

\begin{document}
\draft
%\preprint{TIT/HEP-???/COSMO-???}
%%%%%%%%%%%%%%%%%%%%%%%%%%%%%%%%%%%%%%%%%%%%%%%%%%%%%%%%%%%%%%%%%%%%%%%%%%%%%%%%
%
%
%%%%%%% In order to include ps files. %%%%%%%%%%%%%%%%%%%%%%%%%%%%%%%%%%%%%%%%%%
%
% NOTE: You need "boxedeps.tex" by Laurent Siebenmann, 
% in order to print out the PS figures correctly.  
% "boxedeps.tex" can be obtained from various ftp cites, 
% including the lanl xxx e-print archives.  
%
  \input boxedeps  %%%%%%%%% Including ``boxedeps.tex.''
  \HideDisplacementBoxes %%%%%%%% Hiding the ``debug mode'' boxes. 
% NOTE: Please comment out or uncomment appropriate lines depending on your 
%   dvi_to_ps type driver, in order to include PS files correctly. 
% The current version is set to work with "dvips". 
%%% Special syntax for several drivers. The macros
   \SetRokickiEPSFSpecial %% ``dvips'' by Tom Rokicki
%% \SetUnixCoopEPSFSpecial%% ``dvi2ps'' early unix
%% \SetTexturesEPSFSpecial %% Textures
%% \SetOzTeXEPSFSpecial %% OzTeX by Andrew Trevorrow
%% \SetPSprintEPSFSpecial%% PSprint by Andrew Trevorrow
%% \SetArborEPSFSpecial %% ArborTeX DVILASER/PS
%% \SetClarkEPSFSpecial%% dvitops by James Clark
%% \SetDVIPSoneEPSFSpecial%% DVIPSONE of Y&Y
%% \SetBeebeEPSFSpecial%% DVIALW by N. Beebe
%% \SetStandardEPSFSpecial%% Nonexistant: Placebo below
%
%%%%%%%%%%%%%%%%%%%%%%%%%%%%%%%%%%%%%%%%%%%%%%%%%%%%%%%%%%%%%%%%%%%%%%%%%%%%%%%%

\def\av#1{{\left\langle #1 \right\rangle}}

%@twocolumnfalse\endcsname
%\wideabs{
\title{Renormalization group and critical behaviour in gravitational collapse%
  }
\author{Takashi Hara\thanks{e-mail: hara@ap.titech.ac.jp}}
\address{Department of Applied Physics, Tokyo Institute of Technology,
  Oh-Okayama, Meguro, Tokyo 152, Japan}
\author{Tatsuhiko Koike\thanks{e-mail: koike@rk.phys.keio.ac.jp}} 
\address{Department of Physics, Keio University,
  Hiyoshi, Kohoku, Yokohama 223, Japan}
\author{Satoshi Adachi\thanks{e-mail: adachi@aa.ap.titech.ac.jp}}
\address{Department of Applied Physics, Tokyo Institute of Technology,
  Oh-Okayama, Meguro, Tokyo 152, Japan}
\date{July 4, 1996; revised: April 16, 1997}
\preprint{}
\maketitle
\begin{abstract}
  We present a general framework for understanding and analyzing 
critical behaviour in gravitational collapse.  
We adopt the method of renormalization group, which has the following
advantages. (1) It provides a natural explanation for various types 
of universality and scaling observed in numerical studies.  
In particular,  universality in initial data space and 
universality for different models are understood in a unified way.  
(2) It enables us to perform a detailed analysis of
time evolution beyond linear perturbation, by providing rigorous
controls on nonlinear terms.
Under physically reasonable assumptions we prove:
  (1) {\em Uniqueness}\/ of the relevant mode around a fixed point 
  implies universality in initial data space.
  (2) The critical exponent $\beta_{\rm BH}$ and the
  unique positive eigenvalue $\kappa$ of the relevant mode
  is {\em exactly}\/ related by $\beta_{\rm BH} = \beta /\kappa$, 
  where $\beta$ is a scaling exponent used in calculating the eigenvalue.
  (3) The above (1) and (2) hold also for discretely self-similar
  case (replacing ``fixed point'' with ``limit cycle''). 
  (4) Universality for different models holds under a certain condition.
  [These are summarized as Theorems~\ref{prop-final}, 
  \ref{prop-final-discr} and \ref{theorem-11} of 
  Sec.~\ref{sec-RGscenario}.]
  
  According to the framework, we carry out a rather complete (though
  not mathematically rigorous) analysis for perfect fluids with
  pressure proportional to density, 
  in a wide range of the adiabatic index $\gamma$.
  The uniqueness of the relevant mode around a fixed point is
  established by Lyapunov analyses.
  This shows that the critical phenomena occurs not
  only for the radiation fluid but also for perfect fluids with 
  $1 < \gamma \lesssim 1.88$. 
  The accurate values of critical exponents are calculated for the models.
  In particular, the exponent for the radiation fluid 
  $\beta_{\rm BH} \simeq 0.35580192$ is also
  in agreement with that  obtained in numerical simulation. 
\end{abstract}
%} %% end of \wideabs
%%]  %% End of strange twocolumn

\narrowtext

 \newtheorem{theorem}    {Theorem} 
 \newtheorem{lemma}      {Lemma}
 \newtheorem{prop}        {Proposition}
 \newtheorem{cor}         {Corollary}
 \newtheorem{defn}        {Definition}
 \newtheorem{conj}        {Conjecture}
 \newtheorem{ass}         {Assumption}
 \newtheorem{claim}       {Claim}

 \newcommand{\Oabs}     {{\bar{O}} }
 \newcommand{\Frel}     {F^{\rm rel}}
\newcommand{\grless} 
{ {\, \raise-.24em\hbox{$>$} \hspace{-0.8em} \raise.31em\hbox{$<$}\, } }

% % rf_full_12.tex
% Part of the full paper on RF.
%%%%%%%%%%%%%%%%%%%%%%%%%%%%%%%%%%%%%%%%%%%%%%%%%%%%%%%%%%%%%%%%%%%%%%%%%%%%%%%%
%%%%%%%%%%%%%%%%%%%%%%%%%%%%%%%%%%%%%%%%%%%%%%%%%%%%%%%%%%%%%%%%%%%%%%%%%%%%%%%%
%%%%%%%%%%%%%%%%%%%%%%%%%%%%%%%%%%%%%%%%%%%%%%%%%%%%%%%%%%%%%%%%%%%%%%%%%%%%%%%%
%%%%%%%%%%%%%%%%%%%%%%%%%%%%%%%%%%%%%%%%%%%%%%%%%%%%%%%%%%%%%%%%%%%%%%%%%%%%%%%%

\section{Introduction}
Gravitational collapse with formation of black holes is one of the  
main problems of classical general relativity. A very important
question here is how formation of a black hole depends on initial data,
for example, which initial data evolve into black holes and which do
not. It is concerned with many important problems such as cosmic
censorship \cite{Pen69}. 

Christodoulou \cite{Chr86a,Chr86b,Chr87,Chr91}
rigorously analyzed such a problem in a relatively
simple system, namely, spherically symmetric scalar field collapse. He
proved that the field disperses at later times if the initial field 
is 
sufficiently weak, and also found a sufficient condition for black hole
formation.  However, the behaviour of the field and the space-time for
initial data sets around the threshold of the black hole formation
remained open, and he posed a question whether the phase transition between
space-times with  and without black holes is of the
first or of the second order, i.e., whether the mass of the 
black hole formed is a continuous function at the threshold.

Choptuik \cite{Cho93} numerically  analyzed the system and
obtained a result which strongly suggests the latter.  Furthermore, 
he discovered 
a ``critical behaviour'' of the gravitational collapse of the system.
His result can be summarized as follows: 
Let the initial distribution of the scalar field be parametrized 
smoothly by a parameter $p$,  
such that the solutions with the initial data 
$p>p_{c}$ contain a black hole 
while those with $p<p_{c}$ do not. 
For several  one-parameter families investigated, 
near-critical solutions ($p \approx p_{c}$) satisfy the following: 
(1) the critical solution (i.e. $p = p_{c}$) is {\em universal}\/
in the sense that it  
approaches the identical space-time for all families,
(2) the critical solution  has a discrete self-similarity, and 
(3) for supercritical solutions ($p > p_{c}$) the black hole mass satisfies
$M_{\rm BH}\propto (p-p_{c})^{ \beta_{\rm BH}}$
and the critical exponent $\beta_{\rm BH}$, which is about 0.37, 
is universal for
all families. Abrahams and Evans \cite{AbEv93} found similar
phenomena in axisymmetric collapse of gravitational wave and obtained
$\beta_{\rm BH} =0.38$.  
Evans and Coleman
\cite{EvCo94}  found similar phenomena with $\beta_{\rm BH} \simeq 0.36$ 
in spherically symmetric collapse of radiation fluid,
in which case the self-similarity is not discrete but {\em
continuous}.
Employing a self-similar ansatz they also found a numerical solution
which fits the inner region of the near-critical solutions very well.
Since the above values of $\beta_{\rm BH}$ were close to each other, 
some people considered that there is a universality over many systems.

For a deeper understanding of these phenomena, an important direction
is to pursue the analogy not only phenomenologically but also
theoretically between the ``critical phenomena'' in 
gravitational collapse and those in phase transitions in
statistical mechanics. 
Argyres \cite{Arg94}
reviewed the above numerical results in terms of renormalization
group (RG).  
Independently, the present authors \cite{KHA95} showed that both
qualitative and quantitative understanding is possible by introducing
renormalization group ideas.  They showed that {\em uniqueness} of the
relevant (unstable) eigenmode of the linear perturbations around the
critical self-similar space-time is essential for the universality to be
observed and that the critical exponent $\beta_{\rm BH}$ is given by
$\beta_{\rm BH} = 1/\kappa$, where $\kappa$ is 
the only eigenvalue with positive real part. 
Emphasized there was that the standard renormalization group argument 
implies that the linear perturbation analysis is sufficient to obtain
the {\em exact}\/ value of $\beta_{\rm BH}$.
They confirmed that their scenario holds in the case of radiation
fluid and obtained a very accurate value of the exponent,
$\beta_{\rm BH} = 0.35580192$.   
The difference {}from those obtained for scalar
field collapse (0.37) and gravitational wave collapse (0.38) was
beyond the possible numerical errors in their simulations. 
This showed that there is no universality between radiation fluid
collapse and other systems.
Maison \cite{Mai96} applied the method to the systems of
perfect fluids with pressure proportional to density and showed
how $\beta_{\rm BH}$ depends on the equation of state, under the assumption
that the same phenomena occur and the same scenario holds as
the radiation fluid.
Gundlach~\cite{Gun96} confirmed $\beta_{\rm BH}=1/\kappa$ in scalar
field collapse.

Some new and intriguing phenomena in the critical behaviour have been
found such as bifurcation~\cite{HiEa95b,LiCh96} and coexistence of first
and second order phase transitions~\cite{CCB96}.
However, in this paper we will pursue a slightly different direction, 
namely concentrate on providing deeper
understanding of basic aspects of the critical behaviour.

This paper is the fully expanded version of Ref.~\cite{KHA95}.  
(Some of the materials here have already been treated in \cite{Koi96}.)
The aim of the paper is twofold.  
First, we present a general, mathematically rigorous, 
framework in which critical 
behaviour observed in various models 
so far can be analyzed and understood in a unified way, 
based on a renormalization group picture.   
Now that there seems to be some consensus on the qualitative 
picture on the simplest case, we in this paper focus our attention to 
two problems which remain to be more clarified.  
(1) We give a sufficient condition under which the critical exponent 
is given {\em exactly} by $\beta_{\rm BH} = \beta /\kappa$.  
In the above, $\beta$ is a scaling exponent which appears in 
calculating the Lyapunov exponent (see Sec.~\ref{sec-RGscenario} 
for details).  Note 
that  in  \cite{KHA95} we only considered the radiation fluid 
collapse, 
where one can take $\beta =1$.  
(2)  We also consider the problem of universality, in the wider sense 
of the word that 
details of the definition of the model is irrelevant for 
the critical behaviour, and we present a sufficient 
condition under which the universality of this kind occurs.  
Both conditions are sufficiently weak that they are 
expected to be applicable to most systems for which the 
critical behaviour has been observed.

Second, we present a 
more thorough analysis of the perfect fluid collapse,
with various adiabatic index $\gamma$. 
In addition to presenting the details of our analysis on the 
radiation fluid collapse reported in  \cite{KHA95}, 
we here report a result of a new method of analysis, the so 
called {\em Lyapunov analysis}, which was performed to 
further confirm the uniqueness of the relevant mode 
around the critical solution.
A Lyapunov analysis and the shooting
method for an ordinary differential equation adopted in \cite{KHA95}
are complementary to each other in the following sense: (1) The
former extracts eigenmodes in the descending order of its real part,
whereas the latter affords information only of a finite region of
complex $\kappa$ plane.  (2) The Lyapunov analysis is useful even if
the eigenmodes do not form a complete set.
(3) It is easier in the latter than in the
former to numerically obtain accurate values of the eigenvalues
$\kappa$ hence the critical exponent $\beta_{\rm BH}$.
Our analysis in the second part is sufficiently convincing for 
physicists, we believe.  But mathematically it is unsatisfactory in 
the sense that it  does not provide rigorous theorems.  Rigorous full 
analysis of the system is a subject of future studies. 

This paper is organized as follows. 
First, we present our general scenario in Sec.~\ref{sec-RGscenario}.  
We define renormalization group 
transformations acting on the phase space (Sec.~\ref{sec-RG}),  
and see how all the aspects of critical phenomena for 
radiation fluid collapse, etc. are deduced {}from  
the behaviour of the RG flow near a fixed point 
(Sec.~\ref{sec-scenario}).  We present a sufficient condition for the 
exactness of the relation $\beta_{\rm BH} = \beta /\kappa$ 
in Sec.~\ref{sub-rgflow}. 
We then consider more general cases, i.e. discrete self-similarity 
(Sec.~\ref{sec-discr.SS}) and problem of universality for 
different systems (Sec.~\ref{sec-univ.class}). 

Then we turn our attention to detailed study of perfect fluid 
collapse.  
After reviewing the equations of motion 
in Sec.~\ref{sec-EOM}, we first study 
self-similar solution in Sec.~\ref{sec-ss}.  
Then we study perturbations around the self-similar solution 
and confirm our picture to an extent in Sec.~\ref{sec-pert} 
by numerical study. 
In Sec.~\ref{sec-lyapunov} we establish uniqueness of the relevant
mode by the Lyapunov analysis. 
The uniqueness of the relevant mode implies that that self-similar
solution is responsible for the critical behaviour. 
So we list in Sec.~\ref{sec-results} the value of the critical
exponent $\beta$ for various $\gamma$.  
Sec.~\ref{conc} is for conclusions and discussions.

%%%%%%%%%%%%%%%%%%%%%%%%%%%%%%%%%%%%%%%%%%%%%%%%%%%%%%%%%%%%%%%%%%%%%%%%%
%%%%%%%%%%%%%%%%%%%%%%%%%%%%%%%%%%%%%%%%%%%%%%%%%%%%%%%%%%%%%%%%%%%%%%%%%
%%%%%%%%%%%%%%%%%%%%%%%%%%%%%%%%%%%%%%%%%%%%%%%%%%%%%%%%%%%%%%%%%%%%%%%%%
%%%%%%%%%%%%%%%%%%%%%%%%%%%%%%%%%%%%%%%%%%%%%%%%%%%%%%%%%%%%%%%%%%%%%%%%%
\section{Scenario based on renormalization group ideas}
\label{sec-RGscenario}
In this section we present a scenario of critical behaviour based on
renormalization group ideas, which gives clear understanding of 
every aspect of the phenomena.
We first introduce renormalization group
transformations acting on the phase space of partial differential
equations (PDEs) in Sec.~\ref{sec-RG}. 
In Sec.~\ref{sec-scenario} we give a (nonrigorous) scenario 
for the simplest case where the system has scale invariance
and a (continuously) self-similar solution plays an important role.
Even in this simplest case we can argue most of the essential features 
of the critical behaviour.  This case covers 
a perfect fluid collapse treated in detail in the subsequent
sections. 
We then proceed in Sec.~\ref{sub-rgflow} to make our scenario into a rigorous 
mathematical framework which starts {}from basic 
assumptions, based on renormalization group 
ideas. 
The case where discrete self-similar solution becomes relevant such as
scalar field collapse is understood essentially in the same way, 
as is explained in Sec.~\ref{sec-discr.SS}.
In general cases where the system is not scale invariant, 
the renormalization group transformation 
drives the equations of motion to a fixed point. 
Renormalization group method then tells us which systems can exhibit the 
same critical behaviour, as will be explained in 
Sec.~\ref{sec-univ.class}.
As examples, we consider perfect fluid with a modified equation of state, 
and also show that scalar fields with any potential 
term (which depends on the scalar field only) 
should show the same critical behaviour.

Currently, there seem to be two kinds of renormalization 
group type analysis of PDEs,  \cite{GMOL90} and \cite{BKL94}.  
Although these two are certainly philosophically related, they 
differ in details.  We here 
adapt the approach of \cite{BKL94}, and extend their work to 
cases where the critical behaviour is observed.  It would also be very 
interesting to consider applications of the method of \cite{GMOL90}.

%%%%%%%%%%%%%%%%%%%%%%%%%%%%%%%%%%%%%%%%%%%%%%%%%%%%%%%%%%%%%%%%%%%%%%%%%
%%%%%%%%%%%%%%%%%%%%%%%%%%%%%%%%%%%%%%%%%%%%%%%%%%%%%%%%%%%%%%%%%%%%%%%%%
\subsection{Renormalization group transformation}
\label{sec-RG}
We give a general 
formalism which deals with time evolution of initial data 
considered as a flow of a 
renormalization group transformation.  
The argument is general but the 
notations are so chosen in order to suit the case 
considered in the subsequent sections.
In particular, we explicitly consider only the case of spherical 
symmetry, although our framework can in principle be applied to more 
general cases in $n$-space dimensions with suitable changes: The 
independent space variable $r$ is increased to $r_1, r_2, \ldots, r_n$, and 
the differential operators $\dot{\cal R}$ and $\dot{\cal T}$ 
become partial differential operators on a space of functions of $n$ 
variables.

We are interested in the time evolution of 
$n$ unknown functions $u=(u_1, u_2,..., u_{n})$, which are 
{\em real-valued}  
functions of time $t$ and spatial coordinate $r$, and which 
satisfy a system of partial differential equations (PDEs): 
\begin{equation}
  \label{eq:evolution}
  L\left(u, \frac{\partial u}{\partial t}, 
  \frac{\partial u}{\partial r}, t, r \right)=0. 
\end{equation} 
This real formulation does not restrict applicable systems in any way, 
because any 
PDEs for complex unknown functions can be rewritten 
in the real form by doubling the number of components.  
We employ this real form, because this makes the 
dimensional counting (e.g. number of relevant modes) simpler and 
clearer.   See a Remark after Claim~1 of Sec.~\ref{sec-sen.critical}.  
(However, in practical problems, e.g. complex scalar 
fields, one could proceed without employing the real formulation, as long as 
some caution is taken in counting dimensions.)  

The time evolution of this system is determined once one specifies the 
values of $u(t, \cdot)$ at the initial time $t = -1$. 
We call the space of functions $u(t, \cdot)$ (at fixed time $t$) 
the {\em phase space} $\Gamma$, which is considered as a real 
vector space.  
We leave the definition of $\Gamma$ (e.g. how smooth functions 
in $\Gamma$ should be) 
rather vague, until we develop more detailed theory of 
time evolution in Sec.~\ref{sec-scenario}.

To introduce a renormalization group transformation, we first define 
a {\em scaling transformation} ${\cal S}(s, \alpha, \beta)$
which depends on real parameters $s, \alpha = (\alpha_{1}, 
\alpha_{2}, \ldots, \alpha_{n})$, and
$\beta$:
\begin{equation}
        {\cal S}(s, \alpha, \beta): 
        u_{i}(t,r) \mapsto u_{i}^{(s, \alpha, \beta)}(t,r) 
        \equiv e^{\alpha_{i} s} u_{i}( e^{- s} t, e^{- \beta s} r)
        \label{eq:sctrdef}
\end{equation}
(As will be clear $\beta$ here is related with, but not equal to, 
$\beta_{\rm BH}$.) 

Now we make the fundamental

\bigskip\noindent {\bf Assumption~S (Invariance under scaling)}
        The system of PDEs (\ref{eq:evolution}) are invariant 
        under the scaling transformation (\ref{eq:sctrdef})  
        in the following sense: 
        If $u$ is a solution of (\ref{eq:evolution}), then 
        for suitably chosen constants $\alpha, \beta$, and 
        for an arbitrary scale parameter $s$ (close to $0$), the new 
        function $u^{(s, \alpha, \beta)}$ is also a solution of 
        the system (\ref{eq:evolution}).  

\bigskip

In the following we denote by $\alpha, \beta$ the constants which 
make the above assumption hold for the system 
of PDEs (\ref{eq:evolution}), and denote 
$u^{(s, \alpha, \beta)}$ simply by $u^{(s)}$ unless otherwise stated. 
We consider only those systems for which the above Assumption~S holds 
in Sec.~\ref{sec-RG} through Sec.~\ref{sec-discr.SS}.  
In Sec.~\ref{sec-univ.class}, we consider 
systems for which Assumption~S does not hold, and address the problem 
of universality in its wider sense. 

We now define a {\em renormalization group transformation (RGT)}\/
acting on the phase space $\Gamma$ as
\begin{eqnarray}
    && {{\cal R}_{s}}: U_{i} (\xi)=u_{i} (-1,\xi) \nonumber \\ 
        && \hspace{2mm} \mapsto 
    U_{i}^{(s)}(\xi)= u_{i}^{(s)} (-1, \xi) = 
    e^{\alpha_{i} s} u_i(-e^{- s}, e^{- \beta s}\xi),
\end{eqnarray}
where $\alpha, \beta$ are constants of Assumption~S, $s > 0$, 
and $u$ is a solution of the PDE system. 
That is, $U^{(s)}$ is given by developing the initial data 
$u(-1, r) \equiv U(r)$ {}from $t=-1$ to $t=-e^{- s}$ by the PDE, 
and rescaling the spatial coordinate $r$ and the unknown functions 
$u_{i}$. 
We have written the argument of 
$U$ as $\xi$ rather than $r$, in order to draw attention to the fact
that the physical (or geometrical) length is given by 
\begin{equation}
r=e^{- \beta s}\xi
\label{eq-xi-def}
\end{equation}
rather than by $\xi$ itself.

Because of our assumption of the invariance of the PDEs under scaling, 
${\cal R}_{s}$ forms a (semi) group:
\begin{equation}
        {\cal R}_{s_{1} + s_{2}} 
        = {\cal R}_{s_{2}} \circ {\cal R}_{s_{1}}
        \label{eq:semigrp} .
\end{equation}
{}From this it follows immediately that the $s$-derivative of 
${\cal R}_{s}$ at $s=0$ is in fact an infinitesimal generator 
of ${\cal R}_{s}$:
\begin{equation}
        {\dot{\cal R}} 
        \equiv \lim_{s \rightarrow 0} \frac{{\cal R}_{s} - 1}{s}, 
        \qquad 
        {\cal R}_{s} = \exp \left ( s {\dot{\cal R}} \right ) .
        \label{eq:DRdef}
\end{equation}

A {\em fixed point}\/ of the renormalization group
$\{{{\cal R}}_{s}\ \,|\, s\in{\Bbb R}\}$ is a point $U^{*}$
in $\Gamma$ satisfying ${\cal R}_{s} (U^*) = U^*$ for 
all $s > 0$, and 
can also be characterized by ${\dot{\cal R}} (U^{*}) =0$. 
A function $u$ is called {\em self-similar}\/ with 
parameters $(\alpha, \beta)$ if it satisfies 
$u(t,r) = u^{(s, \alpha, \beta)}(t,r)$. 
Each self-similar solution $u_{\rm ss}$ of the PDEs 
with parameters $(\alpha, \beta)$ of Assumption~S is related
to a fixed point
$U^{*}$ of ${\cal R}_{s}$ by
$u_{\rm ss}(t,r) = (-t)^{\alpha} U^{*}(r (-t)^{-\beta})$.

The {\em tangent map}\/ (or {\em Fr\'echet derivative}\/) 
${{\cal T}}_{s, U}$ of ${\cal R}_s$ at $U$ is defined by
\begin{equation}
        {{\cal T}}_{s, U} ( F) \equiv \lim_{\epsilon \rightarrow 0} 
        \frac{ {\cal R}_{s} (U + \epsilon F ) - {\cal R}_{s} U }
        {\epsilon} .  
        \label{eq:ind.tan}
\end{equation}
Note that ${{\cal T}}_{s, U}$ is a {\em linear}\/ operator for 
fixed $s$ and  $U$, in contrast with $ {\cal R}_{s}$.  
The above definition suggests the following formal relation: 
\begin{equation}
        {\cal R}_{s} ( U + F)  
        = {\cal R}_{s} (U) + {\cal T}_{s,U} (F) 
        + O(F)^{2}
        \label{eq:RTrel}
\end{equation}
where $O(F)^{2}$ denotes a term whose norm is of order $\| F\|^{2}$ 
(with some suitable norm $\|\cdot\|$). 

We define the $s$-derivative of ${{\cal T}}_{s, U}$ 
at $U$ as
\begin{equation}
        {\dot{\cal T}}_{U} \equiv \lim_{s \rightarrow 0} 
        \frac{{\cal T}_{s, U} - 1}{s}, 
        \label{eq:DTRdef}
\end{equation}
which is formally related with ${\dot{\cal R}}$ by 
\begin{equation}
        {\dot{\cal R}} ( U + F) = {\dot{\cal R}} (U) 
        + {\dot{\cal T}}_{U} (F) + O(F)^{2} .
        \label{eq:DRDTrel}
\end{equation}

At a fixed point $U^*$ of ${\cal R}_{s}$, (\ref{eq:ind.tan}) 
simplifies to 
\begin{equation}
       {{\cal T}}_{s} ( F) \equiv \lim_{\epsilon \rightarrow 0} 
        \frac{ {\cal R}_{s} (U^*  + \epsilon F ) - U^* }
      {\epsilon} , 
        \label{eq:ind.tan2}
\end{equation}
and we denote the tangent map at $U^{*}$ simply by  
${{\cal T}}_{s}$ in the following.
${{\cal T}}_{s}$ (at $U^{*}$) forms a (semi)group 
like $ {\cal R}_{s}$, 
and can be represented in terms of its infinitesimal generator 
${\dot{\cal T}}$ as
\begin{equation}
        {\dot{\cal T}} \equiv {\dot{\cal T}}_{U^{*}} 
        = \lim_{s \rightarrow 0} 
        \frac{{\cal T}_{s} - 1}{s}, 
        \qquad 
        {\cal T}_{s} = \exp \left ( s {\dot{\cal T}} \right ) .
        \label{eq:DTdef}
\end{equation}

An {\em eigenmode}\/ $F$ of $\displaystyle {\dot{\cal T}}$ with eigenvalue 
$\kappa$ 
is a (complex-valued) function satisfying ($\kappa \in {\Bbb C}$) 
\begin{equation}
        {\dot{\cal T}} F = \kappa F.
        \label{eq:pert0}
\end{equation} 
Note that we here consider eigenmode (and spectral) problem on  
complexification of $\Gamma$, i.e. we 
allow complex eigenvalues and eigenvectors.   
This is necessary for our later analysis.
Although complex eigenvectors are not elements of $\Gamma$, we
can express  
vectors in $\Gamma$ by taking the real part of a linear combination of 
complex eigenvectors. 

These modes determine the flow of the RGT near the fixed point $U^{*}$%
\footnote{%
Mathematically, in order to determine the 
behaviour of the flow around $U^{*}$, we have to know not only 
the point (discrete) spectrum but also the continuous spectrum 
of $\dot{\cal T}$. See Assumption~L1 of Sec.~\ref{sub-rgflow} for details.
}.
A mode with ${\rm Re \,} \kappa>0$, a {\em relevant mode}, 
is tangent to a flow diverging {}from $U^*$, 
and one with ${\rm Re \,} \kappa<0$, an {\em irrelevant mode}, 
to a flow converging to it. 
A ${\rm Re \,} \kappa=0$ mode is called {\em marginal}.

\bigskip\noindent 
{\bf Remark.} 
Many equations of gravitational systems, including those of 
the perfect fluid, are not exactly of the 
form of (\ref{eq:evolution}).  More precisely, we have two kinds of 
unknowns $u = (u_{1}, u_{2}, \ldots, u_{n})$ and $c = (c_{1}, c_{2}, 
\ldots, c_{m}$), which satisfy the following PDEs of two different 
classes:
(1)the evolution equations: 
\begin{equation}
  \label{eq:evolution2}
  L_E\left(u, c, \frac{\partial u}{\partial t}, 
  \frac{\partial u}{\partial r},  \frac{\partial c}{\partial r}, r\right)=0, 
\end{equation}
and (2) the constraint equations: 
\begin{equation}
  \label{eq:constr}
  L_C\left(u, c, \frac{\partial c}{\partial r}, r \right)=0 .  
\end{equation}
In this case, the well posed Cauchy problem is to specify the initial 
values of $u_{i}(-1, \cdot)$ for $i= 1, 2, \ldots, n$ (and 
appropriate boundary conditions for $c_{i}$'s).  
In this sense, the phase space variables are $u_{i}$'s, while 
$c_{i}$'s should be considered as constraint variables.  

Solving (\ref{eq:constr}) for constraint variables $c_{i}$'s 
in terms of integrals of $u_{i}$'s and substituting them into 
(\ref{eq:evolution2}) yields 
evolution equations for $u_{i}$'s of the type of (\ref{eq:evolution}), 
with the difference that $u$ in the argument of $L$ now 
represents some complicated functionals (integrals) of $u_{i}$'s.  
In this form, we can apply our formalism of the renormalization 
group transformation.

\bigskip\noindent 
{\bf Remark.}
In the above, we have presented the formalism in $(t,r)$ coordinate  
with unknowns $u_{i}$, in order to 
directly reflect our physical interpretation.
However, mathematical treatment can in principle be made much 
simpler by introducing new variables
\begin{eqnarray}
        && x \equiv \ln r - \beta \ln(-t), \qquad 
        s \equiv - \ln (-t), \nonumber \\
        && v_{i}(s,x) \equiv e^{\alpha_{i} s} u_{i}(t,r) .
        \label{eq-newvar}
\end{eqnarray}
In terms of these, the scaling transformation 
${\cal S}(s', \alpha, \beta)$ becomes a simple translation in $s$: 
\begin{equation}
        v_{i}(s,x) \mapsto v_{i}^{(s')}(s,x) 
        \equiv  v_{i}(s+s', x), 
        \label{eq:new-sc-tr}
\end{equation}
and the parameters $\alpha, \beta$ disappear. 
One can thus identify the scale parameter $s'$ with the new ``time'' 
coordinate $s$.  
The assumed scale invariance of the system implies that
the EOM do not include $s$ explicitly.
Now ${\dot{\cal R}}$ and  ${\dot{\cal T}}$ 
can be expressed by  usual $s$-derivatives on the functions $v_{i}$.  
In \cite{KHA95}, we formulated our renormalization group approach directly 
in terms of these variables. 

% Since relativistic systems have general covariance,
% by suitable choice of $u$'s,
% we can write the EOM to be {\em invariant}\/ under transformation
% of the time coordinate, $u(t,r)\mapsto u(e^{-s'}t,r)$,
% as is shown with a concrete example in Sec. \ref{sec-EOM}.
% (Note however that boundary conditions for constraint variables, or
% equivalently, the corresponding functionals,  are altered.)
% This together with the scale invariance implies invariance
% under transformation $u(t,r)\mapsto u(t,e^{\beta s'}r)$. 
% In terms of coordinates $(s,x)$, the EOM are invariant under 
% $v(s,x)\mapsto v(s,x+x')$, provided that $\beta\ne0$.
% It follows that the EOM in $(s,x)$ are {\em autonomous}, i.e., 
% they do not contain $s$ nor $x$ explicitly.
% To summarize, scale invariance together with general covariance imply
% that the EOM in coordinates $(s,x)$ comprise an autonomous system.

%%%%%%%%%%%%%%%%%%%%%%%%%%%%%%%%%%%%%%%%%%%%%%%%%%%%%%%%%%%%%%%%%%%%%%%%%
%%%%%%%%%%%%%%%%%%%%%%%%%%%%%%%%%%%%%%%%%%%%%%%%%%%%%%%%%%%%%%%%%%%%%%%%%

\subsection{Scenario}
\label{sec-scenario}

In this subsection we present our scenario of critical phenomena in
gravitational collapse based on the renormalization group ideas.
To describe our claims clearly, we first classify numerically 
observed (U) {\em universality}\/ and (S) {\em scaling}\/ into two: 
(1) those of the
critical space-time and (2) those of the critical exponent. 
Concerning the critical space-time, it has been observed that 
\begin{description}
\item[(S1)] 
near-critical (but not exactly critical) solutions always once approach, 
but eventually deviate {}from, a self-similar space-time, 
\item[(U1)] which is unique. 
\end{description}

And concerning the critical exponent, it has been observed that 
\begin{description}
\item[(S2)] For slightly supercritical solutions,  the mass of the
black hole formed  is expressed by a scaling law 
$M_{\rm BH} \propto (p-p_{c})^{\beta_{\rm BH}}$, and 
\item[(U2)] the value of the critical exponent $\beta_{\rm BH}$ 
is the same for all
generic one-parameter families in $\Gamma$. 
\end{description}
We could add another kind of universality, universality over different
systems. 
It has been found \cite{Cho94talk} that the system for some
potentials such as $V(\phi)=\mu \phi^2/2$, etc., also have the above
properties with the same $\beta$ as the minimally coupled massless
scalar field.  
This suggests that the following universality holds:
\begin{description}
\item[(U3a)]
  There are some models which show the same critical behaviour 
  (same self-similar solution, same critical exponent). In particular,
  in scalar field collapse, the critical phenomena is the same for
  any potential $V$, or at least, for some potentials $V$.
\end{description}

Note that the term ``universality'' is used in two ways.  In (U1) 
and (U2), it refers to the independence on the initial data (or 
a family of initial data), while in (U3a), it refers to the 
independence on the system's EOM (i.e. details in the definition of 
the model considered).  In statistical mechanics, where the term was 
used for the first time in a similar context, 
``universality'' refers to the latter (although there is no ``initial 
data'' in equilibrium statistical mechanics and thus it is impossible 
to have a counterpart for the former).  It is one of our main purpose 
in this paper to give sufficient conditions under which the two kinds 
of universality hold.  

Historically, because of closeness of numerical values of the critical
exponents of a scalar field~\cite{Cho93}, a radiation
fluid~\cite{EvCo94}, and gravitational waves~\cite{AbEv93}, 
the following universality in still wider sense,  though stated 
less precisely, was speculated. 
\begin{description}
\item [(U3b)]  
        The value of $\beta_{\rm BH}$ is the same over many systems,
        including a scalar field, a radiation fluid, and gravitational waves.
\end{description}
The precise value 0.35580192... in Ref.~\cite{KHA95} and 
the values for scalar
field collapse (0.37) and gravitational wave collapse (0.38) 
shows that this is not true.

We now present our scenario through two claims, and their 
derivations.  We do this on two different levels.  First, in 
Sec.~\ref{sec-sen.critical}  (resp. Sec.~\ref{sec-sen.exponent}) we 
present claims concerning qualitative (resp. quantitative) pictures on 
the critical behaviour, and present their (nonrigorous) 
`derivation' based on naive linear perturbation ideas.  
Presentation in this Sec.~\ref{sec-scenario} is intended to provide 
basic ideas, so the following Claims and their `proofs' are 
{\em not} intended to be precise mathematical statements.  
In particular, naive perturbation 
is not sufficient to deal with long time asymptotics rigorously, and 
this is where the ``renormalization group'' idea comes in.  
We thus present  in Sec.~\ref{sub-rgflow} 
more detailed, rigorous derivation of Theorem~\ref{prop-final},  
a rigorous version of Claims~1 
and 2,    based on 
renormalization group philosophy.

A mathematical remark about the phase space is in order.  
In order to  formulate the time evolution, 
we consider our phase space $\Gamma$ to
be a {\em real}\/ Banach space of functions\footnote{%
If there are some gauge degrees of freedom, which is often the case for
general relativistic systems, we should consider $\Gamma$ as a space of 
gauge equivalence classes of functions.  Or we could fix the gauge and 
consider $\Gamma$ to be a usual function space.  
In the latter case,  we should 
be careful in gauge-fixing so as not to count ``gauge modes'' as physically
meaningful eigenmodes in linear perturbation analysis.
} 
$U(\xi)$'s  which 
satisfy certain decay and smoothness properties, 
with a norm $\| \cdot \|$ under which  the phase space is complete. 
Concrete choice of the decay and smoothness properties should be 
determined on physical grounds for each models, together with the 
requirement that $\Gamma$ is invariant under the time evolution. 
The term ``approach'' and ``deviate'' in the statement (S1) should be 
interpreted in terms of the norm of the phase space, although 
other notion of convergence, e.g. pointwise convergence, could 
equally work.  Precise final definition should be given based on 
rigorous mathematical analysis, which is beyond the scope of the 
present paper. 

In the concrete cases discussed 
later, the fixed point $U^*$ does not represent an 
asymptotically flat space-time.  
Though one is usually interested in asymptotically flat space-time in
gravitational collapse, we define our phase space $\Gamma$ to
be sufficiently large to include $U^*$.
Asymptotically flat data are included in the phase space (if one
chooses the functions $U_i$ appropriately) and form  a subspace of
$\Gamma$.

%%%%%%%%%%%%%%%%%%%%%%%%%%%%%%%%%%%%%%%%%%%%%%%%%%%%%%%%%%%%%%%%%%%%%%%%%%
\subsubsection{Critical space-time} 
\label{sec-sen.critical}

We first consider the qualitative behaviour of \hbox{(near-)} critical
solutions.   

% \bigskip\noindent  {\bf Claim 1.} 
%   {\em 
\begin{claim}
\label{claim-1}
  Suppose there is a fixed point $U^*$,  
  with no marginal modes.  If (S1) and (U1) 
  hold with this $U^*$, then there should be a {\em unique} relevant 
  mode with a {\em real} eigenvalue 
  for this $U^*$.  Conversely, if $U^*$ has 
  a unique relevant mode, then (S1) and (U1) hold with 
  this $U^*$, at least for all the initial data 
  close to $U^*$.  (Except, of course, for the exactly critical 
  initial data.) 
\end{claim} 
%  }%end of \em 
%
%\bigskip 

\noindent {\bf Remark.}
Because we are considering real PDEs, the ODEs
\footnote{%
PDEs, when one is considering 
in more than two dimensions, as remarked at the beginning of 
Sec.~\ref{sec-RG}
} for eigenmodes are 
also real, and thus all eigenmodes with complex 
eigenvalues (if any) appear in complex conjugate pairs.  This 
means that if such a system has a unique relevant mode, 
the eigenvalue must be real. 

\bigskip \noindent {\bf Remark.} (Reason for employing the 
real formulation.) 
One could proceed without employing the real formulation.  However, 
in this case, dimension counting becomes more complicated; because 
the desired critical behaviour is observed only when the relevant 
subspace is (real) one-dimensional, one has 
to see whether a given eigenvector spans a subspace of (real) 
dimension 1 or (real) dimension 2.  By reducing the system of PDEs in 
its real form, we can avoid such subtleties.  We emphasize again that 
this does not restrict the applicability of our claims in any way.  
For example, a system of PDEs (for complex unknowns) which 
has exactly one relevant mode with a complex eigenvalue 
is easily seen to have a relevant invariant subspace which 
is (real) two-dimensional, thus leading to the same conclusion 
(i.e. this does not show usual critical behaviour) as is 
obtained {}from the real formulation. Based on this observation, we 
in this paper drop ``Re'' in the exponent relation 
$\beta_{\rm BH} = \beta / \kappa$.  

\bigskip

%%%%%FIGFIGFIGFIGFIGFIGFIGFIGFIGFIGFIGFIGFIGFIGFIGFIGFIGFIGFIG
%%%%%FIGFIGFIGFIGFIGFIGFIGFIGFIGFIGFIGFIGFIGFIGFIGFIGFIGFIGFIGFIG
\begin{figure}
\begin{center}
{\BoxedEPSF{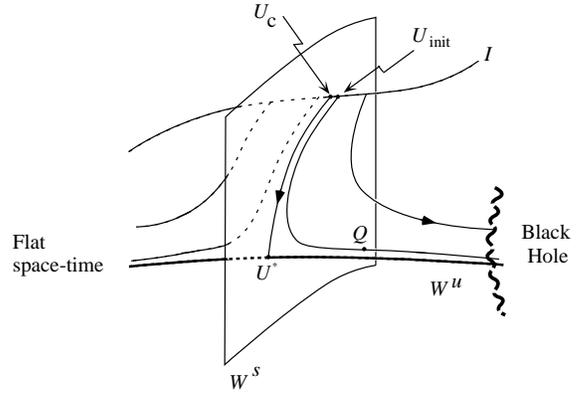 scaled 450} }
%\vspace*{20mm} 
%{\bf // PLACE Fig.~\ref{fig-flow0} HERE. //} 
\end{center}
\caption{Schematic view of the global renormalization group flow.}
\label{fig-flow0}
\end{figure}
%%%%%FIGFIGFIGFIGFIGFIGFIGFIGFIGFIGFIGFIGFIGFIGFIGFIGFIGFIGFIGFIG
%%%%%FIGFIGFIGFIGFIGFIGFIGFIGFIGFIGFIGFIGFIGFIGFIGFIGFIGFIGFIGFIG
%

{\em `Proof.'}\/
Suppose $U^*$ has a unique relevant mode. Then 
the renormalization group flow around the fixed point $U^*$  is 
contracting 
except for the direction of the relevant mode (Fig.~\ref{fig-flow0}), at
least in some neighbourhood ${\cal N} \subset \Gamma$ 
around $U^*$.  
Thus there will be a {\em critical surface}\/ or 
a {\em stable manifold}\/ $W^{\rm s}(U^*)$ of the fixed point 
$U^*$, of codimension one, 
whose points will all be driven towards $U^*$. 
And there will be an {\em unstable manifold}\/ $W^{\rm u}(U^*)$ 
of dimension one, whose 
points are all driven away {}from $U^*$.  
A one-parameter family of initial data $I \subset {\cal N}$ 
will in general intersect 
with the critical surface,  and the intersecting point $U_{\rm c}$ 
will be driven to $U^*$ under the RGT: 
\begin{equation}
        \lim_{s\rightarrow\infty}
        \| U_{\rm c}^{(s)}-  U^* \| =0 .  
        \label{eq:crit}
\end{equation}
Thus, $U_{\rm c}$ is the initial data with critical parameter 
$p_{c}$.  
An initial data $U_{\rm init}$ in 
the one-parameter family $I$, which is close to $U_{\rm c}$
will first be driven towards $U^*$ along the stable manifold, 
but eventually be driven away along the unstable manifold.
Therefore, claims (S1) and (U1) hold.

To show the converse, first suppose $U^*$  had no relevant mode. 
Then any initial data (sufficiently close to $U^*$) would all 
be driven towards $U^*$ until 
 finally be absorbed into $U^*$, which contradicts (S1).  
Next suppose $U^*$ had more than one 
relevant modes.  Then  the stable manifold of 
$U^*$ would  have codimension more than one, 
and thus a generic one-parameter 
family of initial data would not intersect with the stable manifold of 
$U^*$. 
This means that a generic critical or near-critical 
solution would not approach the 
self-similar solution $U^*$; instead, they would be 
driven towards a different self-similar solution with a 
unique relevant mode. Thus claims (S1) and (U1) do not hold. 
\hfill$\Box$

\bigskip\noindent 
{\bf Remark.}  
Eq. (\ref{eq:crit}) states that 
the fixed point $U^*$ is on the closure of the subspace of $\Gamma$ 
which consists of asymptotically flat data.
This explains, in rather precise terms, 
that the self-similar solution plays such an important
role that its perturbation gives an exact critical exponent
even when one is interested only in asymptotically flat collapses.
Asymptotically flat data can converge to the self-similar fixed point 
with infinite mass in the above sense, 
even though the ADM mass conserves.  
It should be emphasized that there is no contradiction 
involved here.  That is, the mass measured in the unit of the 
{\em scaled} coordinate becomes infinite as the space-time approaches 
the self-similar solution, while the physical mass
measured in the unit of the {\em original} coordinate remains finite.

\bigskip\noindent 
{\bf Remark.}  
Some reader might doubt the validity of our claim (\ref{eq:crit}), 
because often 
the critical space-time satisfies $\lim_{r \rightarrow \infty} a(r) = 
a_{\infty} > 1$.  However, it is not difficult to find norms 
$\| \cdot \|$, for which (\ref{eq:crit}) holds; e.g. take 
\begin{equation}
        \| a \| \equiv \sup_{r >0} (1+r)^{-1} | a(r) | , \qquad \mbox{or} 
        \quad \equiv \sup_{r>0} e^{-r} |a(r) | .
        \label{eq:norm-ex}
\end{equation}
The remaining, highly nontrivial, question is whether the 
Assumptions in Sec.~\ref{sub-rgflow} are satisfied in this norm, because 
the spectrum of perturbations can depend on 
the choice of the norm (and thus on the function space)\footnote{%
Although this dependence, especially on the behaviour as 
$r \rightarrow \infty$, may not be so strong in the concrete 
case of perfect fluid, because the sonic point places an 
effective boundary condition
}.  In other words, 
the norm and the function space $\Gamma$ 
should be so chosen that the Assumptions 
hold (and that $\Gamma$ includes $U^{*}$): if we can find such a norm and 
$\Gamma$, the critical behaviour should be observed; this is what is 
proved in Sec.~\ref{sub-rgflow}.  
The question of the right choice of the norm must be answered 
for each models under proper mathematical consideration. 
We emphasize that our goal is to present a {\em general} framework 
(physically reasonable sufficient conditions) 
under which the critical behaviour is bound to happen, which is 
independent of the details of specific models.  

\bigskip\noindent 
{\bf Remark.} 
Fixed points with more than one relevant modes are
responsible for the so-called {\em multicritical behaviour} in statistical 
mechanics.  We give examples of self-similar solutions with more 
than one relevant modes (and thus are not responsible for generic 
critical behaviour) in section~\ref{subsec-pp.other-ss}. 
Another example is provided by the self-similar solutions of 
complex scalar fields \cite{HiEa95}.  
In this model the continuous self-similar solution 
has three relevant modes, and thus is irrelevant for a generic 
critical behaviour.  In other words, we have to adjust three 
parameters (instead of one) in the initial condition, to observe a 
critical behaviour governed by this self-similar solution.

%%%%%%%%%%%%%%%%%%%%%%%%%%%%%%%%%%%%%%%%%%%%%%%%%%%%%%%%%%%%%%%%%%%%%%%%%%
\subsubsection{Critical exponent} 
\label{sec-sen.exponent}

We now turn our attention to the quantitative aspect of the 
critical behaviour, i.e. the critical exponent $\beta_{\rm BH}$.

%   {\bf Claim 2.}
%   {\em 
\begin{claim}
\label{claim-2}
  If the relevant mode is unique with a real eigenvalue $\kappa$, 
  the black hole mass $M_{\rm BH}$ satisfies 
    \begin{equation}
        M_{\rm BH} \propto (p-p_{c})^{\beta_{\rm BH}} 
        \label{eq:eq-exp}
    \end{equation}
    with 
    \begin{equation}
        \beta_{\rm BH} = \frac{\beta}{\kappa} 
        \label{eq:eq-exp0} 
    \end{equation}
    for slightly supercritical solutions, 
    where $\beta$ is the scaling exponent of the scaling transformation 
    (\ref{eq:sctrdef}). 
    Here $f(p) \propto (p-p_{c})^{\beta_{\rm BH}}$ means there exist 
    $p$-independent positive constants $C_{1}, C_{2}$ such that 
    $C_{1} < f(p) (p-p_{c})^{-\beta_{\rm BH}} < C_{2}$ holds for $p$ 
    sufficiently close to $p_{c}$.
\end{claim}
% }%end of \em 

\bigskip

{\em `Proof.'}\/
The claim, in particular the exactness of the relation 
(\ref{eq:eq-exp0}), is one of the main conclusions of this paper.  
The derivation of this claim is rather lengthy, although it is 
a variation on rather 
standard treatment of similar problems in (rigorous) renormalization
group analysis of critical behaviour in 
statistical mechanics.  We first explain the rough 
idea, and then explain each step in detail. 
 
We first consider the fate of an initial data $U_{\rm init}$ in 
the one-parameter family, which is  close to $U_{\rm c}$ 
 ($\epsilon = p - p_{c}$): 
\begin{equation}
        U_{\rm init} = U_{\rm c} + \epsilon F , \qquad \| F \| =1.  
\end{equation}
The data will first be driven 
towards $U^*$ along the critical surface, but eventually be 
driven away along the unstable manifold, until 
it finally blows up and forms a black hole, 
or leads to a flat space-time.

%%%%%FIGFIGFIGFIGFIGFIGFIGFIGFIGFIGFIGFIGFIGFIGFIGFIGFIGFIGFIGFIG
%%%%%FIGFIGFIGFIGFIGFIGFIGFIGFIGFIGFIGFIGFIGFIGFIGFIGFIGFIGFIGFIG
\begin{figure}
%\vspace*{60mm} 
\begin{center}
{\BoxedEPSF{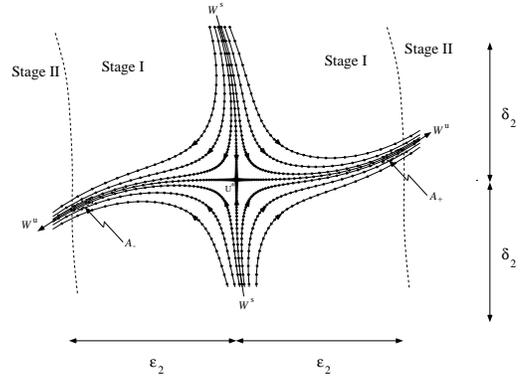 scaled 400} }
\end{center}
%{\bf // PLACE Fig.~\ref{fig-flow2} HERE. //} 
\caption{Schematic view of the flow.  
Each dot denotes approximate location of the 
flow under iterations of ${\cal R}_{\sigma}$, starting {}from several 
different initial data. Note that the points  are most densely 
distributed around the fixed point $U^{*}$. 
The decomposition into two 
stages I and II is shown by vertical dotted lines.  
The set $A_{\pm}$ will be used in our detailed 
analysis in Sec.~\protect\ref{sub-rgflow}. 
}
\label{fig-flow2}
\end{figure}
%%%%%FIGFIGFIGFIGFIGFIGFIGFIGFIGFIGFIGFIGFIGFIGFIGFIGFIGFIGFIGFIG
%%%%%FIGFIGFIGFIGFIGFIGFIGFIGFIGFIGFIGFIGFIGFIGFIGFIGFIGFIGFIGFIG
%
%

To trace the time evolution, we divide the whole evolution into 
two stages: I ($s \leq s_{\rm I}$)  and II ($s > s_{\rm I}$).  
See Fig.~\ref{fig-flow2}:%
\footnote{%
More general situation, where the initial data is not close to $U^{*}$ 
but is close to $W^{\rm s}(U^{*})$,  can be treated in a similar way  
as long as the flow along $W^{\rm s}(U^{*})$ is regular.  
This is because the time spent before it reaches Stage~I 
is finite.} 
the stage I corresponds to 
$U_{\rm init} \rightarrow Q$,  
while stage II to $Q \rightarrow$(black hole). 
The stage I is the portion 
of the journey where $U_{\rm init}^{(s)}$ is close to $U^*$ and 
we can use the perturbation.   Stage II is the region where 
perturbation breaks down, and $s_{\rm I}$ is so chosen to make this 
separation possible.  
Concretely, we choose $s_{\rm I}$ as the smallest $s$ such that 
$\| U_{\rm init}^{(s)} - U^{*} \|$ becomes of order 1 
(after $U_{\rm init}^{(s)}$ begins to deviate {}from $U^{*}$).  
As will be explained in detail, near-critical 
solution $U_{\rm init}^{(s)}$ spends most of its time in stage I, 
and the time of stage II is relatively short.  

Based on the analysis of the flow, we then derive the exponent 
by interpreting the result in our original coordinate. 
As will be explained, what 
matters in determining the exponent is the behaviour of the flow in 
stage I.  

\medskip \noindent {\em Step~1. Flows in stage I:}

The analysis of this stage is essential, because the flow spends 
most of its time at this stage, and gives the value of the exponent. 

For Stage I, we can rely on linear perturbations to get
\begin{eqnarray}
        U^{(s_{\rm I})}_{\rm init} & = &  {\cal R}_{s_{\rm I}} U_{\rm init}
        = {\cal R}_{s_{\rm I}} ( U_{\rm c} + \epsilon F ) 
        \nonumber \\ 
        & = & U_{\rm c}^{(s_{\rm I})} + \epsilon\, {{\cal T}}_{s_{\rm I}} F 
        + O(\epsilon^2) \nonumber \\ 
        & = & U_{\rm c}^{(s_{\rm I})} + \epsilon\, e^{\kappa s_{\rm I}}  
        F_{\rm rel} 
        + O(\epsilon^2),
        \label{eq:lin1}
\end{eqnarray}
where  $F_{\rm rel}$ is roughly the component of the relevant mode 
in $F.$  
The derivation of (\ref{eq:lin1}) requires some calculations, together 
with suitable assumptions on invariant manifolds of ${\dot{\cal T}}$.  
For the ease of reading, we postpone this derivation to 
section~\ref{subsub-rg.quan2}, and proceed assuming (\ref{eq:lin1}) 
is correct.

Due to (\ref{eq:crit}), we have (for large $s_{\rm I}$)
\begin{equation}
        U^{(s_{\rm I})}_{\rm init}  \simeq 
        U^*(\xi) + \epsilon e^{\kappa s_{\rm I} }F_{\rm rel} . 
        \label{eq:RG.result}
\end{equation}
Our choice of  $s_{\rm I}$ implies 
\begin{equation}
        \label{eq:s-0value}
        | \epsilon | e^{ \kappa s_{\rm I}} \approx 1.  
\end{equation}
(Here $1$ means a positive small which is independent of 
$\epsilon$.)  Note that $s_{\rm I}$ diverges logarithmically in $\epsilon$ 
as $\epsilon \rightarrow 0$.

\bigskip\noindent {\em Step~2. Flows in stage II:} 

We further evolve $U^{(s_{\rm I})}_{\rm init}$ until it finally diverges 
or leads to a flat space-time.  
At this point, we {\em assume} that the phase space $\Gamma$ has a 
rather simple structure, i.e. all points  (at least in 
a neighbourhood of $U^{*}$) on one side of 
$W^{\rm s}(U^{*})$ blow up and lead to a black hole, while those 
on the other side disperse and lead to a flat space-time.  
Once one admits this 
nontrivial assumption on the {\em global} structure of $\Gamma$, 
stage II becomes  relatively 
trivial, because of the large second term of (\ref{eq:RG.result}).  
That is,  
the data $U^{(s_{\rm I})}_{\rm init}$ differs {}from $U^*$ so much that 
one can tell the fate of this data depending on the sign of $\epsilon$. 
In particular, if a 
black hole is formed, the solution will blow up in a finite, 
rather short,  time (uniform in $\epsilon$).  
This means the radius of its apparent horizon, 
and thus its mass,  will be $O(1)$ measured in $\xi$.

\bigskip\noindent {\em Step~3. Going back to the original scale:} 

Finally, we translate the above result back into our original coordinate 
$(t,r)$. 
The relation $r = \xi e^{-\beta s}$ implies  
that the radius of the apparent horizon, which is $O(1)$ measured in
$\xi$, is in fact $O(e^{-\beta s_{\rm I}})$ measured in $r$.  
So we have {}from (\ref{eq:s-0value})
\begin{equation}
         M_{\rm BH} = O(e^{-\beta s_{\rm I}}) =
        O(| \epsilon |^{ \beta / \kappa}).  
         \label{eq:MBH}
\end{equation}
Therefore the critical exponent is given {\em exactly}\/ by 
\begin{equation}
\label{eq:exp-exp}
         \beta_{\rm BH} =  \frac{\beta}{ \kappa}  . 
         \label{eq:beta.exp} 
\end{equation}
This concludes our `derivation.'
\hfill$\Box$

{\bf Remark.} 
As can be seen {}from the above 'proof,'
an intuitive understanding of the exactness of $\beta_{\rm BH}$ 
is as follows.
In the limit $\epsilon\rightarrow0$, 
the time spent in a finite region around $U^*$ diverges
whereas the time spent outside is usually finite before a black hole
is formed, if there are no other fixed points, etc near the flow.  
So the change of the profiles of the fields are dominated by the
linear stage so that the exact critical
exponent is given in terms of the eigenvalue of linear perturbation
around $U^*$.
The rigorous argument is presented in Sec.~\ref{sub-rgflow}.

%{\em Remark 1.} 
It follows  immediately {}from the above claim that conditions (S2) and
(U2) hold and the critical exponent $\beta_{\rm BH}$ is exactly given by 
(\ref{eq:beta.exp}).  The above `proofs' show that 
our claims are true not only in asymptotically 
flat cases but also in cases with any definition of mass
proportional to the typical scale of the functions $U(\xi)$.

%{\em Remark 2.} it should be noted that 
If one admits Claim 2, it would in fact  be a big surprise if the condition 
(U3b) did hold, since the critical solution of other systems, 
such as that of a scalar field system and that of a gravitational
wave, are so different that one cannot expect the eigenvalues
of the perturbations around them to be the same.  An explicit example 
of nonuniversality is given by the critical exponent $\beta_{\rm BH}$ for 
various parameter values of $\gamma$ for the perfect fluid: 
see section~\ref{sec-results}.

%%%%%%%%%%%%%%%%%%%%%%%%%%%%%%%%%%%%%%%%%%%%%%%%%%%%%%%%%%%%%%%%%%%%%%%%%%%%%%%%
%%%%%%%%%%%%%%%%%%%%%%%%%%%%%%%%%%%%%%%%%%%%%%%%%%%%%%%%%%%%%%%%%%%%%%%%%%%%%%%%
\subsection{Detailed analysis of renormalization group flows} 
\label{sub-rgflow}

Simple linear perturbations are not sufficient for 
rigorous study of the long time asymptotics of the solution.  
In this section, we describe the ``renormalization group'' 
type analysis in detail, 
which goes beyond the linear perturbation and provides firm 
grounds on our claims.  Our goal is to derive  
Theorem~\ref{prop-final}, which is a rigorous version of 
Claims~1 and 2, {}from a few ``assumptions.''  
The ``assumptions'' {}from which we start are grouped into two: 
(1) assumptions on the {\em linear} time evolution of the 
system, defined by ${\dot{\cal T}}$ (or equivalently 
${\cal T}_{s, U^{*}} = e^{s  {\dot{\cal T}} }$), and (2) certain regularity 
(smoothness) properties of ${\cal R}_{s}$ and ${\cal T}_{s}$ which 
physically look innocent. 

In the following $\sigma$  is a positive constant 
suitably chosen for both of the following assumptions to hold.
The first assumption specifies the properties of the {\em linear} part 
of the time evolution, generated by ${\cal T}_{\sigma, U^{*}}$: 

\bigskip\noindent
{\bf Assumption~L1  (Invariant subspaces of ${\cal T}_{\sigma, 
U^{*}}$)}  
        For a suitably chosen $\sigma$, 
        the whole tangent space $T_{U^*}\Gamma$ 
        is a direct sum of invariant subspaces 
        of ${\cal T}_{\sigma, U^{*}}$: 
        \begin{equation}
                T_{U^*}\Gamma = E^{\rm u}(U^{*}) \oplus E^{\rm s}(U^{*}) .
                \label{eq:Ass.1.decp}
        \end{equation}
        Here $E^{\rm u}(U^{*})$ and $E^{\rm s}(U^{*})$ are invariant 
        subspaces of ${\cal T}_{\sigma, U^{*}}$, of dimension $N$ and of 
        codimension $N$, respectively. 
%         Any vector $F \in \Gamma$ can be decomposed as 
%         $F = F^{\rm u} + F^{\rm s}$ with $F^{\rm u} \in E^{\rm u}(U^{*})$ 
%         and $F^{\rm s} \in E^{\rm s}(U^{*})$, which satisfies  
%         \begin{equation}
%                 \| F^{\rm s} \| , \| F^{\rm u} \| \leq K_{0} \| F \| 
%                 \label{eq:Ass.1.decp2}
%         \end{equation}
%         with some $K_{0} >0$. 
        There exists $\bar{\kappa} >0$ such that the restrictions of 
        ${\cal T}_{\sigma, U^{*}}$ on these invariant subspaces satisfy: 
        \begin{eqnarray} 
                \| {\cal T}_{\sigma, U^{*}} (F) \| & \geq & 
                e^{\bar{\kappa} \sigma} \| F \| \qquad (F \in E^{\rm u}(U^{*}))
                \nonumber \\ 
                \| {\cal T}_{\sigma, U^{*}} (F) \| & \leq & 
                e^{- \bar{\kappa} \sigma} \| F \| \qquad (F \in E^{\rm s}(U^{*})) 
                \label{eq:Ass.1.onEs}
        \end{eqnarray}
\bigskip

The estimates (\ref{eq:Ass.1.onEs}) in Assumption~L1 
could be further deduced {}from 
suitable assumptions on the spectrum of ${\dot{\cal T}}_{U^{*}}$, because 
${\cal T}_{\sigma, U^{*}} = \exp ( s {\dot{\cal T}}_{U^{*}} )$.  
One possibility of 
such an assumption  would be: 
\begin{quote}
  The spectrum\footnote{%
    Here we of course mean both continuous and discrete 
  spectrum.}  
  of ${\dot{\cal T}} \equiv {\dot{\cal T}}_{U^{*}}$ is a union 
  $\Sigma_{-} \cup \Sigma_{+}$. Here (i) $\Sigma_{+}$ is a finite set 
  with total multiplicity $N$, (ii)
  $\Sigma_{-} \subset \{z \in {\Bbb C}\,|\, 
  {\rm Re} z \leq - \bar{\kappa}' \}$ 
  and 
  $\Sigma_{+} \subset \{z \in {\Bbb C}\,|\, {\rm Re} z \geq \bar{\kappa}' \}$ 
  for some $\bar{\kappa}' >0$, and (iii) ${\dot{\cal T}}$ 
  is a sectorial operator. 
\end{quote} 
For a related estimate, see e.g. Henry \cite[Theorem~1.5.3]{Henr81}.

Before stating the second assumption, we set up the following 
convention on the choice of the norm $\| \cdot \|$. 

\bigskip\noindent
{\bf Convention on the choice of the norm $\| \cdot \|$.}  
For a vector $F$, we denote by $F^{\rm s}$ (resp. $F^{\rm u}$) the 
$E^{\rm s}(U^{*})$ (resp. $E^{\rm u}(U^{*})$) component of $F$: 
\begin{equation}
        F = F^{\rm u} + F^{\rm s}
        \label{eq:def-dcmpFsFu} 
\end{equation} 
with $F^{\rm s} \in E^{\rm s}(U^{*})$,  
$F^{\rm u} \in E^{\rm u}(U^{*})$.  
Assumption~L1 guarantees the possibility of the decomposition.  
We define the ``box norm'' 
\begin{equation}
        \| F \|_{\rm box} \equiv \sup \left \{ 
         \| F^{\rm s} \| , \| F^{\rm u} \| \right \}  .
        \label{eq:def-boxnorm}
\end{equation}
Note that this box norm is 
equivalent to the original $\| \cdot \|$, because 
\begin{equation}
         K_{0} \| F^{\rm s} \| , K_{0} \| F^{\rm u} \| \leq \| F \| 
         \leq \| F^{\rm s} \| + \| F^{\rm u} \|
         \label{eq:Ass.1.decp2}
\end{equation}
holds with some ($F$-independent) $K_{0} >0$.  
The first inequality of (\ref{eq:Ass.1.decp2}) is due to 
completeness, while the second is nothing but the triangle inequality. 
Because this box norm simplifies our presentation (e.g. 
conditions on various constants in Propositions), we 
extensively use  this box norm in the following, writing it 
simply  as $\| \cdot \|$.  
Sometimes, we write $\| F \|_{\rm s} \equiv \| F^{\rm s} \|$, 
$\| F \|_{\rm u} \equiv \| F^{\rm u} \|$.

\bigskip\noindent
{\bf Notation}  
We extensively use following abbreviations throughout 
Sec.~\ref{sub-rgflow} to Sec.~\ref{sec-univ.class}. 
\begin{itemize} 
\item We write $a \vee b \equiv \max \{ a, b \}$. 
\item $\Oabs (a)$ denotes a scalar or vector depending on the 
        context,  whose magnitude (measured in absolute value $| \cdot |$ 
        for a scalar, in the box norm $\| \cdot \|$ for a vector) 
        is bounded by $|a|$. 
\item We also write  $f \grless g \pm h$
        to simultaneously represent $f < g+h$ and $f > g-h$.  That is, 
        $f \grless g \pm h$ is 
        equivalent to $f = g + \Oabs(h)$ when $h \geq 0$. 
\end{itemize}

\medskip

The second assumption refers to regularity properties of the flow, and 
essentially requires that the map ${\cal R}_{\sigma}$ has a Lipschitz 
continuous Fr\'echet derivative.  Although it 
is sufficient to require only H\"older continuity (of the derivative)
with a positive exponent in the sequel, 
We here require Lipschitz continuity, to make 
the presentation as simple as possible.

\bigskip\noindent 
{\bf Assumption~L2 (Smoothness of ${\cal R}_\sigma$)}  
        There exist positive constants $\delta_{0}, K_{1}, K_{2}$ 
        such that \\
        (i) the formal relation (\ref{eq:RTrel}) holds for 
        $\| U - U^{*} \| ,  \| F \|\leq \delta_{0}$: 
        \begin{equation}
                        {\cal R}_{\sigma} (U + F) = 
                         {\cal R}_{\sigma}(U) +  {\cal T}_{\sigma,U} (F) 
                                +  \Oabs (K_{1} \| F \|^{2} )
                        \label{eq:RTrel2}
        \end{equation}
        (ii) ${\cal T}_{\sigma,U}$ is uniformly bounded for 
        $\| U - U^{*} \| \leq \delta_{0}$: 
        \begin{equation}
                \| {\cal T}_{\sigma,U} (F) \| \leq K_{2} \| F \| . 
                \label{eq:Ass.1.bddT}
        \end{equation}
        We take $\delta_{0}$ sufficiently small so that 
        $\delta_{0} K_{1} \leq K_{2}$. 

\bigskip

Note that Assumption~L2 implies the following Lipschitz-type  continuity for 
${\cal R}_{\sigma}$ and ${\cal T}_{\sigma,U}$ (for 
$\| U_{i} - U^{*} \| \leq \delta_{0}$ ($i=1,2$)): 
\begin{eqnarray} 
        && 
        {\cal R}_{\sigma} (U_{1}) - {\cal R}_{\sigma} (U_{2})
        = \Oabs (  2 K_{2} \, \| U_{1} - U_{2} \| ) , 
        \label{eq:Ass.Lipshitz} \\ 
        && 
        \left ( {\cal T}_{\sigma, U_{1}} - {\cal T}_{\sigma, U_{2}} \right ) F 
        = \Oabs ( 6 K_{1} \cdot  \| U_{1} - U_{2} \| \cdot \| F \| )  . 
        \label{eq:A.pert}
\end{eqnarray} 
[To prove (\ref{eq:A.pert}) {}from Assumption~L2, first note it suffices 
to consider the case of $\| F \| = \| U_{1} - U_{2} \|/2$, because 
${\cal T}_{\sigma,U}$ is a linear operator.  Then evaluate 
$\Delta R \equiv {\cal R}_{\sigma} (U_{2}+F) - {\cal R}_{\sigma} (U_{2}) 
- {\cal R}_{\sigma} (U_{1}+F) + {\cal R}_{\sigma} (U_{1})$ for 
$\| F \| = \| U_{1} - U_{2} \|/2$ in 
two ways using (\ref{eq:RTrel2}): (1) express it in terms 
of ${\cal T}_{\sigma, U_{i}}$, (2) expand each ${\cal R}_{\sigma}(\cdot)$ at 
$(U_{1}+ U_{2})/2$ and show that $\Delta {\cal R}$ is 
$\Oabs (10 K_{1} \, \|F\|^{2})$.]

Although Assumption~L2 physically looks rather innocent, 
its verification is a nontrivial mathematical problem depending 
on individual cases.
We emphasize that  $K_1$ term in Assumption~L2 
is not necessarily bounded 
uniformly in $\sigma$, and can diverge for large $\sigma$.  
For example, they will certainly diverge (as $\sigma \rightarrow 
\infty$) for non-critical initial data which will lead to a black hole. 
What is claimed in the Assumption is that one can find finite 
nonzero $\sigma$ and $\delta_{0}$ for which the above claims hold. 

We follow the time evolution by considering 
a discrete dynamical system generated by ${\cal R}_{n \sigma}$ 
($n \in {\Bbb Z},  n \geq 0$).  
The reason for considering the discrete system is twofold.  First, it 
is naturally generalized to the case with discrete self-similarity, 
as discussed in Sec.~\ref{sec-discr.SS}.  Second, the operator 
${\cal R}_{\sigma}$ 
could be more regular than ${\dot{\cal R}}$, thanks to a smoothing effect of 
the time evolution over a nonzero period $\sigma$.  (I.e. the 
linearization ${\dot{\cal T}}$ may not be bounded, while ${\cal T}$ 
can be.)

%%%%%%%%%%%%%%%%%%%%%%%%%%%%%%%%%%%%%%%%%%%%%%%%%%%%%%%%%%%%%%%%%%%%%%%%%%
\subsubsection{Local behaviour of the flow: Existence of the critical 
surface (stable manifold)} 
\label{subsub-rg.qual}

Based on the above assumptions, we now start our renormalization group 
study of flows.  We here consider the qualitative aspects of the flow 
in a vicinity of the fixed point $U^{*}$. 

We remark that there are considerable number of mathematical 
literature concerning the local structure of  invariant manifolds.  
Proposition~\ref{prop-3} is a celebrated ``invariant manifold 
theorem''; see e.g. Shub \cite[Theorem~5.2, Theorem~II.4]{Shub87} 
or \cite[Theorem~III.8]{Shub87}, 
for precise statements and a proof 
by sophisticated methods under slightly different assumptions. 
Although the cited theorems are sufficient for our needs in this 
section, they do not cover the case considered in 
Sec.~\ref{sec-univ.class}, where we consider the universality over 
different models.  We therefore  
present our version of an elementary proof, 
in order to make the presentation more self-contained, and 
also to pave a natural way to our later analysis in 
Sec.~\ref{sec-univ.class}.  

\begin{figure}
%\vspace*{40mm} 
\begin{center}
{\BoxedEPSF{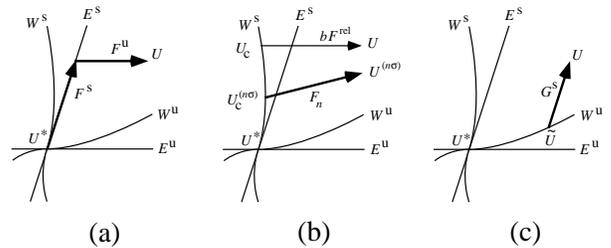 scaled 450} }
\end{center}
%{\bf // PLACE Fig.~\ref{fig-dcmp} HERE. //} 
\caption{Schematic view of three decompositions:  
(a) the decomposition of (\protect\ref{eq:rg.qual.dcmp}), where $F$ 
measures the difference between $U$ and $U^{*}$, and is decomposed 
respect to $E^{\rm s}$ and $E^{\rm u}$, 
(b) the decomposition of (\protect\ref{eq:Udiff-dcmp-init}), and 
(\protect\ref{eq:Udiff-dcmp-non0}), where $F_{n}$ 
measures the difference between $U^{(n\sigma)}$ and $U^{(n\sigma)}_{\rm c}$, 
and then is decomposed with respect to $E^{\rm s}$ and $E^{\rm u}$, 
(c) the decomposition of (\protect\ref{eq:Lem6.dcmp1}), which decomposes 
with respect to $E^{\rm s}$ and $W^{\rm u}$.
}
\label{fig-dcmp}
\end{figure}

We consider the time evolution of an initial data $U$ (close to $U^{*}$) 
under ${\cal R}_{n\sigma}$, by expressing it as ($n \geq 0$) 
\begin{equation}
        U^{(n\sigma)} \equiv {\cal R}_{n\sigma}  (U ) 
        = U^{*} + F_{n} 
        \label{eq:rg.qual.dcmp}
\end{equation}
where the second equality defines the difference $F_{n}$ between 
${\cal R}_{n\sigma}  (U )$ 
and $U^{*}$.
See Fig.~\ref{fig-dcmp}(a). 
Assumption~L1 guarantees that we can always decompose this way.

%%%%%FIGFIGFIGFIGFIGFIGFIGFIGFIGFIGFIGFIGFIGFIGFIGFIGFIGFIGFIGFIG
%%%%%FIGFIGFIGFIGFIGFIGFIGFIGFIGFIGFIGFIGFIGFIGFIGFIGFIGFIGFIGFIG
%%%%%FIGFIGFIGFIGFIGFIGFIGFIGFIGFIGFIGFIGFIGFIGFIGFIGFIGFIGFIGFIG
%%%%%FIGFIGFIGFIGFIGFIGFIGFIGFIGFIGFIGFIGFIGFIGFIGFIGFIGFIGFIGFIG
%

We begin with the following lemma, which claims that an initial data 
with $\| F^{\rm s}_{0} \|  < \| F^{\rm u}_{0} \|$ is driven away {}from 
$U^{*}$ (at least until $\| F^{\rm u}_{n} \|$ becomes somewhat 
large). 

\begin{lemma} 
\label{lem-Ftriangle}
Define $\delta_{1} >0$ by 
\begin{equation}
        \delta_{1} \equiv 
        \min \left \{ \delta_{0}, 
        \frac{1 - e^{-\bar{\kappa}\sigma} }{2 K_{1}}
        \right \} ,  
        \label{eq:delta1cond} 
\end{equation} 
and consider the time evolution of an initial data 
$U = U^{*} + F_{0}$ with 
$\| F_{0} \|_{\rm s} < \| F_{0} \|_{\rm u} \leq \delta_{1}$.   
Then as long as $\| F_{n} \|_{\rm u} \leq \delta_{1}$, 
we have 
\begin{eqnarray}
         \| F_{n} \|_{\rm s} & < & \| F_{n} \|_{\rm u}, 
         \label{eq:Ftriangle1} \\
         e^{\bar{\kappa} \sigma /2}  \, \| F_{n} \|_{\rm u}  
         & < & \| F_{n+1} \|_{\rm u} 
         < 2 K_{2}  \| F_{n} \|_{\rm u}
        \label{eq:Ftriangle2}
 \end{eqnarray}
\end{lemma}

{\em Proof.} 
We proceed by induction in $n$. We start by obtaining 
recursion relations for $F^{\rm s}_{n}$ and $F^{\rm u}_{n}$. 
Assumption~L2 reads  
\begin{equation}
        {\cal R}_{\sigma} (U^{*} + F_{n}) = 
        {\cal R}_{\sigma} (U^{*}) + {\cal T}_{\sigma, U^{*}} (F_{n}) 
        + \Oabs (K_{1} \| F_{n} \|^{2}) 
\end{equation}
which implies (note: ${\cal R}_{\sigma} (U^{*}) = U^{*}$) 
\begin{eqnarray} 
        F_{n+1} & = & {\cal T}_{\sigma, U^{*}} (F_{n}) 
        + \Oabs (K_{1} \| F_{n} \|^{2}) 
        \nonumber \\ 
        & = & {\cal T}_{\sigma, U^{*}} (F^{\rm s}_{n}) 
        + {\cal T}_{\sigma, U^{*}} (F^{\rm u}_{n}) 
        + \Oabs (K_{1}  \| F \|^{2}) ,  
        \label{eq:detail.phys.ex2}
\end{eqnarray}
where in the second step we used the linearity of 
${\cal T}_{\sigma, U^{*}}$  and the fact that $F^{\rm s}$ 
and $F^{\rm u}$ belong to invariant subspaces of 
${\cal T}_{\sigma, U^{*}}$. 
Decomposing as in (\ref{eq:rg.qual.dcmp}),  
we have recursion relations for $F^{\rm i}_{n}$ 
(${\rm i} = {\rm s} \; {\rm or}\; {\rm u}$): 
\begin{equation}
        F^{\rm i}_{n+1} = {\cal T}_{\sigma, U^{*}} (F^{\rm i}_{n})  
                +  \Oabs ( K_{1} \| F_{n} \|^{2} ) .
        \label{eq:rg.qual.4} 
\end{equation}
Our inductive assumption $\| F^{\rm s}_{n} \| < \| F^{\rm u}_{n} \|$ 
implies $\| F \| = \| F^{\rm u}_{n} \|$,  
by the definition of the box norm (\ref{eq:def-boxnorm}). 
So taking the norm of (\ref{eq:rg.qual.4}) and 
utilizing Assumption~L1 and Assumption~L2 to bound 
$\| {\cal T}_{\sigma, U^{*}} (F^{\rm i}_{n}) \|$,  
we have recursions: 
\begin{eqnarray} 
        \| F^{\rm u}_{n+1} \| & \leq & K_{2} 
        \| F^{\rm u}_{n} \| +  K_{1} \| F^{\rm u}_{n} \|^{2}
        \label{eq:qual.Frec.1} \\ 
        \| F^{\rm u}_{n+1} \| & \geq & e^{\bar{\kappa} \sigma} 
        \| F^{\rm u}_{n} \| -  K_{1} \| F^{\rm u}_{n} \|^{2}
        \label{eq:qual.Frec.2} \\ 
        \| F^{\rm s}_{n+1} \| & \leq & e^{-\bar{\kappa} \sigma} 
        \| F^{\rm s}_{n} \| + K_{1} \| F^{\rm u}_{n} \|^{2}  . 
        \label{eq:qual.Frec.3} 
\end{eqnarray} 

Now the recursion (\ref{eq:qual.Frec.1}) immediately implies 
(for 
$\| F^{\rm u}_{n} \| \leq \delta_{1} \leq \delta_{0} \leq K_{2}/K_{1}$)
\begin{equation}
        \| F^{\rm u}_{n+1} \|  
        \leq (K_{2} + K_{1} \| F^{\rm u}_{n} \|) \| F^{\rm u}_{n} \|  
        \leq 2 K_{2} \| F^{\rm u}_{n} \|  .
        \label{eq:qual.us.5}
\end{equation}
On the other hand,  (\ref{eq:qual.Frec.2}) implies 
(for $\| F^{\rm u}_{n} \| \leq \delta_{1}$; use also the definition 
of $\delta_{1}$)
\begin{equation}
        \| F^{\rm u}_{n+1} \|  
        \geq (e^{\bar{\kappa} \sigma}  - K_{1} \| F^{\rm u}_{n} \|) 
        \| F^{\rm u}_{n} \|  
        \geq \left ( \frac{ e^{\bar{\kappa} \sigma} + 1}{2} \right ) 
        \| F^{\rm u}_{n} \|  .
        \label{eq:qual.us.6}
\end{equation}
These two prove (\ref{eq:Ftriangle2}).  [Note: $(1+a)/2 \geq 
\sqrt{a}$ for positive $a$.] 
 
Finally, (\ref{eq:qual.Frec.3}) implies 
\begin{eqnarray}
        \| F^{\rm s}_{n+1} \| & \leq & 
        e^{-\bar{\kappa} \sigma} \| F^{\rm u}_{n} \| 
        + K_{1} \| F^{\rm u}_{n} \|^{2} 
        \nonumber \\   
        & = & ( e^{-\bar{\kappa}\sigma}  + K_{1} \| F^{\rm u}_{n} \| ) 
        \| F^{\rm u}_{n} \|.  
        \label{eq:qual.us.8}
\end{eqnarray}
This,  combined with 
$\| F^{\rm u}_{n} \| < \| F^{\rm u}_{n+1} \|$  
just proven,  implies (\ref{eq:Ftriangle1}), 
because $e^{-\bar{\kappa}\sigma}  + K_{1} \| F^{\rm u}_{n} \| 
\leq e^{-\bar{\kappa}\sigma}  + K_{1} \delta_{1} <1$. 
\hfill$\Box$

%\bigskip\noindent {\bf Proposition~3.} 
%  {\em 
\begin{prop}
\label{prop-3}
Under Assumptions 1 and 2, we have  \\
  (i) [Existence of critical solutions.] 
        For any $F^{\rm s} \in E^{\rm s}(U^{*})$ 
        with $\| F^{\rm s} \| \leq \delta_{1}$, there exists 
        $F^{\rm u}_{\rm c}(F^{\rm s}) \in E^{\rm u}(U^{*})$  such that 
        \begin{equation}
                U^{*}+ F^{\rm s} + F^{\rm u}_{\rm c}(F^{\rm s}) 
                \rightarrow U^{*} 
                \qquad {\rm as} \quad n \rightarrow \infty .
                \label{eq:defcrit.sol}
        \end{equation}
        Moreover, 
        \begin{equation}
                \| F^{\rm u}_{\rm c}(F^{\rm s}) \| 
                \leq C_{1} \| F^{\rm s} \|^{2}, 
                \qquad {\rm with} \quad 
                C_{1} = \frac{2 K_{1}} {1 - e^{-\bar \kappa \sigma}} .
                \label{eq:Fu-Fs} 
        \end{equation}
        We define the ``local stable manifold'' as
        \begin{equation}
                W^{\rm s}_{\rm loc}(U^{*}) \equiv 
                \left \{ U^{*} + F^{\rm s} + F^{\rm u}_{\rm c}(F^{\rm s}) 
                \bigl | \, 
                \| F^{\rm s} \| \leq \delta_{1} \right \} .
                \label{eq:defcrit.sol1}
        \end{equation}
  (ii) [Motion on $W^{\rm s}_{\rm loc}(U^{*})$.] 
        There exists $\bar{\bar{\kappa}} > 0$ such that 
        \begin{equation}
                \left \| {\cal R}_{n\sigma} (U) 
                - U^{*} \right \|  
                \leq 
                e^{- n \bar{\bar{\kappa}}\sigma} 
                \| U - U^{*} \| 
                \label{eq:norm-bound-BS}
        \end{equation}
        and 
        \begin{equation}
                \left \| {\cal R}_{(n+1)\sigma} (U) - U^{*} \right \| 
                \leq e^{-\bar{\bar{\kappa}} \sigma} 
                \left \| {\cal R}_{n\sigma} (U) - U^{*} \right \| 
                \label{eq:Ws-behave}
        \end{equation}
        for any vector $U \in W^{\rm s}_{\rm loc}(U^{*})$ which satisfies 
        $\| U - U^{*} \| \leq \delta_{1}$.   
        We can in fact 
        take $\bar{\bar{\kappa}} \approx \bar{\kappa}$ by taking 
        $\| U - U^{*} \|$ small. 
        \\
  (iii) [Invariant manifolds.] 
        Above 
        $W^{\rm s}_{\rm loc}(U^{*})$ in fact forms a manifold of 
        codimension $N$. 
\end{prop}
%  } % end of \em 

We only present a proof for (i), (ii) of the Proposition.  Detailed 
information presented in (iii) is not necessary for our later 
analysis, so we only make brief comments on their proof at the end of 
the proof of (i) and (ii).  
Our proof is modeled after the argument by Bleher and Sinai  
\cite{BS73}, which first appeared in a similar context in 
statistical mechanics.

\medskip
\noindent{\em Proof of Proposition~\ref{prop-3}, (i) and (ii)} \/

We mainly consider the case of $N=1$, which is of our main interest.  
For the other cases, see the end of this proof. 
We denote the normalized relevant mode of 
${\cal T}_{\sigma, U^{*}}$ by $\Frel$: 
${\cal T}_{\sigma, U^{*}} \Frel = e^{\kappa\sigma} \Frel$, 
$\| \Frel \| = 1$. 

The proof is done in several steps.

{\em Step~1. The goal.} \/  
We are interested in the time evolution 
of vectors  $U^{(0)} = U^{*} + F^{\rm s}_{0} + a_{0} \Frel$, 
for fixed $F^{\rm s}_{0}$ and  various 
$a_{0}$.  Our goal is to show that there 
is at least one value of $a_{0}$ (called $a_{c}$, as a function of 
$F^{\rm s}_{0}$) for which $\lim_{n\rightarrow \infty} 
{\cal R}_{n\sigma}(U^{(0)}) = U^{*}$.

{\em Step~2. Recursion relations for $a_{n}$ and $F^{\rm s}_{n}$.}\/  
We obtained recursions for $F^{\rm s}_{n}$ and 
$F^{\rm u}_{n} = a_{n} \Frel$ in the proof of 
Lemma~\ref{lem-Ftriangle}: (\ref{eq:rg.qual.4}).   
Utilizing Assumption~L1, and using 
$\| F \| \leq \max  \{| a_{n}| , \| F^{\rm s} \| \}$ 
which follows {}from the definition of the box norm (\ref{eq:def-boxnorm}),  
we can convert (\ref{eq:rg.qual.4})  into recursion  
relations for $a_{n}$ and $h_{n} \equiv \| F^{\rm s}_{n} \|$.  
The result is (recall: $a \vee b \equiv \max \{ a, b \}$) 
\begin{eqnarray}
        a_{n+1} & = & e^{\kappa \sigma} a_{n} 
        + \Oabs ( K_{1} (a_{n} \vee h_{n} )^{2} ) 
        \label{eq:rg.qual.5} \\
        h_{n+1} & \leq  & e^{-\bar{\kappa} \sigma} h_{n} 
        +  \Oabs ( K_{1} (a_{n} \vee h_{n} )^{2} )  . 
        \label{eq:rg.qual.6} 
\end{eqnarray}

Roughly speaking, the recursion shows that the map 
for $a$ is expanding, while that for $h$ is contracting, with 
``shifts'' of $\Oabs ( K_{1} (a_{n} \vee h_{n} )^{2} ) $.  
If we neglect the second term, 
$a_{n}$ would blow up.  Thus the stable set $W^{\rm s}_{\rm loc}(U^{*})$ 
is located by carefully balancing 
the inhomogeneous $\Oabs ( K_{1} (a_{n} \vee h_{n} )^{2} )$ 
term with the first expanding term, 
so that $a_{n}$ will remain bounded (in fact goes to zero) 
as we iterate the recursion.    
That we can in fact find at least one such $a_{0} (= a_{c})$ for 
given $h_{0}$ can be shown by using the argument due to 
Bleher and Sinai \cite{BS73}, as explained in Steps~3---5.

{\em Step~3. Continuity of $a_{n}$ and $h_{n}$ as functions of $a_{0}$.} \/ 
{}{}from (\ref{eq:Ass.Lipshitz}) 
${\cal R}_{\sigma}$ is a continuous function 
of its arguments.  This implies in particular that 
$a_{n+1}$ and $F^{\rm s}_{n+1}$ are continuous 
functions of $a_{n}$ and $F^{\rm s}_{n}$.  
Iterating this 
for $n$ times, we see that $a_{n}$ and $F^{\rm s}_{n}$ are continuous 
functions of the initial data, $a_{0}$, for fixed $F^{\rm s}_{0}$. 
This implies $h_{n} \equiv \| F^{\rm s}_{n} \|$ is also continuous in 
$a_{0}$.

{\em Step~4. Solving the recursions.} \/ 
We now solve the recursion, following the technique of Bleher and 
Sinai.  We will show that there are positive sequences 
$r_{n}, t_{n}$ such that 
\begin{description}
\item[(a)] $r_{n}, t_{n}\rightarrow 0$ as $n \rightarrow \infty$
\item[(b)] $0 \leq h_{n+1} < r_{n+1}$ holds, 
        as long as $|a_{n}| < t_{n}$ and $0 \leq h_{n} < r_{n}$. 
\item[(c)] As long as $0 \leq h_{n} < r_{n}$, 
        the set of all $a_{n+1}$ 
        when $a_{n}$ sweeps the interval $(-t_{n}, t_{n})$ contains the 
        interval $(-t_{n+1}, t_{n+1})$ inside. 
\end{description}
The proof of (a)---(c) is postponed to ``Step~5'' for the ease of 
reading.  

Now given (a)---(c), we can show the existence of the critical $a_{c}$ 
as follows.  Let  $I_{n}$ be the interval 
$I_{n} \equiv \{ z \in {\Bbb R}: |z| < t_{n}\}$. 
Because $a_{n}$ is a continuous function of $a_{0}$, we can consider 
the inverse image  ${\rm Inv}_{n}(I_{n})$ of the interval $I_{n}$ under the 
$n$-iterations of the recursion.  But (a)---(c) show that 
${\rm Inv}_{n}(I_{n})$ are nonincreasing (in $n$) sets, 
with nonempty intersections.  
So, taking an  initial condition $a_{c} \in \cap_{n} {\rm Inv}(I_{n})$, we 
can guarantee for all $n \geq 0$ 
\begin{equation}
        |a_{n}| < t_{n} \qquad {\rm and } \quad  h_{n} < r_{n}.  
        \label{eq:lem3.ahbd}
\end{equation}
Because $r_{n}, t_{n} \rightarrow 0$ as $n\rightarrow \infty$, 
(\ref{eq:lem3.ahbd}) shows $U^{(n\sigma)} \rightarrow U^{*}$ for this 
special $a_{c}$,  
i.e. $U^{*} + F^{\rm s} + F^{\rm u}_{\rm c} \in W^{\rm s}(U^{*})$, 
with $F^{\rm u} = a_{\rm c} \Frel$: 
existence of a critical solution for given $F^{\rm s}$.

{\em Step~5. Proof of (a)---(c).} \/ 
Now we give an example of sequences $r_{n}, t_{n}$ which satisfy  
(a)---(c) of Step~4.  We define $r_{n}, t_{n}$ according to 
(for $0 < r_{0} < \delta_{1}$) 
\begin{eqnarray}
        r_{n} & \equiv & 
        \left ( e^{-\bar{\kappa}\sigma} + K_{1} r_{0} \right )^{n} 
        r_{0} 
        \label{eq:rn-bound}
        \\
        t_{n} & \equiv & C_{1} r_{n}^{2} \qquad {\rm with} \quad 
        C_{1} \equiv \frac{2 K_{1}} {1 - e^{-\bar{\kappa}\sigma}} . 
        \label{eq:rntn-rel}
\end{eqnarray}
Then (a) is easily seen to be satisfied, because 
$e^{-\bar{\kappa}\sigma} + K_{1} r_{0} \leq 
(1 + e^{-\bar{\kappa}\sigma})/2 < 1$ 
by the choice of 
$\delta_{1}$.  To prove (b) and (c), 
first note the following sufficient condition: 
\begin{eqnarray}
        r_{n+1} & \geq & e^{-\bar{\kappa}\sigma} r_{n} 
        + K_{1}(r_{n} \vee t_{n})^{2} 
        \label{eq:BScond21} \\
        t_{n+1} & \leq & e^{\kappa\sigma} t_{n} 
        - K_{1}(r_{n} \vee t_{n})^{2} . 
        \label{eq:BScond22}
\end{eqnarray}
The choice of $\delta_{1}, C_{1}$ guarantees $C_{1} r_{n} \leq 1$, 
and thus $t_{n} \leq r_{n}$.  This, together with $\kappa \geq 
\bar{\kappa}$, immediately shows that (\ref{eq:BScond21}) 
and (\ref{eq:BScond22}) are satisfied. 
Eq.(\ref{eq:rntn-rel}) for $n=0$ proves (\ref{eq:Fu-Fs}). 

{\em Step~6. Motion of the critical solution.} \/ 
Now that we have proven the existence of a critical solution, we 
concentrate on its time evolution.   We first note 
$| a_{n} | \leq h_{n}$ holds for the critical 
solution for all $n \geq 0$, because otherwise the solution is shown 
to be driven away from $U^{*}$ by Lemma~\ref{lem-Ftriangle}. 
But then, the recursion (\ref{eq:rg.qual.6}) is reduced to 
\begin{equation}
        h_{n+1} \leq \bigl [ e^{-\bar{\kappa}\sigma} + K_{1} h_{n} \bigr ] 
        h_{n}
        \label{eq:hnrecred2}
\end{equation}
which, in view of $h_{n} \leq r_{n} \leq r_{0}$,  implies 
\begin{displaymath}
        \left \| {\cal R}_{(n+1)\sigma} (U) \! - \! U^{*} \right \|  
        \leq e^{-\bar{\bar{\kappa}}\sigma} 
        \left \| {\cal R}_{n\sigma} (U) \! - \! U^{*} \right \| 
\end{displaymath}
with 
\begin{equation}
        e^{-\bar{\bar{\kappa}}\sigma} 
        \equiv e^{-\bar{\kappa}\sigma } +  K_{1} r_{0} 
        \qquad \left ( \leq \frac{1 + e^{-\bar{\kappa}\sigma}}{2} 
        < 1 \right ) . 
        \label{eq:kappababagiven}
\end{equation}
\hfill$\Box$

\medskip
\noindent{\em Cases of $N \neq 1$ and a comment on 
Proposition~\ref{prop-3}, (iii) } \/  
All the above proves the existence of $W^{\rm s}_{\rm loc}(U^{*})$ 
and the estimate (\ref{eq:Ws-behave}) for the case of $N=1$.  
For $N \geq 2$ we 
can proceed similarly and find the critical surface (this time, we 
have to adjust $N$ parameters). For $N=0$, similar argument shows that the 
map ${\cal R}_{\sigma}$ is a contraction in a neighbourhood of $U^{*}$, thus 
proving that all the initial data close to $U^{*}$ is attracted to 
$U^{*}$.  

Proof of Proposition~\ref{prop-3}, (iii) can be performed, e.g., as 
in \cite[Theorem~II.4]{Shub87}.
\hfill$\Box$

%%%%%%%%%%%%%%%%%%%%%%%%%%%%%%%%%%%%%%%%%%%%%%%%%%%%%%%%%%%%%%%%%%%%%%%%%%%%%%
\subsubsection{Local behaviour of the flow: Unstable manifold} 
\label{subsub-rg.quan1}

In the previous section, we established the existence of the critical 
surface (stable manifold).  We also showed that initial data which 
is sufficiently away 
{}from the critical surface (characterized by 
$\| U - U^{*} \|_{\rm s} < \| U - U^{*} \|_{\rm u}$) 
is driven away {}from $U^{*}$.  
In this and the following section, 
we concentrate on the fate of near (but not exactly) 
critical solutions.   We in this section define an ``unstable 
manifold.''  Then in Sec.~\ref{subsub-rg.quan2}, 
we show that all the near-critical initial data (starting 
{}from an neighbourhood of $U^{*}$) are at first driven towards $U^{*}$ 
along the critical surface, but are eventually driven away along an 
unstable direction.  We also give quantitative estimates on time 
intervals until the deviation of the data {}from $U^{*}$ becomes 
large.

We concentrate on the case where there is a unique relevant mode 
($N=1$ in Assumption~L1).  More precisely we assume, in addition to 
Assumptions~L1 and L2 of Sec.\ref{subsub-rg.qual}, 

\bigskip\noindent {\bf Assumption~L1A (Uniqueness of the relevant mode)}  
        The relevant mode of ${\cal T}_{\sigma,U^{*}}$ is unique, i.e. 
        $N=1$ in Assumption~L1.  We denote the relevant 
        mode by $\Frel$ and 
        the relevant {\em real} eigenvalue by $\kappa$ ($\geq \bar\kappa$): 
        \begin{equation}
                {\cal T}_{\sigma,U^{*}} \Frel = e^{\kappa \sigma} \Frel .
                \label{eq:F1def}
        \end{equation}

\bigskip

Throughout this section, we use following notation. 

\begin{itemize}
\item $\epsilon_{2}$, $\delta_{2}$ are positive constants chosen so 
        that 
        \begin{equation}
                \delta_{2}, \epsilon_{2} \leq \min 
                \left \{ \delta_{1}, 
                \frac{1 - e^{-{\bar\kappa}\sigma}}{12 K_{1}} 
                \right \} , 
                \label{eq:ep2del2def1}
        \end{equation}
        \begin{equation}
                \delta_{2} + \epsilon_{2} \leq 
                \frac{e^{\kappa\sigma} - 1}{12 K_{1}} .
                \label{eq:ep2del2def2}
        \end{equation}
\item Given $U \in \Gamma$, define $n_{1}(U; \epsilon)$ to be the 
        largest integer such that 
        $\| {\cal R}_{k\sigma} (U) - U^{*} \| \leq \epsilon$ holds for 
        all $0 \leq k \leq n_{1}(U; \epsilon)$.  I.e. 
        \begin{equation}
                n_{1}(U;\epsilon) \equiv \max \bigl \{ n \bigl | 
                \| {\cal R}_{k\sigma} (U) \! - \! U^{*} \| \leq \epsilon 
                \;\; {\rm for } \;\; k \leq n \bigr \}
                \label{eq:def-n1-gen} 
        \end{equation}
\end{itemize}

To trace the time evolution, we have to specify the unstable manifold. 
If ${\cal R}_{\sigma}$ is invertible, we can define the unstable 
manifold as
\begin{equation}
        W^{\rm u} (U^{*}) = \left \{ 
        U \bigl | {\cal R}_{n\sigma} (U) \rightarrow U^{*} \; {\rm as} \; 
        n \rightarrow -\infty 
        \right \} , 
        \label{eq:Wudef-wouldbe}
\end{equation}
interpreting ${\cal R}_{n\sigma} = ( {\cal R}_{-n\sigma})^{-1}$ for 
negative $n$. 
However, we here do not assume the invertibility of ${\cal R}_{\sigma}$.  So 
the definition of $W^{\rm u}(U^{*})$ requires some additional work.  
Although our definition is somewhat artificial, it suffices our need.

We define several sets, including the local unstable manifold 
$W^{\rm u}_{\rm loc}(U^{*})$.  
Although all of these are defined with respect to the 
fixed point $U^{*}$, we do not explicitly 
write $U^{*}$ except for $W^{\rm u}_{\rm loc}(U^{*})$, to simplify 
notation. 
We first introduce a line segment $D_{r}$ for $r \in {\Bbb R}$ as 
\begin{equation}
        D_{r} = \left \{ 
        (1-t) (U^{*} \! + \! r \Frel) 
        + t {\cal R}_{\sigma} (U^{*} \! + \! r \Frel)
        \bigl | 0 \leq t \leq 1 
        \right \} 
        \label{eq:Drdef}
\end{equation}
and define $\tilde{W}_{r}$ as [$n_{1}$ is defined in 
(\ref{eq:def-n1-gen})] 
\begin{equation}
        \tilde{W}_{r} = \left \{ {\cal R}_{n\sigma} (U) \bigl | 
        U \in D_{r},  n \leq n_{1}(U^{*}+ r \Frel; \epsilon_{2}) 
        \right \} .  
        \label{eq:Wtildedef}
\end{equation}
Then we define the local unstable manifold at $U^{*}$ as 
\begin{equation}
        W^{\rm u}_{\rm loc} (U^{*}) = 
        \lim_{r \searrow 0} 
        \left [ \tilde{W}_{r} \cup \tilde{W}_{-r} \right ] . 
        \label{eq:Wulocdef}
\end{equation}

Lemma~\ref{lem-6} below guarantees that 
$W^{\rm u}_{\rm loc}(U^{*})$ is well defined, and all 
flows close to $U^{*}$ are squeezed around this 
$W^{\rm u}_{\rm loc}(U^{*})$.

\begin{lemma}
\label{lem-6}
(i) [Structure of $W^{\rm u}_{\rm loc}(U^{*})$] 
$W^{\rm u}_{\rm loc}(U^{*})$ of (\ref{eq:Wulocdef}) 
is a well-defined set, and is 
``one-dimensional'' in the sense that its intersection 
with a plane $U^{*} + E^{\rm s}(U^{*}) + a \Frel$ 
exists as a unique point for $|a| \leq \epsilon_{2}$. 
Moreover, any $U, U' \in W^{\rm u}_{\rm loc}(U^{*})$ satisfy 
\begin{equation}
        \| U - U' \|_{\rm s} \leq C_{4} \| U - U' \|_{\rm u} 
        \label{eq:onWu-tangent}
\end{equation}
with 
\begin{equation}
        C_{4} \equiv \frac{7 K_{1} \epsilon_{2}} 
        {e^{\kappa\sigma} - e^{-\bar{\kappa}\sigma} - 7 K_{1} \epsilon_{2}} 
        \qquad \bigl ( < 1 \bigr ) 
        \label{eq:onWu-tanconst}
\end{equation}
\\
(ii) [Contraction around $W^{\rm u}_{\rm loc}(U^{*})$ in 
$E^{\rm s}$-plane] 
Consider the time evolution of a vector 
$U$ under ${\cal R}_{n\sigma}$, with the initial data  
satisfying $\| U - U^{*} \| \leq \epsilon_{2}$.  We can find 
$\bar{\kappa}'' >0$ (depending on $\epsilon_{2}$) 
such that 
\begin{equation}
        {\rm dist}({\cal R}_{n\sigma}(U), W^{\rm u}_{\rm loc}(U^{*})) 
        \leq e^{-n\bar{\kappa}'' \sigma} 
        \| U - U^{*} \|
        \label{eq:Lem6.1}
\end{equation}
holds as long as $\| {\cal R}_{n\sigma}(U) - U^{*} \| \leq \epsilon_{2}$.
\end{lemma}
For a proof, see Appendix~\ref{proof-lem-6}. 

Finally we define sets of vectors 
\begin{eqnarray}
        A_{\pm}(\epsilon, \delta) & \equiv & 
        \Bigl \{ U_{\pm}  \in \Gamma \Bigl |  \;  
        \epsilon e^{-2\kappa\sigma} \leq 
        \| U_{\pm} - U^{*} \|_{\rm u} \leq \epsilon,         
        \nonumber \\ 
        & & {} \; \; 
        {\rm dist} (U_{\pm}, W^{\rm u}_{\rm loc}(U^{*})) \leq \delta 
        \Bigr \} , 
        \label{eq:deApm}
\end{eqnarray}
where for a given set $A \subset \Gamma$ and a vector 
$U \in \Gamma$, we denote their ``distance'' as
\begin{equation}
        {\rm dist}(U, A) \equiv \inf_{U' \in A} \| U - U' \| .
        \label{eq:distdef0}
\end{equation}
The set $A_{\pm}$ is schematically shown in Fig.~\ref{fig-flow2} of  
Sec.~\ref{sec-sen.exponent}. 

%%%%%%%%%%%%%%%%%%%%%%%%%%%%%%%%%%%%%%%%%%%%%%%%%%%%%%%%%%%%%%%%%%%%%%%%%%%%%%
\subsubsection{Local behaviour of the flow: Near-critical solution} 
\label{subsub-rg.quan2}

We now trace the time evolution of a 
slightly super-critical initial data $U$.  Our final 
goal is to trace it until it finally blows up and forms a black 
hole, but in this section we trace it  until 
its deviation {}from $W^{\rm s}_{\rm loc}(U^{*})$ becomes of 
order one (Stage I of 
Sec.~\ref{sec-sen.exponent}) as in 
(\ref{eq:RG.result}), so that we can tell that it is not critical.  
As shown, this part plays the essential role in determining the 
critical exponents.  Our presentation is modeled after a standard 
approach in applications of 
rigorous renormalization group techniques in statistical mechanics, 
see e.g. \cite{GK85,Hara87a}. 
  The rest of the flow (Stage II) 
is considered in Sec.~\ref{subsub-rg.global}.  

Time evolution of an initial data sensitively depends on its location 
{\em relative} to $W^{\rm s}_{\rm loc}(U^{*})$.  So we decompose the 
{\em initial data} $U$ as 
\begin{equation}
        U = b \Frel + U_{\rm c}, 
        \label{eq:Udiff-dcmp-init}
\end{equation}
where ${U_{\rm c}} \in W^{\rm s}_{\rm loc}(U^*)$ [i.e. 
${\cal R}_{n \sigma} (U_{\rm c}) \rightarrow U^{*}$ as $n 
\rightarrow \infty$],   
$\Frel$ is the normalized relevant eigenmode defined in 
(\ref{eq:F1def}), 
and $b \in {\Bbb R}$ is its coefficient.  See Fig.~\ref{fig-dcmp}(b) 
in Sec.~\ref{subsub-rg.qual}. 
Proposition~\ref{prop-3} (i) guarantees that we can find $U_{\rm c}$ 
and decompose this way for $\| U - U^{*} \| \leq \delta_{1}$. 
Generically $b = O(p - p_{c})$, because $b$ measures the 
component of $\Frel$ in $U -  U_{\rm c}$. 

Now, for $n > 0$, we define $F_{n}$ in terms of the decomposition
\footnote{
We could trace the time evolution by employing the decomposition 
(\ref{eq:Udiff-dcmp-init}) also for $n > 0$, i.e. 
${\cal R}_{n\sigma} (U) = U_{c,n} + b_{n} \Frel$ with 
$U_{c,n} \in W_{\rm loc}^{\rm s}(U^{*})$. 
However, we here 
employ (\ref{eq:Udiff-dcmp-non0}), because it also allows us to deal 
with the case of Sec.~\ref{sec-univ.class} in essentially the same way.
}
\begin{equation}
        {\cal R}_{n\sigma} (U) = {\cal R}_{n \sigma} (U_{\rm c}) + F_{n} , 
        \label{eq:Udiff-dcmp-non0}
\end{equation}
and further decompose $F_{n}$ in the spirit of (\ref{eq:def-dcmpFsFu}): 
\begin{equation}
        F_{n} = b_{n} \Frel + F^{\rm s}_{n}
        \label{eq:Udiff-dcmp-F}. 
\end{equation}
See Fig.~\ref{fig-dcmp}(b) in Sec.~\ref{subsub-rg.qual}. 
We do emphasize that the decomposition (\ref{eq:Udiff-dcmp-non0}) 
differs {}from that of (\ref{eq:rg.qual.dcmp}) in 
Sec.~\ref{subsub-rg.qual}, in that $F$ there measures the difference 
between $U$ and the fixed point $U^{*}$, while $F$ here  
measures the difference between $U$ and the (time-evolved) 
critical solution $U^{(n\sigma)}$.  Also note that 
$F_{n}$ ($n > 0$) is not necessarily parallel to $\Frel$. 

The initial condition implies $b_{0} = b, F^{\rm s}_{0} = 0$.

After these preparations, concerning the flow in Stage I, we have: 

%\bigskip \noindent {\bf Proposition~4.} 
%{\em 
\begin{prop}[Local behaviour of non-critical flows] 
\label{prop-4}
   Consider the time evolution of an initial data 
   $U_{\rm init} \equiv b_{0} \Frel + U_{\rm c}$, 
   with $U_{\rm c} \in W^{\rm s}_{\rm loc}(U^{*})$, 
   $\| U_{\rm c} - U^{*} \| \leq \delta_{2}$, 
   $0<|b_{0}| \leq \epsilon_{2}$, and 
   at each $n\sigma$ we decompose 
   as in  (\ref{eq:Udiff-dcmp-non0}): 
   ${\cal R}_{n\sigma} (U_{\rm init}) 
   = {\cal R}_{n\sigma} (U_{\rm c}) + b_{n} \Frel + F^{\rm s}_{n}$. 
   Let $n_{2} \equiv \max \bigl \{ n \in {\Bbb Z} \bigl |  \;  
   | b_{k} | \leq \epsilon_{2} \,\mbox{ for }\, 0 \leq k \leq n 
   \bigr \}$. 
   Then under Assumptions~L1, L2, and L1A, we can 
   find  $C_{2} > 1$ (independent of $b_{0}$ and $\epsilon_{2}$)  
   such that the following holds: \\ 
   (i) $n_{2}$ is finite and satisfies ($n \leq n_{2}$) 
        \begin{eqnarray}
           &&   \left ( \frac{2}{3 e^{\kappa\sigma} -1} \right ) \epsilon_{2} 
                \leq | b_{n_{2}} | \leq \epsilon_{2}, 
                \label{eq:p4.a2estimate2}
        \\
        &&
                \| F^{\rm s}_{n} \| \leq | b_{n} | 
        \\
        && 
                \| {\cal R}_{n\sigma} (U_{\rm c}) - U^{*} \| 
                \leq \epsilon_{2} 
                \left ( \frac{2}{1 +e^{\kappa\sigma}} \right )^{n_{2}-n}
%                 \| U_{\rm c}(n_{2}\sigma) - U^{*} \| \leq 
%                 \delta_{2} \Lambda_{2}^{n_{2}} + C_{2} \epsilon_{2}^{2}, 
%                 \label{eq:StageI-perp}
        \\
        && 
                \frac{\epsilon_{2}}{C_{2}} \leq | b_{0} | 
                e^{n_{2} \sigma \kappa}  
                \leq C_{2} \epsilon_{2} .
                \label{eq:p4.a2estimate}
        \end{eqnarray}
        (ii)  For sufficiently small $| b_{0} |$, we can find 
        $\delta >0$ such that 
        \begin{equation}
                U_{\rm init}^{(n_{2}\sigma)} \in  
                A_{\pm}(\epsilon_{2}, \delta) 
                \label{eq:prop4.Uend}
        \end{equation}
        where plus or minus sign is chosen according to the sign of $b_{0}$. 
        In fact we can take $\delta \searrow 0$ as $|b_{0}| \searrow 0$. 
\end{prop}
%} % end of \em 
%  

The above Proposition gives an estimate on the time interval spent by 
the flow in Stage~I.  It also provides information on the exit of 
Stage~I, by showing that the exit is essentially 
squeezed into the tiny tube% 
\footnote{%
With an additional assumption on the smoothness on ${\cal T}_{s, U}$ 
for $s \in (0, \sigma)$, we can reduce the tube $A_{\pm}$ to a 
disk, given by 
        $\bigl \{ 
        U_{\pm} \in \Gamma \bigl |  \;  
        \| U_{\pm} - U^{*} \|_{\rm u} = \epsilon_{2},  
        {\rm dist} (U_{\pm}, W^{\rm u}(U^{*})) \leq \delta 
        \bigr \}$. 
}
$A_{\pm}$ around $W^{\rm u}_{\rm loc}(U^{*})$ (see Fig.~\ref{fig-flow2}).

\bigskip \noindent {\em Proof of Proposition~\ref{prop-4}.}
(i) We first derive recursion equations for $b_{n}$ and $F^{\rm s}_{n}$. 

Using (\ref{eq:RTrel2}) with $U = U_{\rm c}$, we have 
\begin{eqnarray}
        && {\cal R}_{(n+1)\sigma}(U) = {\cal R}_{\sigma} (U^{(n\sigma)}) = 
        {\cal R}_{\sigma} (U^{(n\sigma)}_{\rm c} + F_{n}) \nonumber \\
        && \;\;
        = {\cal R}_{\sigma} ( U^{(n\sigma)}_{\rm c} )
        + {\cal T}_{\sigma, U^{(n\sigma)}_{\rm c} } (F_{n}) 
        +  \Oabs ( K_{1} \| F_{n} \|^2 ).
        \label{eq:ap.ev1}
\end{eqnarray}
This implies, according to our decomposition 
(\ref{eq:Udiff-dcmp-non0}), 
\begin{eqnarray} 
        F_{n+1} & = & {\cal T}_{\sigma, U^{(n\sigma)}_{\rm c} } (F_{n}) 
        +  \Oabs (K_{1} \| F_{n} \|^2 )
        \nonumber \\ 
        & = & {\cal T}_{\sigma, U^{*} } (F_{n}) 
        + \Oabs(6 K_{1} \| U^{(n\sigma)}_{\rm c} - U^{*} \| \, 
                \| F_{n} \| ) 
        \nonumber \\
        && {} + \Oabs (K_{1} \| F_{n} \|^{2})
        \label{eq:ap.ev11}
\end{eqnarray} 
where in the second step we used (\ref{eq:A.pert}).  
This corresponds to the first line of (\ref{eq:detail.phys.ex2}); the 
difference is we here have 
$\Oabs(6 K_{1} \| U_{\rm c} - U^{*} \| \, \| F \| )$. 
Arguing as we did in deriving 
(\ref{eq:rg.qual.4})---(\ref{eq:qual.Frec.3}), i.e. decomposing 
$F$ into $F^{\rm s}$ and $F^{\rm u}$ as in (\ref{eq:Udiff-dcmp-F}), 
we this time obtain 
\begin{eqnarray}
        && f_{n+1} \leq 
        e^{-\bar{\kappa} \sigma} f_{n} 
        \nonumber \\
        && \qquad + K_{1} (| b_{n} | \vee f_{n}) 
        \left [ \Oabs (6 g_{n} ) 
                + \Oabs(| b_{n} | \vee f_{n}) \right ]
        \label{eq:ap.ev33} \\ 
        && b_{n+1} =  e^{\kappa\sigma} b_{n} 
        \nonumber \\
        && \qquad + K_{1} (| b_{n} | \vee f_{n}) 
        \left [ \Oabs (6 g_{n} ) 
                + \Oabs (| b_{n} | \vee f_{n}) \right ] . 
        \label{eq:ap.ev34}
\end{eqnarray}
where we have introduced 
$g_{n} \equiv \| U^{(n\sigma)}_{\rm c} - U^{*} \|$, 
$f_{n} \equiv \| F^{\rm s}_{n} \|$ to simplify the notation. 

\smallskip 

Our remaining task is to solve the recursions.  This is an elementary 
exercise in calculus, and is summarized as Lemma~\ref{lem-5} in 
Appendix~\ref{proof-lem-5}: 
the recursive inequalities (\ref{eq:ap.ev33}) and (\ref{eq:ap.ev34}) are 
of the form of (\ref{eq:ap.ev42}) and (\ref{eq:ap.ev41}) there, with
$\Lambda \equiv e^{\kappa \sigma}$, 
$\bar{\Lambda} \equiv e^{-\bar{\kappa}\sigma}$, 
and $C_{3} = 6 K_{1}$.  And that $g_{n}$ satisfies the assumption of 
the Lemma follows from Proposition~\ref{prop-3}, (\ref{eq:norm-bound-BS}). 
Part (i) of Proposition~\ref{prop-4} thus 
follows immediately  {}from Lemma~\ref{lem-5}. 

(ii) We apply Lemma~\ref{lem-6} to the time evolution of the vector 
$U_{2} \equiv U^{(n_{2}\sigma/2)}$.  Here $n_{2}/2$ should be 
interpreted as its integer part.  
We have,  by Lemma~\ref{lem-5}  and Proposition~\ref{prop-3}, 
\begin{eqnarray}
        &&      \| U^{(n\sigma)} - U^{*} \| 
         \leq  | b_{n} | + \| U^{(n\sigma)}_{\rm c} - U^{*}\| 
        \nonumber \\ 
        && \hspace{10mm} \leq  \epsilon_{2} 
        \left( \frac{ 1 + e^{\kappa\sigma}}{2} \right )^{-(n_{2}-n)} 
        + \delta_{2} e^{- n \bar{\bar{\kappa}}\sigma}.  
        \label{eq:U-U*}
\end{eqnarray}
For $n \in [n_{2}/2, n_{2}]$, the above is bounded uniformly by 
$\delta_{2} + \epsilon_{2}$.  This also shows that 
the initial data $U_{2}$ for this part of time evolution satisfies 
$\| U_{2} - U^{*} \| = \| U^{(n_{2}\sigma/2)} - U^{*} \| \searrow 0$ 
as $n_{2} \nearrow \infty$.  So taking 
$\delta_{2} + \epsilon_{2}$ sufficiently small, we can apply 
Lemma~\ref{lem-6} to conclude that 
${\rm dist}(U^{(n_{2}\sigma)}, W^{\rm s}_{\rm loc}(U^{*}))$ 
goes to zero 
as $n_{2}$ goes to infinity (i.e. $b_{0}$ goes to zero)
\footnote{%
Precisely speaking, $|b_{n_{2}}| < \epsilon_{2}$ does not guarantee 
$\| U^{(n_{2}\sigma)} - U^{*} \|_{\rm u} < \epsilon_{2}$, because 
$b_{n_{2}}$ measures $\| U^{(n_{2}\sigma)} - U^{(n_{2}\sigma)}_{\rm c} 
\|$, and 
in general 
$\| U^{(n_{2}\sigma)}_{\rm c} - U^{*} \| > 0$.  However, because 
$\| U^{(n_{2}\sigma)}_{\rm c} - U^{*} \|$ goes to zero as $n_{2}$ goes to 
infinity, the difference can be neglected as $b_{0}$ goes to zero. 
}.
\hfill$\Box$

%%%%%%%%%%%%%%%%%%%%%%%%%%%%%%%%%%%%%%%%%%%%%%%%%%%%%%%%%%%%%%%%%%%%%%%%%%%%%%
\subsubsection{Global behaviour and physical consequences} 
\label{subsub-rg.global}

We are now at the stage of presenting the rigorous version of 
Claims~1 and 2, together with necessary assumptions.  After the analysis in 
the previous section, that is, the analysis of the flow at Stage~I, 
we in this section consider Stage~II of the flow. 
We want to trace the flow further until it 
blows up, and then calculate the mass of the Black hole.  This stage 
involves nontrivial physical and mathematical processes whose detailed 
analysis is beyond the scope of this paper.  In this sense, this 
section is essentially a repetition of Sec.~\ref{sec-sen.exponent}, 
with a difference 
that we here explicitly  list up all the assumptions 
(which are physically reasonable in view of numerical simulations) 
on the global behaviour of the flow,  under which we can present 
the rigorous version of Claims 1 and 2, i.e. 
Theorem~\ref{prop-final}. 

\bigskip\noindent 
{\bf Assumption G1 (Flows in Stage II) } 
For sufficiently small $\delta_{3} >0$ and 
for $A_{\pm} (\epsilon_{2}, \delta_{3})$ defined in (\ref{eq:deApm}) 
we have: 
\begin{itemize}
\item $U_{-}^{(s)} \equiv {\cal R}_{s}(U_{-})$ 
        tends to a flat space-time as 
        $s \rightarrow \infty$, for all $U_{-} \in 
        A_{-} (\epsilon_{2}, \delta_{3})$, 
\item $U_{+}^{(s)} \equiv {\cal R}_{s}(U_{+})$ 
        blows up at some finite time $s = s_{\rm AH} >0$, 
        $\xi = \xi_{\rm AH}$; $s_{\rm AH}$ and $\xi_{\rm AH}$ are uniformly bounded 
        above and below for all $U_{+} \in A_{+} (\epsilon_{2}, \delta_{3})$. 
        [Recall that $\xi$ is related to $r$ by $r=\xi e^{-\beta s}$
        as in (\ref{eq-xi-def}).] 
\end{itemize}

\bigskip\noindent 
{\bf Assumption G2 (Mass inside the apparent horizon)} 
The mass $M_{\rm AH}$ inside the apparent horizon at   
its formation is related with its radius $r_{\rm AH}$ by 
$C^{-1} r_{\rm AH} \leq M_{\rm AH} \leq  C r_{\rm AH}$, where $C$ is  
a finite positive constant independent of  $r_{\rm AH}$. 

\bigskip\noindent
{\bf Assumption G3 (Relation between masses)} 
The final mass of the black hole $M_{\rm BH}$,  and the mass at the 
formation of the apparent horizon $M_{\rm AH}$,  are 
of the same order.  More precisely, there exists a positive constant $C'$ such 
that $C'^{-1} M_{\rm AH} \leq M_{\rm BH} \leq C' M_{\rm AH}$ holds for all the 
near-critical solutions. 

\bigskip

Remarks on the meaning of these assumptions are in order. 

\bigskip\noindent 
{\bf Remark.} 
\begin{enumerate}
\item Assumption~G1 specifies the behaviour of the flow in Stage~II. 
Thanks to our analysis in Stage~I, we have only to know the fate of 
the flow which is emerging {}from a very small neighbourhood of 
$W^{\rm u}(U^{*})$ (recall Fig.~\ref{fig-flow2} of 
Sec.~\ref{sec-sen.exponent}).   
This means that physically it 
is sufficient to trace the time evolution of two vectors which is 
located on $W^{\rm u}(U^{*})$. 
\item Assumption~G2 is valid for a spherically symmetric 
gravitational collapse (in fact, $M_{\rm AH} = r_{\rm AH}/2$), 
which is the main topic in this paper.  
However we termed it as an Assumption, in order to make our presentation 
more general. 
\item Assumption~G3 involves a subtle physical process after the 
formation of an apparent horizon.  
Strictly speaking, the radius of the apparent horizon is not exactly 
related with the final Bondi--Sachs mass which gives the mass of the
black hole.   The final mass $M_{\rm BH}$ 
differs {}from (greater than,  
when the dominant energy condition is satisfied,)
the mass at the apparent
horizon $M_{\rm AH}$ by the matter absorbed into the black hole after the 
formation of the apparent horizon.  However, in our case, it is 
expected {}from the form of the relevant mode and the blow-up profile in 
numerical simulations 
that this effect can only make the black hole slightly 
heavier, say by a factor of two or so.  
(It should be noted that usual numerical simulation assumes 
Assumption~G3, and identifies the radius of apparent horizon with the 
final mass of the black hole.) 
\end{enumerate}

Under these preparations, we can now state our rigorous version of 
Claim~1 and Claim~2, which is the main result of this section.

\begin{theorem}
\label{prop-final}
Under Assumptions L1, L2, L1A and Assumptions G1, G2, G3, the 
universality (U1, U2) and scaling (S1, S2) 
of Sec.~\ref{sec-scenario} hold.  More precisely, \\
(i) 
All the initial data sufficiently close to $U^{*}$ (but not exactly 
on $W^{\rm s}_{\rm loc}(U^{*})$) once approach, but finally deviate {}from, 
$U^{*}$.  The final fate of these data are either a flat space-time or 
a black hole, depending on which side of $W^{\rm s}_{\rm loc}(U^{*})$ lies the 
initial data. \\
(ii) 
Consider the 
time evolution of the initial data 
\begin{equation}
        U_{\rm init} = b \Frel + U_{\rm c}
        \label{eq:Propfinal.init}
\end{equation}
with $\| U_{\rm c} - U^{*} \|\leq \delta_{2}, 
0<|b| \leq \delta_{2}$, 
which is on the side of black hole of $W^{\rm s}_{\rm loc}(U^{*})$. 
Then as $b$ goes to zero, the mass  of the black hole 
formed satisfies 
\begin{equation}
        C_1 | b |^{\beta/\kappa} \leq
        \frac {M_{\rm BH}}{M_0} \leq C_2 | b |^{\beta/\kappa} 
        \label{eq:Prop.final}
\end{equation}
with $b$-independent positive constants $C_1, C_2$. 
Here $M_0$ is the initial (say, ADM or Bondi-Sachs) mass.
\footnote{%
We have included $M_0$ in the expression (\ref{eq:Prop.final})
to avoid a nongeneric situation in which $M_0$ behaves like power of $b_{0}$
and $\beta_{\rm BH}$ becomes different {}from $\beta/\kappa$. 
}
\end{theorem}

\bigskip \noindent {\bf Remark.} 
By refining the above argument leading to Theorem~\ref{prop-final}, 
especially refining the proof of Lemma~\ref{lem-5}, we can see the following:
if there is a marginal mode, but we still have a stable manifold of 
$U^{*}$ of codimension one, it can happen that the scaling of mass has 
a logarithmic correction term, like 
\begin{equation}
        M_{\rm BH} \propto |p^*-p|^{\beta_{\rm BH}} 
        \left ( \ln | p^*-p | \right )^{\beta_{\rm ln}}
        \label{eq:ap.log}
\end{equation}
with some exponents $\beta_{\rm BH}$ and $\beta_{\rm ln}$.

\medskip
\noindent {\em Proof.}  
Proposition~\ref{prop-4} is a rigorous version of 
Step~1 of the `proof' of Claim~2 of Sec.~\ref{sec-sen.exponent}.  
Assumptions~G1, G2 and G3 enable us to perform Step~2, 
proving (i) of the Theorem.  This also shows 
\begin{equation}
        r_{\rm AH}/C \le M_{\rm BH} \le C r_{\rm AH}, \qquad  
        \epsilon_{2}/C\le |b|e^{\kappa\tau} \le C \epsilon_{2}, \qquad  
        \label{eq:thm.prf1}
\end{equation}
where $C$ is a positive constant, and
$\tau$ is the total time until the formation of the apparent 
horizon.   
Since $r_{\rm AH}/M_0=O(e^{-\tau})$ by the definition of scaling 
transformation (\ref{eq:sctrdef}), we get (ii) of the Theorem.
\hfill$\Box$

\def\aav#1{{\left\langle #1 \right\rangle}}
\def\av#1{{\langle #1 \rangle}}
%%%%%%%%%%%%%%%%%%%%%%%%%%%%%%%%%%%%%%%%%%%%%%%%%%%%%%%%%%%%%%%%%%%%%%%%%
%%%%%%%%%%%%%%%%%%%%%%%%%%%%%%%%%%%%%%%%%%%%%%%%%%%%%%%%%%%%%%%%%%%%%%%%%
%%%%%%%%%%%%%%%%%%%%%%%%%%%%%%%%%%%%%%%%%%%%%%%%%%%%%%%%%%%%%%%%%%%%%%%%%
\subsection{Discrete self-similarity}
\label{sec-discr.SS}

Significance of discretely self-similar solutions can be
well understood with minor modifications to the scenario presented in
Sec.~\ref{sec-scenario}. 

%%%%%%%%%%%%%%%%%%%%%%%%%%%%%%%%%%%%%%%%%%%%%%%%%%%%%%%%%%%%%%%%%%%%%%%%
\subsubsection{The picture}
\label{subsec-discr.SS-pict}

Let us begin with some definitions.

A {\em limit cycle}\/ of a renormalization group 
$\{{{\cal R}}_{s} \,|\, s\in{\Bbb R}\}$ with fixed 
$\alpha$ and $\beta$
is a periodic orbit of the flow 
$U_{\rm lc} = \{U_{\rm lc}^{(s)}\,|s\in{\Bbb R}\}$
where $U_{\rm lc}^{(s+\Delta)} = U_{\rm lc}^{(s)}$ with 
fundamental periodicity $\Delta>0$. 
With a slight abuse of notation, we use $U_{\rm lc}$ to denote the orbit 
itself, while $U_{\rm lc}^{(s)}$ denotes a point on it.
Let us call a function $u$ {\em discretely self-similar}\/ 
with parameters $(\alpha, \beta)$ 
if it satisfies 
$u(t,r) = u^{(\Delta, \alpha, \beta)}(t,r)$
for some $\Delta>0$.
Each discretely self-similar solution $u$ of the PDEs 
with parameters $(\alpha, \beta)$ of the Assumption~S in
Sec.~\ref{sec-scenario} is related to a limit 
cycle $U_{\rm lc}$ by
$u(t,r) = (-t)^{\alpha} U_{\rm lc}^{(\ln (-t))}(r (-t)^{-\beta})$.

%%%%%FIGFIGFIGFIGFIGFIGFIGFIGFIGFIGFIGFIGFIGFIGFIGFIGFIGFIGFIGFIG
%%%%%FIGFIGFIGFIGFIGFIGFIGFIGFIGFIGFIGFIGFIGFIGFIGFIGFIGFIGFIGFIG
\begin{figure}
\begin{center}
{\BoxedEPSF{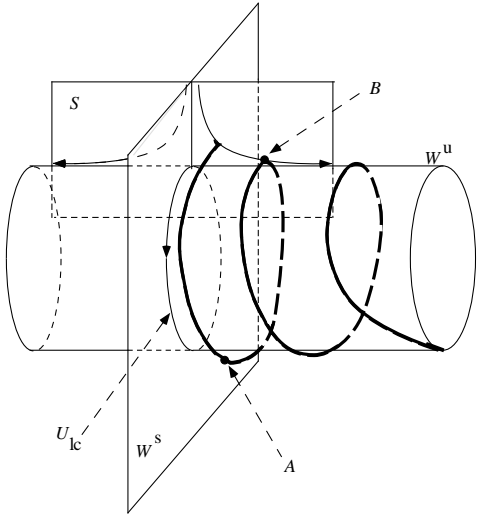 scaled 400} }
\end{center}
%\vspace*{60mm} 
%{\bf // PLACE Fig.~\ref{fig-poin} HERE. //} 
\caption{Schematic view of flows and the Poincar\'e section $S$.
}
\label{fig-poin}
\end{figure}
%%%%%FIGFIGFIGFIGFIGFIGFIGFIGFIGFIGFIGFIGFIGFIGFIGFIGFIGFIGFIGFIG
%%%%%FIGFIGFIGFIGFIGFIGFIGFIGFIGFIGFIGFIGFIGFIGFIGFIGFIGFIGFIGFIG
%

For a limit cycle $U_{\rm lc}$, 
it is convenient to consider
eigenmodes of 
$\displaystyle {\cal T}_{\Delta,U_{\rm lc}^{(0)}}$, where 
${\cal T}_{s, U}$ is defined in (\ref{eq:ind.tan}). 
An {\em eigenmode}\/ $F$ of 
$\displaystyle {\cal T}_{\Delta,U_{\rm lc}^{(0)}}$ 
is a function satisfying
\begin{equation}
        {\cal T}_{\Delta,U_{\rm lc}^{(0)}} F 
        = e^{\kappa \Delta}F.
        \label{eq:pert0:discr}
\end{equation}
The exponent $\kappa$, which is equal to the Lyapunov exponent,  
gives the mean growth rate of the perturbation $F$. 
{\em Relevant, irrelevant}, and {\em marginal modes} are defined 
according to the sign of ${\rm Re}\kappa$ as in Sec. \ref{sec-RG}.
It is easily seen that if $F$ is an eigenmode of 
$\displaystyle {\cal T}_{\Delta,U_{\rm lc}^{(0)}}$
then
$\displaystyle {\cal T}_{s,U_{\rm lc}^{(0)}}F$ 
is an eigenmode of 
$\displaystyle {\cal T}_{\Delta,U_{\rm lc}^{(s)}}$
with the {\em same} $\kappa$.  
Note that for a limit cycle $U_{\rm lc}$
there is a trivial marginal mode $F_0$ of 
$\displaystyle {\cal T}_{\Delta,U_{\rm lc}^{(0)}}F$ 
which is the tangent vector of $U_{\rm lc}^{(0)}$ in $\Gamma$: 
\begin{equation}
  F_0\equiv\lim_{s\rightarrow0}\frac{U_{\rm lc}^{(s)}-U_{\rm lc}^{(0)}}{s}
  = {\dot{\cal R}} (U_{\rm lc}^{(0)}).
  \label{eq:lc-marginal}
\end{equation}

To measure the `distance' between a vector  $U$ and the 
limit cycle $U_{\rm lc}$, we introduce 
\begin{equation}
  {\rm dist}(U, U_{\rm lc}) \equiv 
  \inf_{s} \| U - U_{\rm lc}^{(s)} \|     
  \label{eq:lc-3normdef} , 
\end{equation}
and we simply say ``$U$ is close to $U_{\rm lc}$'' 
when ${\rm dist}(U, U_{\rm lc})$ is small. 

We can now show the discrete versions of the claims in
Sec.~\ref{sec-scenario}:

\bigskip\noindent
{\bf Claim 1${}'$}
{\em Suppose there is a limit cycle $U_{\rm lc}$ of
  periodicity $\Delta$,  with no nontrivial marginal modes.  
  If (S1) and (U1) 
  hold with this limit cycle $U_{\rm lc}$, then there should be a {\em 
    unique} relevant mode for this $U_{\rm lc}$.
  Conversely, if $U_{\rm lc}$ has a unique relevant mode, then
  (S1) and (U1) hold with this $U_{\rm lc}$, at least for
  all the initial data  sufficiently close to $U_{\rm lc}$.  
  (Except, of course, for the exactly critical initial data.) 
  }

\bigskip\noindent
{\bf Claim~2${}'$}
{\em If the relevant mode is unique, the black hole mass satisfies 
  \begin{equation}
    M_{\rm BH} \propto (p-p^*)^{\beta/\kappa} 
    \label{eq:eq-expprime}
  \end{equation}
  for slightly supercritical solutions, 
  where $\kappa$ is the eigenvalue of the unique relevant mode of
  $\displaystyle {\cal T}_{\Delta,U_{\rm lc}^{(0)}}$. 
  }

Here ``self-similarity'' in (S1) and (S2) should be understood as 
``discrete self-similarity.''

Rough idea of the `proof' is as follows (see the following section
for rigorous results). 
We reduce the problem of a limit cycle to that of a fixed point%
\footnote{%
By reducing in this way, we are losing some information on detailed 
behaviour of the flow.  For example, if one is interested in a `wiggle' 
predicted in \cite{Gun96}, one should study the continuous time evolution 
directly.}
\footnote{%
After the submission of this paper we learned that the ``wiggle''
has been studied and numerically confirmed by Hod and Piran~\cite{HP96}. 
} 
by considering the Poincar\'e section $S,$ 
a codimension-one submanifold transverse to the limit cycle,
and the Poincar\'e map ${\cal P}$ on it induced by the renormalization 
group flow ${\cal R}$. (See Fig.~\ref{fig-poin}.)
$(S,{\cal P})$ defines a discrete dynamical system, for which the 
limit cycle in $\Gamma$ corresponds to a fixed point in $S$.
The behaviour of the map ${\cal P}$ on the Poincar\'e section $S$ is 
the same as that of the map ${\cal R}_\sigma$ on the whole phase space
$\Gamma$ in the continuously self-similar case.
Thus it follows {}from Claims~1 and 2 that Claims~1$'$ and 2$'$ hold.

%%%%%%%%%%%%%%%%%%%%%%%%%%%%%%%%%%%%%%%%%%%%%%%%%%%%%%%%%%%%%%%%%%%%%%%%
\subsubsection{The Poincar\'e map  and the local behaviour of the flow}
\label{subsec-discr.SS-Poin}

In this and next sections, we give rigorous results under
certain assumptions.  In this section, we consider the local structure 
of the phase space and show how the Poincar\'e map reduces the
problem of the limit cycle to that of the fixed point.

Let us first list assumptions used in the derivation, which
are similar to those in the case of continuous self-similarity.
To simplify the notation, we often write ${\cal T}_\Delta$ for 
${\cal T}_{\Delta, U^{(0)}_{\rm lc} }$ in the following. 

\noindent{\bf Assumption~L1${}'$}
  (Invariant subspaces of ${{\cal T}_\Delta}$) 
The tangent space $T_{U_{\rm lc}^{(0)}}\Gamma$ of $\Gamma$ at
$U_{\rm lc}^{(0)}$ is a direct sum of invariant subspaces of 
${\cal T}_{\Delta}$: 
\begin{equation}
  T_{U_{\rm lc}^{(0)}}\Gamma =  
  E^{\rm u}(U_{\rm lc}^{(0)}) \oplus E^{\rm c}(U_{\rm lc}^{(0)})
  \oplus E^{\rm s}(U_{\rm lc}^{(0)});
\end{equation}
where $E^{\rm c}(U_{\rm lc}^{(0)})$ is a one-dimensional eigenspace 
spanned by $F_{0}$ satisfying
\begin{equation}
  {\cal T}_{\Delta} (F_{0}) = F_{0};
  \label{eq:rg.Ass.1B.0}
\end{equation}
and $E^{\rm u}(U_{\rm lc}^{(0)})$ and $E^{\rm s}(U_{\rm lc}^{(0)})$ 
are of dimension $N$ and of codimension $N+1$, respectively, 
and there exists $\bar{\kappa}> 0$ such that 
\begin{eqnarray}
  \| {\cal T}_\Delta (F) \| \geq 
  e^{\bar{\kappa}\Delta} \| F \|, 
  && \quad F \in E^{\rm u}(U_{\rm lc}^{(0)}),
  \nonumber\\
  \| {\cal T}_\Delta (F) \| \leq 
  e^{- \bar{\kappa}\Delta} \| F \|, 
  && \quad F \in E^{\rm s}(U_{\rm lc}^{(0)}).
\end{eqnarray}

\noindent{\bf Assumption~L2${}'$ (Smoothness of ${\cal R}$.)}
The renormalization group transformation ${\cal R}$ considered as a
function 
${\Bbb R}\times\Gamma\ni(s,U)\mapsto{\cal R}_{s}U\in\Gamma$ 
is of class $C^{1,1}$
in a tubular neighborhood of the limit cycle $U_{\rm lc}$.
Namely, 
for $| s' - s |, \| U' - U \| < \delta_{0}$
and
$\sup_{0\le s''\le s}
{\rm dist}( {\cal R}_{s''} (U), U_{\rm lc})  < \delta_{0}'$,
\begin{eqnarray} 
  {\cal R}_{s'}U - {\cal R}_{s}U
  &=& (s'-s) {\dot{\cal R}} ({\cal R}_{s}U)  + O( |s'-s|^2),
  \nonumber\\
  &&\qquad 
  \label{eq:A2'.R-DR}
  \\
  {\cal R}_{s}U' -{\cal R}_{s}U
  &=& 
  {\cal T}_{s,U} (U'-U) +  O( \| U'-U \|^{2} ).
  \nonumber\\
  &&\qquad 
  \label{eq:A2'.R-T}
\end{eqnarray} 

\noindent{\bf Remark.}
Assumption L1${}'$ together with the existence of ${\dot{\cal R}}$ in
Assumption L2${}'$ is a natural limit cycle version of Assumption L1
of Sec.~\ref{sub-rgflow}. 
Assumption L2${}'$ is slightly stronger than Assumption L2. 
However, we consider it to be also physically reasonable
and adopt it in order to rule out subtle and unexpected behaviour of the
flow near the limit cycle $U_{\rm lc}$. 
Assumption L2${}'$ implies the following:
\begin{eqnarray} 
  {\dot{\cal R}} (U') - {\dot{\cal R}} (U) 
  &=& O( \| U' - U \| ) , 
  \label{eq:A2'.DR} 
  \\
  \left ( {\cal T}_{s', U} - {\cal T}_{s, U} 
  \right ) F 
  &=& O( | s'-s | \cdot \| F \| ) ,
  \label{eq:A2'.Ts}
  \\
  \left ( {\cal T}_{s, U'} - {\cal T}_{s, U} 
  \right ) F 
  &=& O( \| U' - U \| \cdot \| F \| ) . 
  \label{eq:A2'.T}
\end{eqnarray} 

Let us define the Poincar\'e map and reduce the problem to that of 
a dynamical system on a submanifold of codimension one.

Let $S$ be a submanifold of $\Gamma$ defined by
\begin{eqnarray}
  S\equiv U_{\rm lc}^{(0)}+Y,
  \quad
  Y\equiv  E^{\rm u}(U_{\rm lc}^{(0)})\oplus E^{\rm s}(U_{\rm lc}^{(0)}).
\end{eqnarray}
A point moving along a flow curve ${\cal R}_sU$ starting {}from point
$U$ on $S$ will return to $S$ for some $s\approx\Delta$, 
and we can define the {\em Poincar\'e map} ${\cal P}:S \to S$ so that 
${\cal P}(U)\in S$.
More precisely, we have the following lemma.

\begin{lemma}[The Poincar\'e map] 
  \label{lem-poin}
  Under Assumptions~L1$'$ and L2$'$, the following (i) and (ii) hold. 
  \\ 
  (i) 
  There is a $\delta$-neighborhood of $U_{\rm lc}^{(0)}$ in $S$ 
  and the Poincar\'e map ${\cal P}$ which is a local
  $C^{1,1}$-diffeomorphism {}from $S$ to $S$ satisfying 
  \begin{equation}
    {\cal P}(U)={\cal R}_{p(U)}(U), 
  \end{equation}
 where $p:S\to {\Bbb R}$ is $C^{1,1}$ and $p(U_{\rm lc}^{(0)})=\Delta$.
  In particular, the tangent map 
  $ {\cal T}^{\cal P}_{U_{\rm lc}^{(0)}}$ of ${\cal P}$ at $U_{\rm lc}^{(0)}$
  is given by 
  \begin{equation}
    {\cal T}^{\cal P}_{U_{\rm lc}^{(0)}}
    = 
    \left . 
      {\cal T}_{\Delta,U_{\rm lc}^{(0)}} 
    \right | _Y.
    \label{eq:TP on S}
  \end{equation}
  \\ \noindent (ii) 
  Furthermore, a tubular $\delta$-neighbourhood of $U_{\rm lc}$ is
  foliated by sections
  \begin{eqnarray}
    &&
    S^{(s)} \equiv U_{\rm lc}^{(s)}+Y^{(s)},
    \quad
    Y^{(s)} \equiv {\cal T}_{s,U_{\rm lc}^{(0)}}Y,
    \quad
    (s\in{\Bbb R}),
    \nonumber\\
    &&
  \end{eqnarray}
  and on each section there exists the Poincar\'e map ${\cal P}^{(s)}$ 
  which satisfies all of (i) with 
  $Y$, ${\cal P}$, $S$ and $U_{\rm lc}^{(0)}$ replaced by
  $Y^{(s)}$, ${\cal P}^{(s)}$, $S^{(s)}$ and $U_{\rm lc}^{(s)}$. 
  In particular, we have
  \begin{equation}
    {\cal T}^{{\cal P}^{(s)}}_{U_{\rm lc}^{(s)}}
    = 
    \left . 
      {\cal T}_{\Delta,U_{\rm lc}^{(s)}} 
    \right | _{Y^{(s)}}.
    \label{eq:TPs on Ss}
  \end{equation}
  The Poincar\'e maps ${\cal P}^{(s)}$ for various $s$
  are related by
  \begin{equation}
    {\cal P}^{(s+s')} \circ \bar{\cal P}_{s'}
    = \bar{\cal P}_{s'} \circ {\cal P}^{(s)},
    \label{eq:PP=PP}
  \end{equation}
  where 
  $\bar{\cal P}_{s'}: 
  U \mapsto {\cal R}_{{\bar p}(U)}(U)$ ($s\in I\equiv[-\Delta, \Delta]$)
  is a foliation--preserving local diffeomorphism on $\Gamma$
  which maps $S^{(s)}$ to $S^{(s+s')}$.
  Here $\bar p$ is a $C^{1,1}$-function determined by $s'$ such that
  ${\cal R}_{\bar p(U)}(U)\in S^{(s+s')}$,  and satisfies
  ${\bar p}(U_{\rm lc}^{(s)})=s'$.
  The tangent map ${\cal T}^{\bar{\cal P}_{s'}}_{U_{\rm lc}^{(s)}}$ of
  $\bar{\cal P}_s$ at $U_{\rm lc}^{(s)}$ satisfies
  \begin{equation}
    {\cal T}^{\bar{\cal P}_{s'}}_{U_{\rm lc}^{(s)}} 
    = {\cal T}_{s',U_{\rm lc}^{(s)}}.
    \label{eq:TPbar}
  \end{equation}
  Moreover $\bar{\cal P}$ is $C^{1,1}$, as a function on 
  $I\times  (\delta{\rm -neibourhood\; of}\; U_{\rm lc})$.
\end{lemma}

\noindent{\bf Remark.}
The choice of $S$ being a linear space, though in general 
it need not to be,  makes the subsequent 
discussion much simpler. In particular, by reducing 
the flow on $\Gamma$ to maps
on $S$, we can use the results in the fixed point case.

\medskip\noindent{\bf Remark.}
In fact, (i) is included in (ii). 
One could define $\bar{\cal P}_s$ first and then the Poincar\'e
map by ${\cal P}^{(s)}=\bar{\cal P}_\Delta|_{S^{(s)}}$.

\bigskip
\noindent{\em Proof.}
(i) The proof here is based on the standard treatment of the
Poincar\'e map~(see, e.g. \cite[Section~8]{Henr81}).
Assumption~L$1'$ immediately implies that  $S$ is {\em transverse} to
$U_{\rm lc}$ at $U_{\rm lc}^{(0)}$,
i.e., the direct sum of the tangent spaces is the whole tangent space, 
since 
$T_{U_{\rm lc}^{(0)}}S= E^{\rm u}(U_{\rm lc}^{(0)})
        \oplus E^{\rm u}(U_{\rm lc}^{(0)})$ 
and $T_{U_{\rm lc}^{(0)}}U_{\rm lc}=E^{\rm c}(U_{\rm lc}^{(0)})$.

Consider a map 
$f:\Gamma\times {\Bbb R}\times S\to \Gamma$,
$f(U,s,U')\equiv{\cal R}_s U - U'$.
We have $f(U_{\rm lc}^{(0)},\Delta,U_{\rm lc}^{(0)})=0$.
We show that we can implicitly define maps
$p:B(U_{\rm lc}^{(0)},\delta_0'', \Gamma) \to {\Bbb R}$ and 
${\cal P}: B(U_{\rm lc}^{(0)},\delta_0'', \Gamma) \to S$ 
by $f(U,p(U),{\cal P}(U))=0$
for some positive $\delta_0''$,
where $B(U_{\rm lc}^{(0)},\delta_0'',\Gamma)$ is an open ball 
of center $U_{\rm lc}^{(0)}$ and radius $\delta_0''$ in $\Gamma$.

Assumption~L$2'$ implies that $f$ is $C^{1,1}$, and the explicit form 
of the derivative of $f$ at ${(U,s,U')}$ with respect to $(s,U')$ is
\begin{eqnarray}
  \left(
    \frac{\partial f}{\partial s}, 
    \frac{\partial f}{\partial U'}
  \right): 
  {\Bbb R}\times Y &\to& T_{f(U,s,U')}\Gamma,
  \nonumber\\
  (t,F) &\mapsto& t {\dot{\cal R}}({\cal R}_sU)-F,
  \label{eq:pds of f}
\end{eqnarray}
where ${\partial f}/{\partial U'}$ is understood as 
a Fr\'echet derivative.
If ${\dot{\cal R}}({\cal R}_sU)\not\in Y$ 
this has a smooth inverse
\begin{eqnarray}
  &&\left(
    \frac{\partial f}{\partial s}, 
    \frac{\partial f}{\partial U'}
  \right)^{-1}(F)
  \nonumber\\
  &&\quad
  =
  \left(
    \frac{\| F-\pi_{f(U,s,U')}(F) \|}
                {\| {\dot{\cal R}}({\cal R}_sU) \|},
    -\pi_{f(U,s,U')}(F)
  \right)
  \label{eq:inv f}
\end{eqnarray}
for $F\in T_{f(U,s,U')}\Gamma$, 
where $\pi_{U}$ is the projection of 
$T_U\Gamma$ into $S$ along ${\dot{\cal R}}(U)$. 
At $(U_{\rm lc}^{(0)},\Delta,U_{\rm lc}^{(0)})$
there exists a smooth inverse (\ref{eq:inv f})
because of the transversality of $U_{\rm lc}$ and $S$, i.e.,
${\dot{\cal R}}({\cal R}_\Delta 
U_{\rm lc})=F_0 \not\in Y$.

Then the implicit function theorem implies that there exist
$\delta_0'' > 0$ and a one-to-one map 
$(p, \widetilde{ {\cal P} }): B(U_{\rm lc}^{(0)},\delta_0'',\Gamma)
\ni U \mapsto (p(U), \widetilde{ {\cal P} }(U))\in {\Bbb R}\times S$ 
satisfying $f(U,p(U), \widetilde{ {\cal P} }(U))=0$, and   
that $(p, \widetilde{ {\cal P} })$ is as smooth as $f$, i.e., of class
$C^{1,1}$. 
Denoting its partial derivative by 
$(q_U,{\cal T}^{ \widetilde{ {\cal P} } }_U)$,
we have
\begin{eqnarray}
  p(U') - p(U) &=& q_{U} (U'-U) +  O( \| U'-U \|^{2} ),
  \label{eq:p-q}
  \\
   \widetilde{ {\cal P} }(U') 
        -  \widetilde{ {\cal P} }(U) &=& 
  {\cal T}^{ \widetilde{ {\cal P} } }_{U} (U'-U) 
        +  O( \| U'-U \|^{2} ),
  \label{eq:P-TP}
  \\
  \left( q_{U'} - q_U \right) F 
        &=& O( \| U' - U \| \cdot \| F \| ), 
  \label{eq:q.conti.} 
  \\
  \left ( {\cal T}^{ \widetilde{ {\cal P} } }_{U'} 
  - {\cal T}^{ \widetilde{ {\cal P} } }_{U} \right ) F 
  &=& O( \| U' - U \| \cdot \| F \| ). 
  \label{eq:TP.conti}
\end{eqnarray}
The explicit form of the derivatives are given by
\begin{eqnarray}
  q_U(\cdot) 
  &=& \frac{\| (1-\pi_{{\cal R}_{p(U)}U}) \circ 
    {\cal T}_{p(U),U} (\cdot) \|}
  {\| {\dot{\cal R}}(U) \|},
  \label{eq:q.explicit} 
  \\
  {\cal T}^{ \widetilde{ {\cal P} } }_U 
  &=& \pi_{{\cal R}_{p(U)}U}\circ {\cal T}_{p(U),U}
  \label{eq:TP.explicit},
\end{eqnarray}
which follows {}from
\begin{eqnarray}
  \left(q_{U}, 
        {\cal T}^{ \widetilde{ {\cal P} } }_{U}\right)
  = 
  -\left(
    \frac{\partial f}{\partial s}, 
    \frac{\partial f}{\partial U'}
  \right)^{-1}
  \circ
  \frac{\partial f}{\partial U},
\end{eqnarray}
where the inverse derivative of $f$, evaluated 
at ${(U,p(U), \widetilde{ {\cal P} }(U))}$, exists and 
is given by (\ref{eq:inv f}).

The Poincar\'e map on $S$ is the restriction of 
$\widetilde{ {\cal P} }$ to  
$S\cap B(U_{\rm lc}^{(0)},\delta_0'',\Gamma)$,
which we  denote by ${\cal P}$.
The transversality of ${\dot{\cal R}} (U)$ and $S$ implies that this
restriction is a one-to-one map.
Eq. (\ref{eq:TP.explicit}) implies that 
\begin{equation}
  {\cal T}^{\cal P}_{U_{\rm lc}^{(0)}}
  = \left . 
  \pi_{U_{\rm lc}^{(0)}} 
        \circ {\cal T}_{\Delta,U_{\rm lc}^{(0)}}
  \right | _Y 
  = 
  \left . 
    {\cal T}_{\Delta,U_{\rm lc}^{(0)}} 
  \right | _Y , 
\end{equation}
where the second equality follows {}from invariance of $Y$ 
under ${\cal T}_{\Delta,U_{\rm lc}^{(0)}}$. 

\smallskip
\noindent%
(ii) It follows {}from Assumption 1$'$ and 
${\cal T}_{\Delta,U_{\rm lc}^{(s)}}
        \circ{\cal T}_{s,U_{\rm lc}^{(0)}}
        ={\cal T}_{s,U_{\rm lc}^{(0)}}
        \circ{\cal T}_{\Delta,U_{\rm lc}^{(0)}}$
that
${\cal T}_{s,U_{\rm lc}^{(0)}}E(U_{\rm lc}^{(s)})$'s are invariant subspaces of 
${\cal T}_{\Delta,U_{\rm lc}^{(s)}}$
so that $S^{(s)}$ and ${\dot{\cal R}}(U_{\rm lc}^{(0)})$ are transverse.
Then by the same argument as (i), one can define $\bar{\cal P}_s$
and the lemma is proved.
(The existence of a tubular neighborhood on which $\bar{\cal P}_s$ is
well defined follows {}from compactness of $I$.)
\hfill$\Box$

\bigskip

Though we will consider qualitative behaviour mainly on the Poincar\'e
section, we show here the qualitative behaviour in the whole phase space, 
which is the limit cycle version of Proposition~\ref{prop-3}(i).

\bigskip
\noindent
{\bf Proposition~\ref{prop-3}${}'$}
{\bf (Invariant manifolds of ${\cal R}$ at $U_{\rm lc}$)}
{\em Under Assumptions L1$'$ and L2$'$, 
        we have the following:
  In a tubular neighbourhood of $U_{\rm lc}$, 
  there exist a stable manifold $W^{\rm s}(U_{\rm lc})$ 
  of  codimension $N$, 
  and an unstable manifold $W^{\rm u}(U_{\rm lc})$ 
  of dimension $N+1$,
  whose tangent spaces at $U_{\rm lc}^{(s)}$ are 
  ${\cal T}_{s,U_{\rm lc}^{(0)}}
  (E^{\rm s}(U_{\rm lc}^{(0)}) 
        \oplus E^{\rm c}(U_{\rm lc}^{(0)}))$ 
  and 
  ${\cal T}_{s,U_{\rm lc}^{(0)}}
  (E^{\rm u}(U_{\rm lc}^{(0)}) \oplus 
        (E^{\rm c}(U_{\rm lc}^{(0)}))$,
  respectively.

  In terms of the Poincar\'e sections, 
  there exist a stable manifold
  $W^{\rm s}(U_{\rm lc}^{(s)},S^{(s)})$ and 
        an unstable manifold 
  $W^{\rm u}(U_{\rm lc}^{(s)},S^{(s)})$ 
  of each $(S^{(s)},{\cal P}^{(s)})$ and are given by
  \begin{eqnarray}
    W^{\rm s}(U_{\rm lc}^{(s)},S^{(s)}) &=& 
        \bar{\cal P}_s W^{\rm s}(U_{\rm lc}^{(0)},S),
    \\
    W^{\rm u}(U_{\rm lc}^{(s)},S^{(s)}) &=& 
        \bar{\cal P}_s W^{\rm u}(U_{\rm lc}^{(0)},S).
  \end{eqnarray}
  The stable and unstable manifolds 
        $W^{\rm s}(U_{\rm lc})$ and 
  $W^{\rm u}(U_{\rm lc})$ of the whole flow are written as
  \begin{eqnarray}
    W^{\rm s}(U_{\rm lc}) 
    &=& 
    \left \{
      \left.
        {\cal R}_s W^{\rm s}(U_{\rm lc}^{(0)},S) \, 
      \right| \, s\in {\Bbb R}
    \right \},
    \label{eq:W vs W on S;s}
    \\
    W^{\rm u}(U_{\rm lc}) 
    &=& 
    \left \{
      \left.
        {\cal R}_s W^{\rm u}(U_{\rm lc}^{(0)},S) \, 
      \right| \, s\in {\Bbb R}
    \right \}.
    \label{eq:W vs W on S;u}
  \end{eqnarray}
  }
\bigskip

\noindent{\em Proof.}
We now show that discrete dynamical system $(S,{\cal P})$ 
satisfies Assumptions L1 and L2 of Sec.~\ref{sub-rgflow}, i.e.,
we show that the assumptions hold under the replacement
\begin{equation}
  (\Gamma, {\cal R}_\sigma, {\cal T}_{\sigma,U})
  \longrightarrow
  (S, {\cal P}, {\cal T}^{\cal P}_{U}).
  \label{eq:replacement}
\end{equation}

Assumption L2 holds for $(S,{\cal P})$ 
because ${\cal P}$ is of class $C^{1,1}$
(see (\ref{eq:p-q})--(\ref{eq:TP.conti})).
Assumption L1 holds for $(S,{\cal P})$ 
because of (\ref{eq:TP on S}) and Assumption L1$'$.
Then Proposition~\ref{prop-3} of Sec.~\ref{sub-rgflow} 
implies that in $S$
there exists a stable manifold of codimension $N$ in $S$ 
and an $N$-dimensional unstable manifold of $U_{\rm lc}^{(0)}$.
Again, more sophisticated argument guarantees their 
smoothness~\cite[Theorem~5.2, Theorem~5.II.4]{Shub87}; 
they are of class $C^{1,1}$.

Next, consider a one-parameter family of discrete dynamical systems 
$(S^{(s)},{\cal P}^{(s)})$.
Lemma~\ref{lem-poin}(ii) states 
that 
${\cal P}_s$ is a diffeomorphism {}from $S$ to $S^{(s)}$ 
which preserves the action of the Poincar\'e maps on them.
This implies that 
$\lim_{n\rightarrow\infty}{\cal P}^{(s)}{}^n(U) = 
U_{\rm lc}^{(s)}$
if and only if
$\lim_{n\rightarrow\infty}{\cal P}^n 
(\bar{\cal P}_{-s}U)=U_{\rm lc}^{(0)}$
so that
$\bar{\cal P}_s$ sends
the stable manifold $W^{\rm s}(U_{\rm lc}^{(0)},S)$
to $W^{\rm s}(U_{\rm lc}^{(s)},S^{(s)})$.
It also implies that $W^{\rm s}(U_{\rm lc}^{(s)},S)$ is tangent to 
${\cal T}_{s,U_{\rm lc}^{(s)}}
E^{\rm s}(U_{\rm lc}^{(s)})$
so that
$W^{\rm s}(U_{\rm lc}^{(s)})$ is tangent to 
${\cal T}_{s,U_{\rm lc}^{(s)}}
(E^{\rm s}(U_{\rm lc}^{(s)}) \oplus E^{\rm c}(U_{\rm lc}^{(s)}) )$.
We have a similar result for the unstable manifold (consider
${\cal P}^{-1}$ in place of ${\cal P}$).
Thus there exist the stable and unstable manifolds 
$W^{\rm s}(U_{\rm lc})$ and $W^{\rm u}(U_{\rm lc})$
of codimension $N$ and dimension $N+1$ in
$\Gamma$, which are given by
(\ref{eq:W vs W on S;s}) and (\ref{eq:W vs W on S;u}),
respectively.
\hfill$\Box$
\bigskip

We now consider quantitative behaviour of the renormalization group
flow.
We assume the following, which, because of
Proposition~\ref{prop-3}$'$, asserts 
that the stable manifold is of codimension 1 and the unstable manifold 
is of dimension 2.  

\bigskip\noindent{\bf Assumption L1$'$A
  (Uniqueness of the relevant mode)}
{  
  \label{Ass-L1'A}
  $E^{\rm u}(U_{\rm lc}^{(0)})$ is one-dimensional, 
        i.e., $N=1$.
  Thus $E^{\rm u}(U_{\rm lc}^{(0)})$ is an 
        eigenspace spanned by $\Frel$
  satisfying
  \begin{equation}
    {\cal T}_{\Delta} (\Frel) = e^{\kappa \sigma} \Frel,
  \end{equation}
  where $\kappa$ is the unique positive eigenvalue of
  ${\cal T}_\Delta$.
  }
\bigskip

Consider the time evolution 
$U_{\rm init}^{(s)} \equiv {\cal R}_{s} (U_{\rm init})$
of an initial data $U_{\rm init}$ in $S^{(s_0)}$.
We are going to trace the evolution on the Poincar\'e section $S$. 
To measure how $U_{\rm init}$ is deviated {}from the critical surface, 
we introduce a decomposition of $U_{\rm init}$ in the tubular 
neighbourhood of 
$U_{\rm lc}$.  Lemma~\ref{lem-poin}(ii) implies that for any
$U_{\rm init}$ 
there is a unique $S^{(s_0)}\ni U_{\rm init}$.  
We can decompose any $U_{\rm init}$ as
(\ref{eq:Udiff-dcmp-init}) on $S^{(s_0)}\ni U_{\rm init}$:
\begin{equation}
  \label{eq:init-decomp}
  U_{\rm init} = U_{\rm init}{}_\perp 
  {}+ b\,\frac{F_1^{(s_0)}}{\|F_1^{(s_0)}\|}.
\end{equation}
As in the continuously self-similar case, we generically have
$b(0)=O(p-p_c)$.
In the special case of  $U_{\rm init} \in S$, 
Proposition~\ref{prop-4} applies for the dynamical
system $(S,{\cal P})$ and we immediately have the estimates in
Proposition~\ref{prop-4}.
In a general case where $U_{\rm init} \not\in S$ (say, the point $A$ of 
Fig.~\ref{fig-poin}), 
all we have to do is to estimate how $U_{\perp}$ and $b$ behave 
before the flow intersects $S$ for the first time (at the point $B$ of 
Fig.~\ref{fig-poin}); after that, we 
can directly apply Proposition~\ref{prop-4}.  
The estimate is given by the following lemma.

\begin{lemma}
\label{lem-discr-step0}
  Under Assumptions L1$'$, L2$'$ and L1$'$A 
  there exist positive constants $\delta$ 
        and $\epsilon$ 
  which satisfy the following:
  Let  $|b|\le \epsilon, \quad 
        \| U_\perp - U_{\rm lc}^{(s_{0})} \|\le\delta$,
  $s'=\Delta-s_0$ and 
  \begin{eqnarray}
    U &=& U_{\perp} + b\,F_1^{(s_0)}, 
    \quad U_{\perp} \in W^{\rm s}(U_{\rm lc}) 
        \cap S^{(s_0)},
    \\
    \bar U &=& \bar{\cal P}_{s'}U = \bar U_{\perp} 
        + \bar b\,F_1, 
    \quad \bar U_{\perp} \in W^{\rm s}(U_{\rm lc})\cap S.
  \end{eqnarray}
  Then we have
  \begin{eqnarray}
    &&
    \| \bar U_{\perp} - U_{\rm lc}^{(0)} \| \leq 
    C (\delta+ \epsilon^{2}), 
    \label{eq:discr-step0-perp}
    \\
    && 
    \frac{b}C \leq \bar b \le  Cb,
    \label{eq:discr-step0-para}
  \end{eqnarray}
  where $C$ is a positive constant.
\end{lemma}

\noindent{\em Proof.}
Let $b'\equiv b/\|F_1^{(s_0)}\|$ and 
$\bar b' \equiv \bar b/\|F_1\|$.
We have, for fixed $s_0$, 
\begin{eqnarray}
  \bar U
  &=& \bar{\cal P}_{s'}(U_{\perp}+b' F_1^{(s_0)}) 
  \nonumber\\
  &=& \bar{\cal P}_{s'}(U_{\perp})
  + b'{\cal T}^{\bar{\cal P}_{s'}}_{U_{\perp}}F_1^{(s_0)}
  + O(b'^2)
  \nonumber\\
  &=& \bar{\cal P}_{s'}(U_{\perp})
  + b'{\cal T}^{\bar{\cal P}_{s'}}_{U_{\rm lc}^{(s_0)}{}_\perp}F_1^{(s_0)}
  \nonumber\\
  &&{}+
  O\left(|b'| 
  \left(|b'| + \left \| U_{\perp} - U_{\rm lc}^{(s_0)} 
        \right \| \right)
  \right)
  \nonumber\\
  &=& \bar{\cal P}_{s'}(U_{\perp})
  + b'F_1
  + O\left(|b'| 
  \left(|b'| + \left \| U_{\perp} - U_{\rm lc}^{(s_0)} 
        \right \| \right)
  \right),
  \nonumber\\
  &&
  \label{eq:Ubar-estimate}
\end{eqnarray}
where we have used (\ref{eq:TPbar}) at the fourth equality.
($O(x)$ above may depend on $s'$ or $s_0$.)
Since $\bar{\cal P}_{s'}(U_{\perp})\in W^{\rm s}(U_{\rm lc}^{(0)})$,
we have
\begin{eqnarray}
  \bar U_{\perp} &=& \bar{\cal P}_{s'}(U_{\perp}) 
  + O\left(\epsilon  ( \epsilon+\delta ) \right),
        \label{eq:lem-9.first}
  \\
  \bar b' &=& b' \left(1+ O( \epsilon+\delta ) \right).
        \label{eq:lem-9.second}
\end{eqnarray}
The first equality gives
\begin{eqnarray}
  \| \bar U_{\perp} -U_{\rm lc}^{(0)} \| 
  &\le& \| \bar{\cal P}_{s'} (U_{\perp}) 
    -\bar{\cal P}_{s'}(U_{\rm lc}^{(s_0)}) \| 
    + O\left(\epsilon(\epsilon+\delta ) \right)
  \nonumber\\
  &=& O(\delta+\epsilon^2),
\end{eqnarray}
where we have used the smoothness of $\bar{\cal P}$ in 
Lemma~\ref{lem-poin}.
Eq. (\ref{eq:lem-9.second}) gives
\begin{eqnarray}
  \bar b = b\cdot{\|F_1^{(s_0)}\|}
  \left(1+ O( \epsilon+\delta ) \right)
  =b\cdot O(1).
\end{eqnarray}
Since $0\le s'\le\Delta$, the $O(x)$ terms above are bounded by 
a constant independent of $s_0$.
\hfill$\Box$

\bigskip

%%%%%%%%%%%%%%%%%%%%%%%%%%%%%%%%%%%%%%%%%%%%%%%%%%%%%%%%%%%%%%%%%%%%%%%%
\subsubsection{Global behaviour of the flow}
\label{subsec-discr.SS-global}
Now that the local flow structures have been analyzed, we can state 
a rigorous theorem which corresponds to  Theorem~\ref{prop-final}, 
under moderate assumptions on the global structures of the flow 
just  as in the case of continuous self-similarity.  
One possibility is:
\begin{theorem}
\label{prop-final-discr}
Suppose Assumptions L1$'$, L2$'$, L1$'$A, G2, and G3 hold, and Assumption G1
holds for the discrete dynamical system $(S, {\cal P})$. Then 
the following hold.  
\\
(i) 
All the initial data sufficiently close to $U_{\rm lc}$ (but not
exactly on $W^{\rm s}(U_{\rm lc}$) once approach, but finally deviate 
{}{}from, $U_{\rm lc}$.  The final fate of these data are either a flat
space-time or a black hole, depending on which side of 
$W^{\rm s}(U_{\rm lc})$ lies the initial data. \\
(ii) 
Consider the 
time evolution of the initial data 
(\ref{eq:init-decomp})
with $\| U_{\perp} - U_{\rm lc}^{(s_0)} \|\leq \delta$, 
$0<|b| \leq \delta$, 
which is on the side of black hole of $W^{\rm s}(U_{\rm lc})$. 
Then as $b$ goes to zero, the mass  of the black hole 
formed satisfies 
\begin{equation}
        C_1 | b |^{\beta/\kappa} \leq
        \frac {M_{\rm BH}}{M_0} \leq C_2 | b |^{\beta/\kappa} 
        \label{eq:Prop.finaldiscr}
\end{equation}
with $b$-independent positive constants $C_1, C_2$. 
Here $M_0$ is the initial (say, ADM or Bondi-Sachs) mass.%
\end{theorem}

\noindent{\em Proof.}
Lemma \ref{lem-discr-step0} reduces the problem into the dynamical
system $(S,{\cal P})$ on the Poincar\'e section $S$, where Theorem
\ref{prop-final} can be applied.
\hfill$\Box$

%%%%%%%%%%%%%%%%%%%%%%%%%%%%%%%%%%%%%%%%%%%%%%%%%%%%%%%%%%%%%%%%%%%%%%%%
%%%%%%%%%%%%%%%%%%%%%%%%%%%%%%%%%%%%%%%%%%%%%%%%%%%%%%%%%%%%%%%%%%%%%%%%
\subsection{Universality class}
\label{sec-univ.class}

In this section, we focus on the problem of the universality in the 
original sense of the term in statistical mechanics, based on 
renormalization  group philosophy.  We have largely benefitted {}from the 
formulation of \cite{BKL94}.
This is the rigorous version of the heuristic argument of universality
class given in \cite{Koi96}.
\footnote{%
After the first draft of this paper was submitted for publication, we 
have learned that a heuristic argument of universality class
has also been presented by Gundlach and Martin-Garcia~
\cite{GM96}, which is at the same level as had been
presented in \cite{Koi96}.}
  
Consider general cases in which the PDEs do not necessarily
satisfy the fundamental assumption of scale invariance, Assumption~S  
in Sec.~\ref{sec-scenario}, or self-similar solutions with the scale 
invariance is not relevant for a critical behaviour (e.g. having more 
than one relevant mode).    
Methods of renormalization group gives us a clear understanding on 
which systems can exhibit the same critical behaviour.
The main point is that the renormalization group transformation drives
the equations of motion to a fixed point, 
where they gain the scale invariance,
and the problem reduce to the cases treated in previous sections.
As a result, we understand why and to what extent condition (U3a)
holds.  Knowing the {\em universality class}  
(the class of models which exhibit the same critical behaviour) 
is very important {}from a physical point of view, 
because it guarantees that certain models exhibit the same critical 
behaviour with a model in the same class, 
whose analysis could be much simpler.  (Imagine we want to deal with a 
realistic matter.  It would be almost impossible to know or deal with 
the exact form of the microscopic interaction.  
Universality could make the problem easier and accessible, by 
reducing it to that for a simpler model, e.g. perfect fluid.)

Suppose unknowns $u$ satisfy the PDEs of the following form, which is
not necessarily scale invariant:
\begin{equation}
  L\left(u, \frac{\partial u}{\partial t}, 
  \frac{\partial u}{\partial r}, t, r \right)
  =0. 
  \label{eq:uc-evol-orig.}  
\end{equation}
Let us define the scaling transformation ${\cal S}(s, \alpha, \beta):
u(t,r)\mapsto u^{(s)}(t,r)=e^{\alpha s}u(e^{-s}t,e^{-\beta s}r)$
as in (\ref{eq:sctrdef}).  Then $u^{(s)}$ satisfies equations
\begin{equation}
  L^{(s)}\left(u^{(s)}, \frac{\partial u^{(s)}}{\partial t}, 
  \frac{\partial u^{(s)}}{\partial r}, t, r \right)
  =0,
  \label{eq:uc-evol-scaled}  
\end{equation}
where 
\begin{eqnarray}
  && L^{(s)}\left(u^{(s)}, 
  \frac{\partial u^{(s)}}{\partial t}, 
  \frac{\partial u^{(s)}}{\partial r}, 
  t, r \right)
  \nonumber \\
  && = e^{\gamma s}L
  \Biggl ( e^{-\alpha s} u^{(s)},\,
  e^{(-\alpha+1)s}
  \frac{\partial u^{(s)}}{\partial t}, \,
   e^{(-\alpha+\beta)s}
  \frac{\partial u^{(s)}}{\partial r},\, 
  \nonumber \\
  && \hspace{20mm} 
  e^{-s}t,  e^{-\beta s}r \Biggr ),
  \label{eq:uc-L-scaled}
\end{eqnarray}
and $\gamma=(\gamma_1, \gamma_2,...,\gamma_m)$ (where $m$ is the
number of component of $L$) is a set of constants which may depend on
$\alpha$ and $\beta$ and are chosen 
so that $L^{(s)}$ remains finite and nonzero when $s\rightarrow\infty$.
It should be noted that if $L^{(s)}=L$ holds with some $\gamma$ 
the system $L=0$ is invariant under scaling transformation 
in the sense of Sec.~\ref{sec-RG}.

%%%%%FIGFIGFIGFIGFIGFIGFIGFIGFIGFIGFIGFIGFIGFIGFIGFIGFIGFIGFIGFIG
%%%%%FIGFIGFIGFIGFIGFIGFIGFIGFIGFIGFIGFIGFIGFIGFIGFIGFIGFIGFIGFIG
\begin{figure}
\begin{center}
{\BoxedEPSF{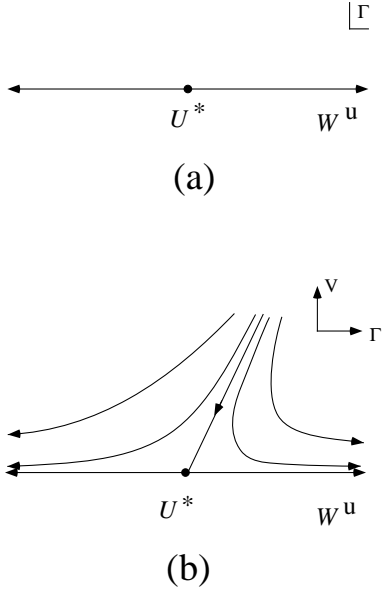 scaled 400} }
\end{center}
%\vspace*{60mm} 
%{\bf // PLACE Fig.~\ref{fig-withV} HERE. //} 
\caption{Conceptual diagram of RGT: (a) the case $V=0$, (b) the case with 
`irrelevant' $V$.  To simplify the figure, only the unstable manifold 
$W^{\rm u}$ is shown.
}
\label{fig-withV}
\end{figure}
%%%%%FIGFIGFIGFIGFIGFIGFIGFIGFIGFIGFIGFIGFIGFIGFIGFIGFIGFIGFIGFIG
%%%%%FIGFIGFIGFIGFIGFIGFIGFIGFIGFIGFIGFIGFIGFIGFIGFIGFIGFIGFIGFIG
%

A renormalization group transformation is now defined to be a 
{\em pair}  of a transformation on the EOM: $L \mapsto L^{(s)}$, and 
the following transformation on the $\Gamma$: 
\begin{eqnarray}
    && {{\cal R}_{s,L}}: U_{i} (\xi)=u_{i} (-1,\xi) \nonumber \\ 
        && \mapsto 
    U_{i}^{(s)}(\xi)= u_{i}^{(s)} (-1, \xi) = 
    e^{\alpha_{i} s} u_i(-e^{- s}, e^{- \beta s}\xi),
\end{eqnarray}
where we added subscript $L$ to show explicitly
the RGT's dependence on $L$.  See Fig.~\ref{fig-withV}.  
We emphasize that the renormalization 
group transformation is now to be considered as a pair of two 
transformations on different spaces: the space of EOM, and the space 
of initial data (phase space $\Gamma$). 
One can easily show the (semi)group property
\begin{equation}
        {\cal R}_{s_{1} + s_{2},L} 
        = {\cal R}_{s_{2},L^{(s_{1})}} \circ {\cal R}_{s_{1},L},
        \label{eq:uc-semigrp} 
\end{equation}
or
\begin{eqnarray}
  & &{\cal R}_{{s_n},L}
  ={\cal R}_{s_n-s_{n-1},{L^{(s_{n-1})}}}\circ\cdots
  {\cal R}_{{s_2-s_1},{L^{(s_1)}}}\circ{\cal R}_{{s_1},L},
  \nonumber\\
  & & \qquad s=s_n>s_{n-1}>...>s_0=0.
\end{eqnarray}
This states that long time evolution problem of the original equation 
defined by $L$ 
is the composition of finite time evolution problems of equations with
(in general) different PDEs $L^{(s)}=0$. 

We now turn to the problem of determining the universality class.  
Concretely, we consider the following situation.  
Suppose, with a certain choice of $\alpha$, $\beta$,  and $\gamma$,  
$L^{(s)}$ approaches a fixed point (i.e. a fixed function of its 
arguments) $L^*$ asymptotically:
\begin{equation}
  L^{(s)}\rightarrow L^* \quad (s\rightarrow\infty),
  \qquad  (L^*)^{(s)}=L^*.
\end{equation}
It follows {}from definition (\ref{eq:uc-L-scaled}) of transformation
of $L$ that the fixed point $L^*=0$ satisfies the scale invariance in
the sense of Sec.~\ref{sec-RG}.
Suppose, in addition,  there exists a fixed point (or a limit cycle)
of ${\cal R}_{s,L^*}$ which has a unique relevant mode.  And the 
question is: under what circumstances do $L$ and $L^{*}$ exhibit the 
same critical behaviour (i.e. in the same universality class)?  

At first glance the question looks rather trivial; a naive guess would  
be that they are always in the same universality class, because 
$L^{(s)}$ approaches $L^{*}$.  However, there is a rather subtle 
problem involved here:  Values of $L^{(s)}$ and 
$L^{*}$ at the {\em same argument} will approach each other as 
$s \rightarrow \infty$;   but this does not necessarily mean that their 
difference at their respective (generally different) $U^{(s)}$,  
which are defined by the time evolution by $L^{(s)}$ and $L^{*}$, goes 
to zero.  In fact, if one starts {}from the {\em 
same initial condition} this usually does not hold (e.g. the critical 
value $p_c(V)$ is shifted due to $V = L - L^*$); we have to prove 
structural stability for these systems.  

In the 
following, we provide a sufficient condition under which the 
above naive guess does hold.  To simplify the notation, we define 
$V^{(s)} \equiv L^{(s)} - L^{*}$, and introduce 
$||| V(U) |||$ which measures the ``strength'' of the difference 
$V(U)$.  The strength $|||\cdot|||$ should be so 
chosen that the following Assumption~2C should hold. 
We write the time evolution operator in terms of  $L$ by 
${\cal R}^{V}$, and that by $L^{*}$ by ${\cal R}$; and their 
tangent maps by the same superscript convention. 
We assume that 
the finite time evolution operators are regular with respect to 
this difference.  
Concretely, we assume (in addition to Assumption~L1 of and Assumption~L2 
Sec.~\ref{sub-rgflow}, which are about time evolutions {\em without} 
$V$) : 

\bigskip
\noindent{\bf Assumption~L2V (Regularity in the presence of $V$)}
There are  positive $K_{1}, K_{3}, \delta_{0}$ and  $\sigma$ 
such that the following hold for 
$\| U - U^{*} \|, \| F \| \leq \delta_{0}$ and 
$||| V(U) ||| \leq \delta_{0}$: 
\begin{eqnarray}
        && {\cal R}^{V}_{\sigma} (U+F) =  
        {\cal R}^{V}_{\sigma} (U) + {\cal T}^{V}_{\sigma, U} (F) 
        + \Oabs(K_{1} \| F \|^{2} ) 
        \label{eq:iniv.reg0} \\
        && 
        \| {\cal R}^{V}_{\sigma} (U) - {\cal R}_{\sigma} (U) \| 
        \leq  K_{3} ||| V(U) ||| 
        \label{eq:univ.reg} \\
        && 
        \| {\cal T}^{V}_{\sigma, U} (F) - {\cal T}_{\sigma, U} (F) \| 
        \leq  K_{3} ||| V(U) ||| \cdot \| F \| 
        \label{eq:univ.regT}
\end{eqnarray}

\bigskip\noindent
These guarantee the structural stability of the flow in the neighbourhood 
of $U^{*}$ and $V=0$.  Concretely, we have the following Proposition, 
which corresponds to Proposition~\ref{prop-3} (i) of 
Sec.~\ref{sub-rgflow}.  We denote by $E^{\rm s}(U^{*})$ and 
$E^{\rm u}(U^{*})$ the invariant 
subspaces of ${\cal T}_{\sigma, U^{*}}$, {\em in the absence of $V$}. 
\begin{prop}
\label{prop-3prime} 
  Suppose there is a fixed point $U^{*}$ of ${\cal R}_{s,L^*}$ 
  with a unique relevant mode, i.e. Assumptions~L1, L2, L1A of 
  Sec.~\ref{sub-rgflow} hold, and $||| V^{(s)} (U) ||| \rightarrow 0$ 
  for $\| U - U^{*} || \leq \delta_{0}$.  Suppose further that 
  Assumption~L2V holds.  Then  there exist small 
  positive constants $\delta_{1}', \delta_{1}''$ (how small is 
  specified in the proof) such that if 
  \begin{equation}
          ||| V^{(s)} (U) ||| \leq \delta_{1}'' \qquad {\rm for} 
          \quad  \| U - U^{*} \| \leq \delta_{1}' 
        \label{eq:cond.on.V}
  \end{equation}
  then for any $F^{\rm s} \in E^{\rm s}(U^{*})$ with 
  $\| F^{\rm s} \| \leq \delta_{1}'$, there exists 
  $F^{\rm u}_{\rm c}(F^{\rm s}) \in E^{\rm u}(U^{*})$ such that 
  \begin{equation}
          {\cal R}_{n \sigma} (U^{*} + F^{\rm s} 
          + F^{\rm u}_{\rm c}(F^{\rm s})) 
          \rightarrow U^{*}
        \label{eq:crit.sol.withV}
  \end{equation}
  as $n \rightarrow \infty$. 
\end{prop} 
  
We define 
\begin{equation}
        W^{\rm c} (U^{*}; V) \equiv \{ U^{*} + F^{\rm s} 
                  + F^{\rm u}_{\rm c}(F^{\rm s}) \} 
        \label{eq:Wcdef-withV}
\end{equation}
as we did in (\ref{eq:defcrit.sol1}).     Note that now the definition 
depends on $V$.

\bigskip\noindent 
{\em Sketch of the Proof of Proposition~\ref{prop-3prime}.} 
The proof is done by Bleher-Sinai argument as was done for 
Proposition~\ref{prop-3}.  We sketch the proof by mainly pointing out 
necessary modifications. 

{\em Step~1. Decomposition.} \/ 
Same as the proof of Proposition~\ref{prop-3}.  We employ the same 
decomposition (\ref{eq:rg.qual.dcmp}), 
and for each $F^{\rm s}$, try to find $a_{c}$ such that 
\begin{equation}
        {\cal R}_{n \sigma} (U^{*} + F^{\rm s} + a_{c} \Frel) 
                  \rightarrow U^{*} .
        \label{eq:goal.quan.V}
\end{equation}

{\em Step~2. Recursion.} \/ 
By (\ref{eq:univ.reg}), we have 
\begin{equation}
        {\cal R}^{V}_{\sigma}(U^{*} +F) = 
        {\cal R}^{V}_{\sigma}(U^{*} +F) + \Oabs(K_{3} ||| V(U^{*} +F) |||) 
        \label{eq:1st-step.V}.  
\end{equation}
Calculating ${\cal R}^{V}_{\sigma}(U^{*} +F)$ just as in 
(\ref{eq:detail.phys.ex2}), we this time obtain 
[cf. (\ref{eq:rg.qual.4}), (\ref{eq:rg.qual.5}), (\ref{eq:rg.qual.6})] 
\begin{eqnarray}
        a_{n+1} & = & e^{\kappa\sigma} a_{n} 
                + \Oabs(K_{1} \| F_{n} \|^{2}) + \Oabs(K_{3} v_{n}), 
        \label{eq:detail.phys.ex3.V} \\
        F^{\rm s}_{n+1} & = & {\cal T}_{\sigma, U^{*}} (F^{\rm s}_{n})  
                + \Oabs(K_{1} \| F_{n} \|^{2}) + \Oabs(K_{3} v_{n}), 
        \label{eq:rg.qual.4.V} 
\end{eqnarray}
where 
$\displaystyle v_{n} \equiv \sup_{\| U - U^{*} \| \leq \delta_{0}} 
||| V^{(n\sigma)} (U) |||$.  Note that the only difference {}from 
Sec.~\ref{subsub-rg.qual} is the presence of $\Oabs(K_{3} v_{n})$.

{\em Step~3. Continuity.} \/ 
Same as the proof of Proposition~\ref{prop-3}.  Assumption~2C 
guarantees the continuity in $a_{0}$, even in the presence of $V$. 

{\em Step~4. Solving the recursion.} \/ 
Same as the proof of Proposition~\ref{prop-3}.  Our goal is to find 
sequences $r_{n}, t_{n}$ for which (a)---(c) of Step~4 of 
the proof of Proposition~\ref{prop-3} hold. 

{\em Step~5. Proof of (a)---(c).} \/ 
This part significantly differs {}from the proof of 
Proposition~\ref{prop-3}, due to the presence of the term 
$\Oabs(K_{3} v_{n})$ 
in the recursion.  This time, we define $t_{n} = r_{n}$ as follows. 
First define $\bar{v}_{n}  \equiv  \sup_{k \geq n} v_{k}$ 
and choose $\delta_{1}'$ and $\delta_{1}''$ as 
\begin{equation}
        \delta_{1}' \leq \frac{1 - e^{-\bar{\kappa}\sigma}}{2 K_{1}}, 
        \qquad 
        \delta_{1}'' \equiv \frac{1 - e^{-\bar{\kappa}\sigma}}{2 K_{3}} 
        \delta_{1}' .
        \label{eq:deltadef.V}
\end{equation}
Then define $r_{n} = t_{n}$ recursively, starting {}from 
$r_{0} = \delta_{1}'$ as 
\begin{equation}
        r_{n+1} = 
        \left ( e^{-\bar{\kappa}\sigma} + K_{1} \delta_{1}' 
        \right ) r_{n} + K_{3} \bar{v}_{n}
        \label{eq:rn-def-V-rec}
\end{equation}
or ($\bar{\bar{\Lambda}}' 
\equiv e^{-\bar{\kappa}\sigma} + K_{1} \delta_{1}' < 1$)
\begin{equation}
        r_{n} = \left ( \bar{\bar{\Lambda}}' 
        \right )^{n} \delta_{1}' 
        + K_{3} \sum_{k=0}^{n-1} \bar{v}_{k} 
        \left ( \bar{\bar{\Lambda}}' 
        \right )^{n-k-1}
        \label{eq:rn-def-V}
\end{equation}

We now show that (a)---(c) is satisfied by this choice of $r_{n}$.  
First, dividing the sum over $k$ in (\ref{eq:rn-def-V}) according to 
$\; k \grless n/2 \; $, 
(\ref{eq:rn-def-V}) implies ($\bar{v}_{k}$ is non-increasing in $k$ 
by definition) 
\begin{eqnarray}
        r_{n} & \leq & \left ( \bar{\bar{\Lambda}}' 
                \right )^{n} \delta_{1}' 
                + K_{3} \bar{v}_{0} \sum_{k=0}^{n/2} 
                \left ( \bar{\bar{\Lambda}}' 
                \right )^{n-k-1} 
                \nonumber \\
        & & {}  + K_{3} \bar{v}_{n/2} 
                \sum_{k=n/2}^{n-1} 
                \left ( \bar{\bar{\Lambda}}' 
                \right )^{n-k-1}.  
        \label{eq:rn-to0.V}
\end{eqnarray}
Three terms on the RHS goes to zero (as $n \rightarrow \infty$), 
either because $e^{-\bar{\kappa}\sigma} + K_{1} \delta_{1}' < 1$ or 
$\bar{v}_{n/2} \rightarrow 0$.  This proves (a).  
Similar reasoning also shows the following uniform bound: 
\begin{equation}
        r_{n} \leq \left ( \bar{\bar{\Lambda}}' 
                \right )^{n} \delta_{1}' 
                + K_{3} \sum_{k=0}^{n-1} \delta_{1}'' 
                \left ( \bar{\bar{\Lambda}}' 
                \right )^{n-k-1}
        \leq 2 \delta_{1}' .
        \label{eq:rn-unibd.V}
\end{equation}

Now we turn to (b) and (c).  First note that it is sufficient to 
prove [cf. (\ref{eq:BScond21}), (\ref{eq:BScond22})] 
\begin{eqnarray}
        r_{n+1} & \geq & 
                e^{-\bar{\kappa}\sigma} r_{n} 
                + K_{1} (r_{n} \vee t_{n})^{2} + K_{3} v_{n} 
                \nonumber \\ 
        t_{n+1} & \leq & e^{\kappa\sigma} t_{n} 
                - K_{1} (r_{n} \vee t_{n})^{2} - K_{3} v_{n} .
                \nonumber 
\end{eqnarray} 
Because we are taking $r_{n} = t_{n}$ and because 
$r_{n} \leq 2 \delta_{1}'$, it is then sufficient to prove 
\begin{eqnarray}
        r_{n+1} & \geq & 
        \left ( e^{-\bar{\kappa}\sigma} + K_{1} \delta_{1}' \right ) r_{n}
        + K_{3} \bar{v}_{n} 
                \label{eq:rcond1.V} \\ 
        r_{n+1} & \leq & 
        \left ( e^{\kappa\sigma}  - K_{1} \delta_{1}' \right ) r_{n}
        - K_{3} \bar{v}_{n} .
                \label{eq:rcond2.V}
\end{eqnarray} 
Eq.(\ref{eq:rcond1.V}) is satisfied, by our definition of $r_{n}$.  
And (\ref{eq:rcond2.V}) is satisfied, if 
\begin{equation}
        ( e^{\kappa\sigma} - e^{-\bar{\kappa}\sigma} - 2 K_{1} \delta_{1}' ) 
        r_{n} 
        \geq 2 K_{3} \bar{v}_{n} .
        \label{eq:rcond3.V}
\end{equation}
It is not difficult to see (by induction) that (\ref{eq:rcond3.V}) is 
in fact satisfied under our choice of $\delta_{1}'$ and 
$\delta_{1}''$.  This completes Step~5.  

\hfill$\Box$

\bigskip

To complete the story, we have to make additional assumptions on the 
global behaviour of the flow in the presence of $V$.  Concretely we assume: 

\bigskip
\noindent{\bf Assumption~GV (Global behaviour in the presence of $V$)} 
Consider the time evolution of initial data 
$U_{\pm} \in A_{\pm}(\epsilon_{2}, \delta_{3})$ in the presence of $V$. 
Then there exist $\epsilon_{2}, \delta_{3}, \delta_{4} > 0$ such that if 
$|||V(U_{\pm})||| \leq \delta_{4}$ then 
Assumptions~G1, G2, G3 hold for suitably chosen $C, C'$. 

\bigskip
\noindent 
In the above, $A_{\pm}(\epsilon_{2}, \delta_{3})$ was 
defined in (\ref{eq:deApm}), in terms of dynamics without $V$.

Under these preparations, we can now state our theorem. 

\begin{theorem}
\label{theorem-11}
Suppose there exist $\alpha, \gamma$ and $\beta>0$ satisfying
  $L^{(s)}\rightarrow L^*\;(s\rightarrow\infty$), where 
  $L^{*}$ is a fixed point of the scaling transformation: 
  $(L^*)^{(s)}=L^*$.  Suppose  
  there is a fixed point or limit cycle $U^{*}$ of ${\cal R}_{s,L^*}$ 
  with a unique relevant mode, i.e. Assumptions~L1, L2, L1A of 
  Sec.~\ref{sub-rgflow} hold.  Moreover, suppose Assumption~2C and 
  Assumption~GV hold.   
  Then, 
  \\ \noindent (i) 
  if there exists $\delta_{0} >0$ such that for 
  $\| U - U^{*} \| \leq \delta_{0}$ we have 
  (1) $||| V^{(s)} (U) ||| \rightarrow 0$ as $s \rightarrow \infty$ and 
  (2)  $||| V^{(s)} (U) ||| < \delta_{5}$ for 
  sufficiently small $\delta_{5}$, 
  then the system $L=0$ exhibits 
  qualitatively the same critical behaviour as the system $L^{*} =0$. 
  (i.e. there is a critical behaviour and near-critical solutions 
  once approach and then deviate {}from $U^{*}$). 
  \\ \noindent (ii) 
  If in addition  $||| V^{(s)}(U) ||| \rightarrow 0$ 
  sufficiently fast (for example, $||| V^{(s)}(U) |||$ integrable 
  in $s$ on $(1, \infty)$ is sufficient), the system $L=0$ exhibits 
  quantitatively the same critical behaviour as the system $L^{*} =0$ 
  (i.e. the same critical exponent). 
\end{theorem}

\bigskip\noindent 
{\bf Remark.} 
In Assumption~GV, $V$ in fact means $V^{(s)}$ at the instant when 
the flow exits the 
perturbative region through the tube $A_{\pm}(\epsilon_{2}, 
\delta_{3})$: 
the local analysis which corresponds to that of 
Proposition~\ref{prop-4} shows 
that the near (but not exactly) critical solutions once 
approach and then deviate {}from $U^{*}$, and they exit 
{}from the perturbative region through the tubes $A_{\pm}$.  
Because $|||V^{(s)}(U) ||| \rightarrow 0$ as $s \rightarrow \infty$, 
the effect of $V$ goes to zero as $p \rightarrow p_c(V)$, and could be 
neglected.  However, we made the Assumption~GV in order to avoid pathological 
situation where the slight effect caused by this small 
$|||V^{(s)}(U) |||$ leads to 
qualitatively different global behaviour.  (Because the solution is 
expected to blow up, we cannot expect nice regularity in $V$ in this region.) 

\medskip

Theorem~\ref {theorem-11} explains how and why claim (U3a) of 
Sec.~\ref{sec-scenario} holds. 
The mere condition 
$L \rightarrow L^{*}$ is not 
sufficient to guarantee the universality.  The following proof in 
fact suggests the deviation in the critical behaviour: 
\begin{equation}
  M_{\rm BH} \propto |p - p_{c} |^{\beta/ \kappa} \, f(p)
  \label{eq:univ-deviation}
\end{equation}
where
\begin{equation}
        f(p) \approx %\frac{1}{\log | p - p_{c}| } 
        \exp \left [ c_{2} \int_{1}^{c_{1} \log | p - p_{c}|} V(s) ds 
        \right ] 
        \label{eq:univ-deviation2} ,  
\end{equation}
with positive constants $c_{1}, c_{2}$. 
Concretely, $f(p)$ will be a power of logarithm 
in $| p - p_{c}|$ for $||| V^{(s)} ||| \propto 1/s$, and 
\begin{equation}
  f(p) \propto 
  \exp \left ( {\rm const.}  \log^{1 -\alpha} |p - p_{c} | \right ) 
  \label{eq:univ-deviation3}
\end{equation}
for $||| V^{(s)} ||| \propto s^{-\alpha}$ ($0 < \alpha < 1$).  
The deviation in (\ref{eq:univ-deviation3}) is greater than 
logarithmic, although it does not change the critical exponent (if we 
define it as the limit of the logarithmic ratio of both hand sides).

\bigskip\noindent 
{\em Proof of Theorem~\ref{theorem-11}.} 
Proposition~\ref{prop-3prime} shows the existence of a critical solution.  
We now trace the time evolution of off-critical solutions. 

We employ the scheme of Sec.~\ref{subsub-rg.quan2}.  That is, 
we decompose the initial data $U$ as [cf. (\ref{eq:Udiff-dcmp-init})] 
\begin{equation}
        U = U_{\rm c} + b \Frel 
        \label{eq:Udiff-dcmp-initV}
\end{equation}
and for $n >0$ decompose [cf. (\ref{eq:Udiff-dcmp-non0})] 
\begin{equation}
        {\cal R}^{V}_{n\sigma} (U) 
        = {\cal R}^{V}_{n\sigma} (U_{\rm c}) + F_{n} 
        \label{eq:Udiff-dcmp-non0V}
\end{equation}
and further decompose $F_{n}$ as in (\ref{eq:Udiff-dcmp-F}): 
\begin{equation}
        F_{n} = b_{n} \Frel + F_{n}^{\rm s} . 
        \label{eq:Udiff-dcmp-FV}
\end{equation}
Note that $U_{\rm c}$ here means the critical solution {\em in the 
presence of $V$} (and its time evolution). 
Then arguing as we did in deriving (\ref{eq:ap.ev1}), we now obtain, 
using (\ref{eq:iniv.reg0}) 
\begin{eqnarray} 
        && {\cal R}^{V}_{(n+1)\sigma} (U) = 
        {\cal R}^{V}_{\sigma} 
        \bigl (U^{(n\sigma)}_{\rm c} + F_{n} \bigr ) 
        \nonumber \\ 
        && \quad = {\cal R}^{V}_{\sigma} (U^{(n\sigma)}_{\rm c} ) 
        + {\cal T}^{V}_{\sigma, U^{(n\sigma)}_{\rm c} } (F_{n}) 
        + \Oabs(K_{1} \| F_{n} \|^{2} ) 
\end{eqnarray}
which implies 
\begin{equation}
        F_{n+1} = {\cal T}^{V}_{\sigma, U^{(n\sigma)}_{\rm c}} (F_{n}) 
                + \Oabs(K_{1} \| F \|^{2} ) . 
        \label{eq:univ.Fnplone}
\end{equation}
Now using (\ref{eq:univ.regT}) and (\ref{eq:A.pert}), 
\begin{eqnarray}
        {\cal T}^{V}_{\sigma, U} (F) 
        & = & 
        \bigl [ {\cal T}^{V}_{\sigma, U} - {\cal T}_{\sigma, U} \bigr ] 
        (F) 
        + \bigl [ {\cal T}_{\sigma, U} - {\cal T}_{\sigma, U^{*}} \bigr ] 
        (F)
        \nonumber \\
        && {} +  {\cal T}_{\sigma, U^{*}} (F) 
        \nonumber \\
        & = & {\cal T}_{\sigma, U^{*}} (F)  
        + \Oabs( K_{3} ||| V (U) ||| \cdot \| F \|) 
        \nonumber \\ 
        && {} + \Oabs(6  K_{1} \cdot \| U - U^{*} \|  \cdot \| F \|) 
\end{eqnarray} 
So decomposing $F_{n}$ into $F^{\rm s}_{n}$ and $F^{\rm u}_{n}$ in 
(\ref{eq:univ.Fnplone}), we have recursions for $b_{n}$ and 
$f_{n} \equiv \| F^{\rm s}_{n} \|$, 
$g_{n} \equiv \| U^{(n\sigma)}_{\rm c} - U^{*} \|$: 
\begin{eqnarray}
        && b_{n+1} = e^{\kappa\sigma} b_{n}     
                + K_{3} (| b_{n} | \vee f_{n}) \Oabs(v_{n}) 
        \nonumber \\ 
        && \qquad +  K_{1} (| b_{n} | \vee f_{n}) 
        \bigl [ \Oabs (| b_{n} | \vee f_{n}) 
                + 6 \Oabs(g_{n}) 
        \bigr ] 
        \label{eq:rec-ineq-V1}
        \\
        && f_{n+1} \leq e^{-\bar{\kappa}\sigma} f_{n} 
        \nonumber \\
        && \quad +  (| b_{n} | \vee f_{n}) 
        \bigl [ K_{1} (| b_{n} | \vee f_{n})
                + 6 K_{1} g_{n}
                + K_{3} v_{n}
        \bigr ] 
        \label{eq:rec-ineq-V2}
\end{eqnarray}

As seen, the only difference here {}from (\ref{eq:ap.ev42}) and 
(\ref{eq:ap.ev41}) is the appearance of $v_{n}$.  The recursion is 
of the form considered in Lemma~\ref{lem-5} 
($6 K_{1} g_{n} + K_{3} v_{n}$ playing the role of $g_{n}$ there), 
and this proves the quantitative 
estimate which corresponds to Proposition~\ref{prop-4}.  In particular,  
Lemma~\ref{lem-5} guarantees that 
the presence of $O(v_k)$ does not affect the critical behaviour. 

The above concludes the local analysis of the flow in the neighbourhood 
of $U^{*}$, which corresponds to that in Sec.~\ref{subsub-rg.qual} and 
\ref{subsub-rg.quan2}.  The analysis of global behaviour, corresponding 
to that of Sec.~\ref{subsub-rg.global}, can be carried out in a similar 
manner under the Assumption~GV. 
\hfill$\Box$

\bigskip\noindent 
{\bf Remark.} 
By refining the the proof of Proposition~\ref{prop-3prime}, we can 
show in general\footnote{
Heuristically, we can go further.  
If $V$ is of definite sign, we can heuristically  
get a lower bound on the difference, and show 
\begin{equation}
        | p_c(V) - p_c(0) |= O( ||| V(U^{*}) |||) . 
        \label{eq:pcVestimate2}
\end{equation} 
This estimate gives an answer to the following question 
considered in \cite{HP96b}: fix the value of 
$p$  to the critical value when $V=0$ (i.e. $p = p_c(0)$), and vary the 
magnitude of $V$.  What kind of critical behaviour can we observe?  
The above estimate, together with Theorem~\ref{theorem-11} shows that 
we will observe the {\em same} critical 
exponent $\beta_{\rm BH} = \beta / \kappa$, because 
\begin{equation}
        M_{\rm BH} (p_c(0); V) \propto |  p_c(0) - p_c(V) | ^{\beta_{\rm BH}} 
        \propto  ||| V(U^{*}) |||^{\beta_{\rm BH}} 
\end{equation}
where we used Theorem~\ref{theorem-11} in the first step, and 
(\ref{eq:pcVestimate2}) in the second. 
}
($p_c(0)$ denotes the critical value 
without $V$) 
\begin{equation}
        | p_c(V) - p_c(0) |\leq  O( ||| V(U^{*}) |||) . 
        \label{eq:pcVestimate1}
\end{equation}

\bigskip
We conclude this section by giving two examples to which the above 
analysis can be applied.  Note that the following arguments are not 
rigorous in contrast to what we have been doing, in the sense that 
no rigorous verification of Assumptions are given.

%%%%%%%%%%%%%%%%%%%%%%%%%%%%%%%%%%%%%%%%%%%%%%%%%%%%%%%%%%%%%%%%%%%%%%%%%
\subsubsection{Example: perfect fluid with a modified equation of state}
\label{subsub-univ.rf}

The first example is a perfect fluid, with equation of state 
($f$ is a given function) 
\begin{equation}
        p = (\gamma-1) \rho + f(\rho) . 
        \label{eq:univ.state}
\end{equation}
The EOM for this system is given by (\ref{eq:EOM-fll.rt}) of the next
section,  
supplemented by the above equation of state. 
When $f \equiv 0$, this is a usual perfect fluid, to be studied 
in detail beginning with the next section.  It will be shown that the 
system with $f \equiv 0$ exhibits a critical behaviour governed by a 
self-similar solution $U^{*}$, which is a fixed 
point of a renormalization group transformation induced {}from a scaling 
transformation ${\cal S}$ whose effect on $\rho$ and $p$ is: 
\begin{eqnarray}
        \rho^{(s)} & \equiv & e^{-2 \beta s} \rho(e^{-s} t, e^{-\beta s} r) \\
        p^{(s)} & \equiv & e^{-2 \beta s} p(e^{-s} t, e^{-\beta s} r), 
\end{eqnarray}
where $\beta >0$ is arbitrary due to gauge invariance. 
The EOM (\ref{eq:EOM-fll.rt}) is invariant under the scale transformation, 
and so is the equation of state with $f \equiv 0$: 
$L^{*}_{6} \equiv p - (\gamma-1) \rho$. (Here we write $L_{6}$ 
because this is the sixth of EOM.)  Thus the system 
with $f \equiv 0$ is an example to which  our scenario in 
Sec.~\ref{sec-scenario} and Sec.~\ref{sub-rgflow} applies.  

Our question is to determine the models (forms of $f$) which belong 
to the same universality class as the $f \equiv 0$ model. 
Now under the scaling transformation the EOM in question,  
$L_{6} \equiv  p - (\gamma-1) \rho - f(\rho)$,  is transformed into 
\begin{equation}
        L^{(s)}_{6} = p^{(s)} - ( \gamma -1) \rho^{(s)} - f^{(s)} (\rho^{(s)}) 
        \label{eq:univ.rf.EOM3}
\end{equation}
where 
\begin{equation}
        f^{(s)} (x) \equiv e^{- 2\beta s} f( e^{2 \beta s} x) . 
        \label{eq:univ.ftr}
\end{equation}
At the fixed point in question, 
$\rho^{*}$ and $p^{*} = (\gamma -1) \rho^{*}$ are finite 
smooth functions of their argument (Sec.~\ref{sec-ss}). 
So we can expect Assumption~2C to 
hold in a neighbourhood of the fixed point, by simply taking 
$||| V ||| \equiv f(\rho)$.  
Then Theorem~\ref{theorem-11} implies 
that the model belongs to the same universality class as that of 
$f \equiv 0$, as long as (for fixed $x$) 
\begin{equation}
        f^{(s)} (x) \rightarrow 0 \qquad (s \rightarrow \infty).
        \label{eq:univ.fcond}
\end{equation}
This is the sufficient condition we are after.  An example of $f$ 
which shows the same critical behaviour as $f \equiv 0$ is given by 
$f(x) = x^{\delta}$ with $0 < \delta < 1$.

%%%%%%%%%%%%%%%%%%%%%%%%%%%%%%%%%%%%%%%%%%%%%%%%%%%%%%%%%%%%%%%%%%%%%%%%%
\subsubsection{Example: scalar field collapse}
\label{subsub-univ.scl}
Let us present another example, i.e. 
spherically symmetric collapse of a scalar field.

We consider a scalar field $\phi$ with energy--momentum tensor
\begin{equation}
  T_{ab}=\nabla_a\phi\nabla_b\phi
  -\frac12g_{ab}g^{cd}\nabla_c\phi\nabla_d\phi
  -g_{ab}V(\phi),
  \label{eq:scalar:en.mom.tsr.}
\end{equation}
where $g_{ab}$ is the metric tensor, $V$ is any, say, $C^1$-,
function, and the abstract index notation is used. 
The line element of a general spherically symmetric
space-time can be written as 
\begin{equation}
  ds^2= - g(u,r)\,{\bar g}(u,r)du^2-2g(u,r)dudr+r^2d\Omega^2.
  \label{eq:sphsym:l.e.}
\end{equation}
We fix the gauge by the condition $g(\cdot,0)=1$.
The equations of motion can be written in terms of retarded time $u$,
area radius $r$ and field variable $h(u,r)$ as
$L_V[h]=0$, where functional $L$ labeled by $V$ 
is given by~\cite{Chr86a}
\begin{equation}
  L_V[h]\equiv D_Vh
  -\frac12 (h-\av h)\;\frac {\partial{\bar g_V}[h]}{\partial r}
  +\frac12 r\, g[h]\, V'(\av h),
\end{equation}
where $\av \cdot$, $g$, $\bar g$, and $D_V$ are functionals defined as
\begin{eqnarray}
  \av f&\equiv& \frac1r\int_0^r dr f,
  \nonumber\\
  g[f]&\equiv& \exp\left({-4\pi r \aav{r^{-1}(f-\av f)^2}}\right),
  \nonumber\\
  {\bar g_V}[f]&\equiv& 
  \left\langle{g[f]\,(1-8\pi r^2V(\av f))}\right\rangle,
  \nonumber\\
  D_V f&\equiv& \frac{\partial f}{\partial u}
  -\frac12{\bar g_V[f]}\frac{\partial f}{\partial r}.
  \label{eq:scalar:Ein.,integro.}
\end{eqnarray}
In the equation of motion, $L_V[h]=0$ above, 
metric variables $g$ and $\bar g$
have been solved as functionals $g$ and $\bar g_V$, 
respectively, of $h$,
and the scalar field is given by $\phi=\av h$.
Hereafter, for a functional $F$,  
$F[h]$ denotes a function determined by a function $h$
and $F[h](a,b)$ denotes the value of function $F[h]$ at $(a,b)$.
For functionals $F=\av\cdot, g, g_V, D_V$,
we rewrite the values $F[h](e^{-s}u,e^{-\beta s}r)$ in terms of 
$h^{(s)}(u,r)$ 
using the relation 
$h^{(s)}(u,r)=e^{\alpha s}h(e^{-s}u,e^{-\beta s}r)$:
\begin{eqnarray}
  \av h (e^{-s}u,e^{-\beta s}r) &=& e^{-\alpha s}\av {h^{(s)}}(u,r),
  \label{eq:fnls-scaled-first}
  \\
  g[h](e^{-s}u,e^{-\beta s}r) &=& 
  \left ( g[h^{(s)}] \right )^{e^{-2\alpha s}}(u,r),
  \\
  {\bar g_V}[h](e^{-s}u,e^{-\beta s}r)
  &=& 
  \Bigl \langle
    \left ( g[h^{(s)}] \right )^{e^{-2\alpha s}}
  \nonumber\\
  && \hspace*{-6em}
  {}\times
    \left ( 1-8\pi r^2 V^{(s)}(e^{-\alpha s}\av {h^{(s)}}) \right )
  \Bigr \rangle\,(u,r),
  \\
  (D_V h)(e^{-s}u,e^{-\beta s}r)
  &=& 
  \Bigl ( e^{(-\alpha+1) s}
    \frac{\partial h^{(s)}}{\partial u}
  \nonumber\\
  && \hspace*{-6em}
    -\frac12
    e^{(-\alpha+\beta) s}
    {\bar g_{V^{(s)}}[h^{(s)}]}
    \cdot
    \frac{\partial h^{(s)}}{\partial r}
  \Bigr ) (u,r).
  \label{eq:fnls-scaled-last}
\end{eqnarray}
where 
\begin{equation}
  {V^{(s)}}(a) = e^{-2\beta s} V(a).
  \label{eq:V-scaled}
\end{equation}
We find the renormalized functional $(L_V)^{(s)}$ with 
parameters $\alpha$, $\beta$ and $\gamma$
by substituting (\ref{eq:fnls-scaled-first})--(\ref{eq:fnls-scaled-last}) into  
$e^{\gamma s}L_V[h](e^{-s}u, e^{-\beta s}r)$ 
and reinterpreting it as $(L_V)^{(s)}[h^{(s)}](u,r)$, the value
of function $(L_V)^{(s)}[h^{(s)}]$ at $(u,r)$.

Let us find fixed points, i.e., $L_V$ satisfying $(L_V)^{(s)}=L_V$. 
It follows {}from (\ref{eq:fnls-scaled-last})--(\ref{eq:fnls-scaled-last}) that 
we must choose $\alpha=0$, $\beta=1$. 
Then (\ref{eq:fnls-scaled-first})--(\ref{eq:fnls-scaled-last})
and (\ref{eq:V-scaled}) read 
\begin{eqnarray}
  \av h (e^{-s}u,e^{-\beta s}r)&=&\av {h^{(s)}}(u,r),
  \\
  g[h](e^{-s}u,e^{-s}r)&=& g[h^{(s)}](u,r),
  \\
  {\bar g_V[h]}(e^{-s}u,e^{-s}r)&=& {\bar g_{V^{(s)}}}[h^{(s)}](u,r), 
  \\
  (D_V h)(e^{-s}u,e^{-\beta s}r)&=& e^{-s}(D_{V^{(s)}} h^{(s)})(u,r),
  \label{eq:av-scaled-2}
\end{eqnarray}
and
\begin{equation}
  {V^{(s)}}(a) =e^{-2s}V (a) . 
  \label{eq:V-scaled-2}
\end{equation}
Transformation of $L$ is thus given by
\begin{eqnarray}
  {(L_V)^{(s)}}[h^{(s)}](u,r)
  &\equiv &e^{(\gamma-1) s}L[h](e^{-s}u,e^{-\beta s}r)
  \nonumber\\
  &=&e^{(\gamma-1) s}L_{V^{(s)}}[h^{(s)}](u,r).
  \label{eq:L-scaled}
\end{eqnarray}
Choosing $\gamma=1$, we have 
${(L_V)^{(s)}}=L_{V^{(s)}}$, and 
{}from (\ref{eq:V-scaled-2}) we find that 
$L_0$ is the only fixed point, i.e., $L_0=0$ is the only model having
scale invariance,  among the models considered here.

Choptuik's original result~\cite{Cho93}, which was further confirmed
by a theoretical work by Gundlach \cite{Gun95,Gun96},   
was that the system $L_0=0$
shows critical behaviour with this choice of parameters,
$\alpha=0$, $\beta=1$, for the RGT. 
The system has a limit cycle which should have a unique relevant mode,
according to our scenario in previous sections.

It follows directly {}from (\ref{eq:V-scaled-2}) and 
(\ref{eq:L-scaled}) that 
\begin{equation}
  V^{(s)}\rightarrow V^*=0,\quad 
  (L_V)^{(s)}\rightarrow L_0 \quad 
  (s\rightarrow\infty),
  \label{eq:L-limit}
\end{equation}
and the convergence is so fast that Theorem~\ref{theorem-11} is applicable.
This shows that potential $V$ is an {\em irrelevant} term which
vanishes asymptotically.
Therefore, all $L_V$'s are in the same universality class as $L_0$,
i.e., all scalar field models with arbitrary smooth potential $V$
exhibits the same critical behaviour with the same critical
exponent as the minimally coupled massless scalar field.

%%%%%%%%%%%%%%%%%%%%%%%%%%%%%%%%%%%%%%%%%%%%%%%%%%%%%%%%%%%%%%%%%%%%%%%%%%
%%%%%%%%%%%%%%%%%%%%%%%%%%%%%%%%%%%%%%%%%%%%%%%%%%%%%%%%%%%%%%%%%%%%%%%%%%
%%%%%%%%%%%%%%%%%%%%%%%%%%%%%%%%%%%%%%%%%%%%%%%%%%%%%%%%%%%%%%%%%%%%%%%%%%
%%%%%%%%%%%%%%%%%%%%%%%%%%%%%%%%%%%%%%%%%%%%%%%%%%%%%%%%%%%%%%%%%%%%%%%%%%
\section{Equations of Motion. }
\label{sec-EOM}

We now begin our concrete analysis of gravitational collapse of 
perfect fluid, as a case study of renormalization group ideas. 

%%%%%%%%%%%%%%%%%%%%%%%%%%%%%%%%%%%%%%%%%%%%%%%%%%%%%%%%%%%%%%%%%%%%%%%%%%
%%%%%%%%%%%%%%%%%%%%%%%%%%%%%%%%%%%%%%%%%%%%%%%%%%%%%%%%%%%%%%%%%%%%%%%%%%

\subsection{Equations of Motion}
\label{sec-EOM.SS}

The line element of any spherically symmetric space-time is written as
\begin{equation}
\label{eq:LE}
        ds^2=-\alpha^2(t,r)dt^2+a^2(t,r)dr^2+r^2 (d\theta^2+\sin^2\theta 
d\phi^2).
\end{equation}

We assume the matter content is a perfect fluid having
energy-momentum tensor
$
        T_{ab}=\rho u_au_b+p(g_{ab}+u_au_b),
$
where $\rho$ is the density, $p$ is the pressure, and 
$u^a$ is a unit timelike vector whose components are given by: 
\begin{equation}
\label{u-exp}
        u_t={- \alpha \over \sqrt{1-V^2}},
        \quad      
        u_r={a V \over \sqrt{1-V^2}}.  
\end{equation}
Here $V$ is the 3-velocity of fluid particles.  Introducing locally 
$U \equiv V / \sqrt{1 - V^{2}}$ and $W \equiv 1 / \sqrt{1 - V^{2}}$, 
the equations of motion (EOM) 
are given in this coordinate by:

\begin{mathletters}
\label{eq:EOM-fll.rt}  %% Label for all the following. Note this location!!
\begin{equation}
        \label{eq:EOM-fll.rt.3}
        \hspace*{-6mm} \frac{a_{,r}}{a}  = \frac{1 - a^2}{2 r} 
        + 4 \pi r a^2 (\rho W^2 + p U^2 ) 
\end{equation}

\begin{equation}
        \label{eq:EOM-fll.rt.4}
        \hspace*{-6mm} \frac{\alpha_{,r}}{\alpha}  
        = \frac{a^2 -1}{2 r} + 4 \pi r a^2 (\rho U^2 + p W^2  ) 
\end{equation}

\begin{equation}
        \label{eq:EOM-fll.rt.5}
        \hspace*{-6mm} \frac{a_{,t}}{a}  
        = - 4 \pi r a \alpha (\rho + p) U W
\end{equation}

\begin{equation}
\label{eq:EOM-fll.rt.1}
        ( a \rho W )_{,t} + p ( a W)_{,t}
        + \frac{ ( r^2 \alpha \rho U )_{,r} + p ( r^2 \alpha U )_{,r} }{r^2} 
        = 0 
\end{equation}

\begin{equation}
\label{eq:EOM-fll.rt.2}
        (a p U )_{,t} + \rho ( a U )_{, t} 
        + (\alpha p W)_{,r} + \rho (\alpha W)_{, r}
        = 0 
\end{equation}
\end{mathletters}
In the above, first three are obtained by taking linear combinations of 
$(0,0)$, $(1,1)$, $(0,1)$ components of 
Einstein equation.  The last two are obtained by taking linear 
combinations of the ``Bianchi identity''  ${T^{\mu \nu}}_{;\nu} = 0$. 
We have to supplement the above by the equation of state for 
the fluid.  We consider the case 
\begin{equation}
\label{eq:EOM-fll.state}
        p=(\gamma-1)\rho , 
\end{equation}
where $\gamma \in (1, 2)$ is a constant (adiabatic index).

In terms of variables $s \equiv -\ln(-t), x \equiv \ln (- r/t)$, 
and introducing  
\begin{equation}
        N \equiv {\alpha\over a e^x},\quad 
        A \equiv a^2,\quad %%d={\rho\over 4\pi r^2 }. 
        %%\Omega = 4 \pi r^2 \rho  . 
        \omega \equiv 4 \pi r^2 a^2 \rho , 
\end{equation}
we can write the 
equations of the system in an autonomous form, which makes the scale 
invariance of the system transparent: 

\begin{mathletters}
\label{eq:EOM-fll}  %% Label for all the following. Note this location!!
\begin{equation}
        \label{eq:EOM-fll.3}
        \hspace*{-6mm} \frac{ A_{,x} }{A}= 
        1 - A 
        + {{2\,\omega\,\left( 1 + (\gamma -1) V^2 \right) }\over {1 - V^2}} 
\end{equation}

\begin{equation}
        \label{eq:EOM-fll.4}
        \hspace*{-6mm} \frac{N_{,x}} {N} = - 2 + A - (2 - \gamma) \omega 
\end{equation}

\begin{equation}
        \label{eq:EOM-fll.5}
        \hspace*{-6mm} \frac{ A_{,s} }{A} + \frac{ A_{,x} }{A} 
        = - \frac{2 \gamma N V \omega} {1 - V^2} 
\end{equation}

\begin{eqnarray}
\label{eq:EOM-fll.1}
        &&\frac{ \omega_{,s} }{\omega}
        + \frac{\gamma V V_{,s}}{1 - V^2} 
        + ( 1 + N V )   \frac{\omega_{,x}}{\omega} 
        + \frac{\gamma (N+V)  V_{,x}}{1 - V^2} \nonumber \\
        & & \;\; = \frac{3(2-\gamma)}{2} N V - \frac{2 + \gamma}{2} A N V 
        + (2 - \gamma) N V \omega 
\end{eqnarray}

\begin{eqnarray}
\label{eq:EOM-fll.2}
        && (\gamma-1) V \frac{\omega_{,s}}{\omega} 
        + \frac{\gamma V_{,s}}{1 - V^2} \nonumber \\
        && \;\; \;\; 
        + (\gamma -1) (N + V) \frac{\omega_{,x}}{\omega} 
        + \frac{\gamma (1 + N V)  V_{,x}}{1 - V^2} \nonumber \\
        &&\;\; = - (\gamma - 2) (\gamma - 1 ) N \omega 
        + \frac{7 \gamma - 6}{2} N 
        + \frac{2 - 3 \gamma}{2} A N 
\end{eqnarray}
\end{mathletters}

Only four out of the above five equations are independent.  More 
precisely, the eqn. (\ref{eq:EOM-fll.5}) is automatically  satisfied 
by solutions of the set (\ref{eq:EOM-fll.3}),  (\ref{eq:EOM-fll.4}),  
(\ref{eq:EOM-fll.1}) and   (\ref{eq:EOM-fll.2}), as long as they 
satisfy a boundary condition 
$A(s, - \infty) =1,  V(s, -\infty) = \omega(s, -\infty) = 0$. 
In view of this, we pick
up the above set of four equations as our basic equations of motion, 
and use (\ref{eq:EOM-fll.5}) as an auxiliary equation at 
appropriate stages in the following.

%%%%%%%%%%%%%%%%%%%%%%%%%%%%%%%%%%%%%%%%%%%%%%%%%%%%%%%%%%%%%%%%%%%%
%%%%%%%%%%%%%%%%%%%%%%%%%%%%%%%%%%%%%%%%%%%%%%%%%%%%%%%%%%%%%%%%%%%%
\subsection{Gauge degrees of freedom}

The only coordinate transformation which preserves the form of 
  (\ref{eq:LE}) is 
\begin{equation}
\label{eq:CT} 
        t\mapsto F^{-1}(t), 
\end{equation}
which corresponds to
\begin{equation}
        (s,x) \mapsto (f^{-1}(s),x-s+f^{-1}(s)),
\label{eq:CT:(s,x)}
\end{equation}
where $F^{-1}(t)=-e^{-f^{-1}(s)}$ with $\dot{f} \equiv df/ds  \neq 0$.
Under this transformation, the variables $h$ transform to $\tilde h$, 
where 
\begin{equation}
\label{eq:h}
        \tilde{h}(s,x) = 
        \left \{ 
        \begin{array}{ll}
        h(s,x) & (h = A, \omega, V) \\
        \dot{f} (s) h(s,x) & (h = N) 
        \end{array} 
        \right .    .
\end{equation}
Eqs.(\ref{eq:EOM-fll})  are 
of course invariant under this transformation. 
Fixing the value of $N$ at a point, for example, $x=0$,
 of each constant $s$ line
determines the coordinate system completely. We shall retain the degree 
of freedom for the time being.

One note:  The gauge degree of freedom present in our system 
enables us to take $\beta =1$ in our formulation of Sec.~\ref{sec-RG}, 
{\em without} introducing new variables.

%%%%%%%%%%%%%%%%%%%%%%%%%%%%%%%%%%%%%%%%%%%%%%%%%%%%%%%%%%%%%%%%%%%%%%%%%%
%%%%%%%%%%%%%%%%%%%%%%%%%%%%%%%%%%%%%%%%%%%%%%%%%%%%%%%%%%%%%%%%%%%%%%%%%%
%%%%%%%%%%%%%%%%%%%%%%%%%%%%%%%%%%%%%%%%%%%%%%%%%%%%%%%%%%%%%%%%%%%%%%%%%%
%%%%%%%%%%%%%%%%%%%%%%%%%%%%%%%%%%%%%%%%%%%%%%%%%%%%%%%%%%%%%%%%%%%%%%%%%%
\section{The critical (self-similar) solution.} 
\label{sec-ss}

To carry out the first step of our scenario, we first have to find 
out fixed points (i.e. self-similar solutions) of RGT.  
In this section, we find out (almost) all of the self-similar solutions 
of the EOM.  
Because this paper is intended to present the ``renormalization group''
approach, we here mildly try to exhaust all possible self-similar solutions.

%%%%%%%%%%%%%%%%%%%%%%%%%%%%%%%%%%%%%%%%%%%%%%%%%%%%%%%%%%%%%%%%%%%%%%%%%%
%%%%%%%%%%%%%%%%%%%%%%%%%%%%%%%%%%%%%%%%%%%%%%%%%%%%%%%%%%%%%%%%%%%%
\subsection{Equations for self-similar space-time}

We first require that the space-time is self-similar, i.e. that $N$ and $A$ 
depend only on $x$: $N=N_{\rm ss}(x), A=A_{\rm ss}(x)$. 
Conversely, it can be shown (see Appendix ??) that any spherically 
symmetric self-similar space-times can be expressed in that 
form if a  freedom of coordinate transformation
(\ref{eq:CT}) is used.    
Then it follows {}from  (\ref{eq:EOM-fll}) 
that $\omega_{\rm ss}$ and $V_{\rm ss}$ are also 
functions of $x$ only: $\omega=\omega_{\rm ss}(x), V=V_{\rm ss}(x)$. 
In this sense, the space time we are interested in 
is a {\em fixed point} of the RGT. 

The equations for a self-similar solution, 
${\dot{\cal R}} (U^{*}) =0$,  are  then given by omitting 
terms containing derivatives of $s$ in (\ref{eq:EOM-fll}): 

%%\widetext
\begin{mathletters}
\label{eq:EOM-ss}  %% Label for all the following. Note this location!!
\begin{equation}
\label{eq:EOM-ss.3}
        \hspace*{-6mm}  \frac{A_{,x}}{A} = 1 - A 
        + {{2\,\omega\,\left( 1 + (\gamma -1) V^2 \right) }\over {1 - V^2}} 
\end{equation}
\begin{equation}
\label{eq:EOM-ss.4}
        \hspace*{-6mm} \frac{N_{,x}}{N} = - 2 + A - (2 - \gamma) \omega 
\end{equation}
\begin{equation}
\label{eq:EOM-ss.5}
        \hspace*{-6mm} \frac{A_{,x}}{A}_{,x} 
        = - \frac{2 \gamma N V \omega} {1 - V^2} 
\end{equation}
\begin{eqnarray}
\label{eq:EOM-ss.1}
        &&   ( 1 + N V ) \frac{\omega_{,x}}{\omega} 
        + \frac{\gamma (N+V)  V_{,x}}{1 - V^2} \nonumber \\
        && \;\; = \frac{3(2-\gamma)}{2} N V - \frac{2 + \gamma}{2} A N V 
        + (2 - \gamma) N V \omega 
\end{eqnarray}
\begin{eqnarray}
\label{eq:EOM-ss.2}
        && (\gamma -1) (N + V)  \frac{\omega_{,x}}{\omega} 
        + \frac{\gamma (1 + N V)  V_{,x}}{1 - V^2} \nonumber \\
        && \;\; = (2 - \gamma) (\gamma - 1) N \omega 
        + \frac{7 \gamma - 6}{2} N 
        + \frac{2 - 3 \gamma}{2} A N 
\end{eqnarray}
\end{mathletters}
%%\narrowtext

{}{}from  (\ref{eq:EOM-ss.3}) and (\ref{eq:EOM-ss.5}), one 
obtains an algebraic identity
\begin{equation}
\label{eq:alg-ss}
        1 - A 
        + {{2\,\omega\,\left( 1 + (\gamma -1) V^2 \right) }\over {1 - V^2}}
        = - \frac{2 \gamma N V \omega} {1 - V^2} . 
\end{equation}
One could eliminate one variable (e.g. $A$) {}from 
(\ref{eq:EOM-ss}) using the 
above identity.  However, to avoid making equations too complicated, 
we keep using four variables ($A, N, \omega, V$) in the following, 
using the above  (\ref{eq:alg-ss}) as a check at appropriate 
stages of our numerical calculation.

Under the self-similar ansatz, the coordinate freedom
(\ref{eq:CT})  
reduces to $t\mapsto\tilde t=k t$ with constant $k$, 
which corresponds to the translation 
of $x$ [the other transformations 
alter the constant $x$ lines in $(t,r)$ space]. 
This freedom of coordinate transformation allows one to adjust the 
value of $N$ at a given point arbitrarily. 
We make use of this freedom, and  fix the coordinate system by 
requiring that the sonic point (see below) be at $x=0$.

%%%%%%%%%%%%%%%%%%%%%%%%%%%%%%%%%%%%%%%%%%%%%%%%%%%%%%%%%%%%%%%%%%%%%%%%%
%%%%%%%%%%%%%%%%%%%%%%%%%%%%%%%%%%%%%%%%%%%%%%%%%%%%%%%%%%%%%%%%%%%%%%%%%

\subsection{Conditions on self-similar solutions} 
\label{sec-ss.cond-ss}

The behaviour of self-similar solutions has been extensively 
discussed by Bogoyavlenskii in \cite{Bog77}, followed by
other works \cite{OP90,FoHe93}.  
We here briefly outline our analysis, in our coordinate system, 
for completeness.

We are interested in self-similar solutions  which 
satisfy the following two properties:
\begin{description}
\item [(i)]  The 
self-similar solution is analytic (or at least smooth, in the 
sense of having an asymptotic expansion to all orders) 
for all $x \in {\Bbb R}$, 
\item [(ii)]  The space-time and the matter are regular,
$A=1$ and $V=0$, at the center ($x = - \infty$).  
\end{description}

The reason why we require these may be summarized as follows. 
In this paper, we are interested in self-similar solutions which 
could represent the space-time structure and radiation fluid profile 
at the edge of the formation of a black hole.  Such a critical space time 
is formed {}from a suitable initial condition through time evolution 
given by the EOM, and we have introduced the self-similar coordinate 
$(s,x)$ in order to absorb any singularity in the process. 
Thus, the solution should be smooth for all $x$.  This is the first 
condition.  

The second (boundary) condition is a physical, natural one; $A=1$ 
at the center guarantees that 
the space time has no physical singularity at the center. $V=0$ at 
the center should 
hold for any spherically symmetric problem, except for the case where 
there is a source of fluid at the center.

Now,   (\ref{eq:EOM-ss}) is a set of ODE's for 
four variables ($A, N, \omega, V$), which satisfies the Lipschitz 
condition except at the so-called {\em sonic point} (explained in 
section~\ref{sec-ss.sonic}).  So, there are three types of possible 
singularities for (\ref{eq:EOM-ss}): $x = - \infty$, $x = \infty$ and 
the sonic point.  We consider each of them in the following, but 
before that, we conclude this subsection with the following 
simple observation, which will be repeatedly used in subsequent 
sections.

\medskip

% \noindent{\bf Lemma} 
% {\em 
\begin{lemma}
\label{lem-3.1}
Any solution of  (\ref{eq:EOM-ss}) is monotonically non-decreasing, 
as long as $A <1$.  
\end{lemma} 
% } % end of \em 
%\medskip

A direct consequence of the Lemma is that any solution, which is 
regular at the center (i.e. $A \rightarrow 1$ as $x \rightarrow -\infty$),  
must satisfy 
$A \geq 1$ for all $x \in {\Bbb R}$. 

\medskip
\noindent {\em Proof of Lemma~\ref{lem-3.1}.} 
Follows {}from  (\ref{eq:EOM-ss.3});  The last term on the RHS 
is nonnegative, so we have 
\begin{equation}
        \bar{A}_{,x} \geq 1 - A. 
        \label{eq:ss3.ineq}
\end{equation}
\hfill$\Box$

%%%%%%%%%%%%%%%%%%%%%%%%%%%%%%%%%%%%%%%%%%%%%%%%%%%%%%%%%%%%%%%%%%%%%
%%%%%%%%%%%%%%%%%%%%%%%%%%%%%%%%%%%%%%%%%%%%%%%%%%%%%%%%%%%%%%%%%%%%%
\subsection{Conditions at the sonic point} 
\label{sec-ss.sonic}

%%%%%%%%%%%%%%%%%%%%%%%%%%%%%%%%%%%%%%%%%%%%%%%%%%%%%%%%%%%%%%%%%%%%%%%%
\subsubsection{The sonic point} 
\label{subsubsec-ss.sonic}

Eqs. (\ref{eq:EOM-ss.1}) and (\ref{eq:EOM-ss.2}) can be written in the 
form: 
\begin{equation}
\label{eq:EOM-ss.12form}
        \left ( 
        \begin{array}{c c}
        a & b \\
        c & d 
        \end{array}
        \right ) 
        \left ( 
        \begin{array}{c}
        {\omega}_{,x} \\
        V_{,x}
        \end{array}
        \right ) 
        = 
        \left ( 
        \begin{array}{c}
        e \\
        f 
        \end{array}
        \right ) 
\end{equation}
where $a$, $b$, $c$, $d$, $e$, $f$ are functions which do not depend on 
derivatives of ${\omega}$ and $V$. 
Solving the above in favor of ${\omega}_{,x}$ and $V_{,x}$, 
the resulting equation violates the Lipschitz condition at 
the {\em sonic point}, 
where the ``determinant'' of the coefficient 
matrix of ${\omega}_{,x}$ and $V_{,x}$ vanishes: 
\begin{eqnarray}
\label{eq:det0}
        {\rm det} & \equiv & a d - b c \nonumber \\ 
        & \propto & \frac
        {\gamma \{ (1 + N V )^2 - ( \gamma -1)  (N+V)^2\}}
        {1 - V^2} = 0. 
\end{eqnarray}
Physically, the sonic point is characterized by the 
fact that the velocity of fluid particles seen {}from the observer on
the constant $x$ line is equal to the speed of sound $\sqrt{\gamma-1}$.  

%%%%%%%%%%%%%%%%%%%%%%%%%%%%%%%%%%%%%%%%%%%%%%%%%%%%%%%%%%%%%%%%%%%
\subsubsection{Expansion at the sonic point} 
\label{subsubsec-ss.sonicexp}

Because generic solutions at the sonic point can be singular, the 
analytic condition (i) required at the sonic point places a severe 
constraint on possible form of solutions and in the end leaves only one 
free parameter, as we now explain. 

The sonic point is characterized by the vanishing of 
the determinant,   (\ref{eq:det0}).  
In order to have finite derivatives, 
${\omega}_{,x}$ and $V_{,x}$,  at the sonic point, 
the rows of  (\ref{eq:EOM-ss.12form}) must be proportional to each other: 
\begin{equation}
\label{eq:ss.rows}
        a f - e c = 0. 
\end{equation}
These two conditions,   (\ref{eq:det0}) and  (\ref{eq:ss.rows}), 
together with the identity (\ref{eq:alg-ss}) between $A, N, \omega, V$, 
enable us to express values of $A$, $N$, $\omega$ 
at the sonic point in terms of that of $V_0 \equiv V_{\rm ss}$ 
(we write $A_0, N_0, V_{0}$, etc. for values of these functions 
at the sonic point).

To find out the explicit forms of these expressions, we first list up  
the following properties to be satisfied at the sonic point: 
\begin{enumerate}
\item $N_0 \geq 0$, because of its definition in terms of lapse function.
\item $A_0 \geq 1$, {}from Lemma~\ref{lem-3.1} of Section~\ref{sec-ss.cond-ss}. 
\item The derivatives ${\omega}_{,x}$ and $V_{,x}$ should 
exist at the sonic point. In other words, first order expansion coefficients, 
in particular those 
of $\omega$ and $V$ must exist as real numbers.  
\end{enumerate}

Note that   (\ref{eq:det0}), being quadratic in $N_0$, 
in general allows two solutions for $N_0$.   
However, one of them leads to $A_0 < 1$, and is excluded {}from the 
second requirement above. 
The first and third requirements together restrict allowed values of 
$V_0$ to $-1 \leq V_0 \leq \sqrt{\gamma-1}$.  
As a result, the zeroth order expansion coefficients are
given by:  
\begin{eqnarray}
        N_0 & = & \frac{1 - \sqrt{\gamma-1} V_0}{\sqrt{\gamma-1} - V_0},  
        \nonumber \\ \quad \\
        A_0 & = & 
                \frac{\gamma^2 + 4 \gamma -4 + 8 (\gamma-1)^{3/2} V_0 
                        - (3 \gamma -2) (2 - \gamma) V_0^2 } 
                {\gamma^2 ( 1 - V_0^2) } 
                \nonumber \\ \quad \\
        \omega_0 & = &  
                \frac{2 \sqrt{\gamma-1} (\sqrt{\gamma-1} - V_0) 
                        (1 + \sqrt{\gamma-1} V_0) } 
                {\gamma^2 ( 1 - V_0^2) } .  \nonumber \\ \quad 
\end{eqnarray}

Once these values at the sonic point are given, we can then determine 
the higher coefficients of power series expansion of $A, N, \omega$, 
and $V$ with respect to $x$ using the above 
conditions and   (\ref{eq:EOM-ss}).  To avoid making presentation 
too complicated, we do not give explicit expressions, but only 
remark that  
the equations for the first order expansion coefficients, given in terms of 
$A_0, N_0, \omega_0, V_0$, have two solutions in general; 
some of which are not allowed for some values of $V_0$ due to the 
third requirement above.

The self-similar 
solution which is regular at the sonic point is thus characterized by a 
single parameter $V_0 \equiv V_{\rm ss}$.

%%%%%%%%%%%%%%%%%%%%%%%%%%%%%%%%%%%%%%%%%%%%%%%%%%%%%%%%%%%%%%%%%%%%%%%%
%%%%%%%%%%%%%%%%%%%%%%%%%%%%%%%%%%%%%%%%%%%%%%%%%%%%%%%%%%%%%%%%%%%%%%%%
\subsection{Asymptotic behaviour of solutions as $x \rightarrow \pm \infty$} 

In searching for desired self-similar solutions, it becomes desirable  
to understand the correct asymptotic behaviour of solutions as 
$x \rightarrow \pm \infty$.

%%%%%%%%%%%%%%%%%%%%%%%%%%%%%%%%%%%%%%%%%%%%%%%%%%%%%%%%%%%%%%%%%%%%%%%%
\subsubsection{Behaviour at the center, $x = -\infty$}

Behaviour of solutions of ODE's (\ref{eq:EOM-ss}), 
as  $x \rightarrow - \infty$, can be most easily understood 
by introducing a new variable 
\begin{equation}
\label{eq:Mdef}
        M \equiv N V 
\end{equation}
and rewriting (\ref{eq:EOM-ss}) in terms of $A, M, \omega, V$.  
It is easily seen that 
\begin{equation}
\label{eq:ss-m-infty}
        A = 1, \quad M = - \frac{2}{3 \gamma}, \quad 
        \omega = V = 0 
\end{equation} 
is a stationary (fixed) point of (\ref{eq:EOM-ss}), i.e. all the 
$x$-derivatives vanish.  Moreover, linear analysis around this 
stationary point shows that there is one unstable direction for this 
stationary point, namely in the direction of $M + \frac{2}{3 \gamma}$. 
Thus, a generic solution which satisfies the analyticity condition 
at the sonic point will diverge as $x \rightarrow -\infty$, 
and only a small number of solutions 
with carefully chosen values of $V_0$ can satisfy the regularity condition 
at $x = -\infty$; thus leaving only discrete acceptable solutions.

By similar reasoning, it is easily seen that 
any self-similar solution {\em which satisfies our boundary condition 
at the center}  (i.e. has a smooth derivative),  
allows a (possibly only asymptotic) 
power series expansion in $\xi \equiv e^{x}$ at $\xi = 0$, and it is 
easily seen that the asymptotic behaviour is
\[
        A_{\rm s.s.} (x) \sim 1 + A_{-\infty} e^{2 x} , 
        \quad 
        N_{\rm ss} (x) \sim N_{-\infty} e^{-x},  
\]
\begin{equation}
\label{eq:ss-asymp1}
        \omega_{\rm ss} \sim  \omega_{-\infty} e^{2 x}, 
        \quad 
        V_{\rm ss} \sim V_{-\infty} e^x , 
\end{equation}
with coefficients 
$A_{-\infty}, N_{-\infty}, \omega_{-\infty}, V_{-\infty}$  
satisfying the relation: 
\begin{equation}
        A_{-\infty} =  \frac{2 \omega_{-\infty}}{3} , 
        \quad 
        N_{-\infty} V_{-\infty} = - \frac{2}{3 \gamma }  .
\end{equation}

%%%%%%%%%%%%%%%%%%%%%%%%%%%%%%%%%%%%%%%%%%%%%%%%%%%%%%%%%%%%%%%%%%%%%%%%
\subsubsection{Behaviour at $x = \infty$}

It is easily seen that 
\begin{equation}
\label{eq:ss-p-inf}
        N = 0, \quad A = 1 
        + \frac{2 \omega \{ 1 + (\gamma -1) V^2 \} } {1 - V^2} 
\end{equation}
is a stationary point (parameterized by two free parameters $\omega, V$) 
of (\ref{eq:EOM-ss}).  
Linear analysis around the stationary point shows that 
this stationary point becomes an attracting node for
\begin{equation}
        \omega \leq \frac{1 - V^{2}}{\gamma ( 1 + V^{2})}.      
        \label{eq:ss.infty}
\end{equation}
As will be explained in Section~\ref{sec-ss.results}, 
most of the self-similar solutions which are analytic for $x < \infty$ 
found numerically 
satisfy this condition, and thus are seen to be attracted to this 
stationary point.

%%%%%%%%%%%%%%%%%%%%%%%%%%%%%%%%%%%%%%%%%%%%%%%%%%%%%%%%%%%%%%%%%%%%%%%%%
%%%%%%%%%%%%%%%%%%%%%%%%%%%%%%%%%%%%%%%%%%%%%%%%%%%%%%%%%%%%%%%%%%%%%%%%%
\subsection{Numerical Method} 
\label{sec-ss.practice}

Based on considerations presented in the above, we proceed 
as follows in order to find out desired self-similar solutions.  

We first fix the value $V_0$ at the sonic point $x=0$, and use the power 
series expansion there (Section~\ref{subsubsec-ss.sonicexp}) 
to derive the derivatives.  
Then starting {}from the sonic point $x=0$, we solve ODE's 
(\ref{eq:EOM-ss}) towards 
$x = -\infty$, using Runge--Kutta fourth order integrator, with  reliable 
error estimates. 
If  at some $x <0$ (1) either $A < 1$ or (2) $det = 0$, 
we stop solving and conclude that 
this $V_0$ does not give rise to a desired 
self-similar solution.
   The first case ($A<1$) is excluded based on the Lemma~\ref{lem-3.1} of 
section~\ref{sec-ss.cond-ss}. The second case implies the existence of 
another sonic point between $x=0$ and $x = -\infty$.  Such a solution with 
more than one (but with finite number of) sonic points can always be 
found by searching for its part between its left-most sonic point and 
$x = -\infty$, and then solving (\ref{eq:EOM-ss}) {}from 
the left-most sonic point to 
the right (towards $x = \infty$).  However, it is quite unlikely that 
such solutions with more than one sonic points exist, because the 
existence of only one fixed point already reduces the freedom of  
parameters to discrete ones, and other possible sonic point will further 
reduce it. 

Our numerical results show that 
each allowed $V_0$ ends in either of these two possibilities.  
Moreover it is seen,  both numerically and mathematically rigorously, that 
a desired self-similar solution exists at those values of $V_0$, 
where $V_0+0$ leads to the case (1) and $V_0-0$ leads to the case (2).

%%%%%%%%%%%%%%%%%%%%%%%%%%%%%%%%%%%%%%%%%%%%%%%%%%%%%%%%%%%%%%%%%%%%%%%
%%%%%%%%%%%%%%%%%%%%%%%%%%%%%%%%%%%%%%%%%%%%%%%%%%%%%%%%%%%%%%%%%%%%%%%
\subsection{(Numerical) results}
\label{sec-ss.results}
We have searched for self-similar solutions which satisfy the above 
conditions (i) and (ii) numerically, for 
all the allowed values of $V_0$. 
As has been extensively studied in \cite{Bog77}, 
there exists a sequence of self-similar solutions, each of which 
is characterized by the number of zeros of $V$.  The number of zeros 
is an integer starting {}from $1$, and the one with exactly one zero is 
the solution cited in  \cite{EvCo94}, which we call the Evans--Coleman 
solution.  Due to a numerical difficulty, we could find only  
self-similar solutions with odd number of zeros of $V$. 

\begin{figure}
\begin{center}
{\BoxedEPSF{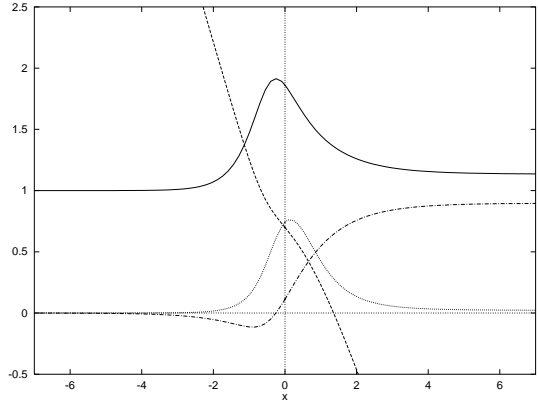 scaled 300} }
\end{center}
%\vspace*{60mm} 
%{\bf // PLACE Fig.~\ref{fig4.1} HERE. //} 
\caption{Profile of a self-similar solution (the Evans--Coleman solution). 
        Curves represent $A_{\rm ss}$ (solid line), $\ln(N_{\rm ss})$ (dashed), 
        $\omega_{\rm ss}$ (dotted), and $V_{\rm ss}$ (dot-dashed).
}
\label{fig4.1}
\end{figure}

In Fig.~\ref{fig4.1}, we show the profile of  the Evans--Coleman solution, 
which will play the central role in later sections. 
This is the same as Fig.~1 of \cite{EvCo94} except for $\alpha$ which is 
due to the coordinate condition for $t$.

%%%%%FIGFIGFIGFIGFIGFIGFIGFIGFIGFIGFIGFIGFIGFIGFIGFIGFIGFIGFIGFIG
%%%%%FIGFIGFIGFIGFIGFIGFIGFIGFIGFIGFIGFIGFIGFIGFIGFIGFIGFIGFIGFIG
%%%%%FIGFIGFIGFIGFIGFIGFIGFIGFIGFIGFIGFIGFIGFIGFIGFIGFIGFIGFIGFIG
%%%%%FIGFIGFIGFIGFIGFIGFIGFIGFIGFIGFIGFIGFIGFIGFIGFIGFIGFIGFIGFIG

%%%%%%%%%%%%%%%%%%%%%%%%%%%%%%%%%%%%%%%%%%%%%%%%%%%%%%%%%%%%%%%%%%%%%%%%%%
%%%%%%%%%%%%%%%%%%%%%%%%%%%%%%%%%%%%%%%%%%%%%%%%%%%%%%%%%%%%%%%%%%%%%%%%%%
%%%%%%%%%%%%%%%%%%%%%%%%%%%%%%%%%%%%%%%%%%%%%%%%%%%%%%%%%%%%%%%%%%%%%%%%%%
%%%%%%%%%%%%%%%%%%%%%%%%%%%%%%%%%%%%%%%%%%%%%%%%%%%%%%%%%%%%%%%%%%%%%%%%%%
\section{Perturbation}
\label{sec-pert}

We studied perturbations around self-similar solutions by two different 
methods.  The first one is to directly find out eigenmodes by solving 
the system of ODE's (\ref{eq:EOM-eigenmodes}) below, 
which characterizes the eigenmodes.  The second one is to apply the 
so--called Lyapunov analysis.  The former has the advantage that it 
reduces the problem to that of solving ODE's, and enables us fairly 
accurate estimate of eigenvalues, but has the disadvantage that 
it is almost impossible to search for {\em every possible} eigenvalues.   
The latter has an advantage that it is quite suited to 
finding out eigenvalues in descending order of their real part, but has 
disadvantages that (1) eigenvalue estimates are not so accurate, and 
(2) we have to solve a PDE.  However, we here emphasize that 
the PDE we have to solve for the Lyapunov analysis is a very regular one, 
and can be handled by standard techniques for solving PDE's.

In this section, we explain the first method, that is, direct search for 
perturbations by shooting method.  To simplify the notation, we 
express perturbations in terms of perturbations of 
$\bar{A} = \ln(A), \bar{N} = \ln(N), \bar\omega = \ln(\omega)$. 
For example, the $\bar{A}$-version of (\ref{eq:EOM-pp}) in fact means 
\begin{equation}
        \bar{A} (s,x) = \bar{A}_{\rm ss} (x) + \epsilon \bar{A}_{\rm var}(s,x)
        \label{eq:EOM-pp0}
\end{equation}
where 
\begin{equation}
        \bar{A}(s,x) \equiv \ln(A(s,x)), \quad 
        \bar{A}_{\rm ss}(x) \equiv \ln(A_{\rm ss}(x)). 
        \label{eq:EOM-pp1}
\end{equation}

%%%%%%%%%%%%%%%%%%%%%%%%%%%%%%%%%%%%%%%%%%%%%%%%%%%%%%%%%%%%%%%%%%
%%%%%%%%%%%%%%%%%%%%%%%%%%%%%%%%%%%%%%%%%%%%%%%%%%%%%%%%%%%%%%%%%%
\subsection{Perturbation equations and the gauge degrees of freedom}

Perturbation equations are obtained by taking the first order 
variation in   (\ref{eq:EOM-fll}) {}from the 
self-similar solution $H_{\rm ss}$ considered:
\begin{equation}
\label{eq:EOM-pp}
        h(s,x) = H_{\rm ss}(x) + \epsilon  h_{\rm var}(s,x), 
        \label{eq:pert.exp1}
\end{equation}
where $h$ represents each of $(\bar{A}, \bar{N}, \bar{\omega}, V)$. 
In the following, to simplify the notation, 
we write $V$ for $V_{\rm ss}$ and write $V_{\rm var}$ 
for perturbations. 

Explicitly, we have:

\begin{eqnarray}
\label{eq:EOM-var}
        && 
        \left [ 
        \left ( 
                \begin{array}{cccc}
                0 & 0 & 0 & 0 \\
                0 & 0 & 0 & 0 \\
                0 & 0 & A_s  & B_s \\
                0 & 0 & C_s & D_s 
                \end{array} 
        \right ) \! \partial_{s} + 
        \left ( 
                \begin{array}{cccc}
                1 & 0 & 0 & 0 \\
                0 & 1 & 0 & 0 \\
                0 & 0 & A_x & B_x \\
                0 & 0 & C_x & D_x 
                \end{array} 
        \right ) \! \partial_x
        \right ] 
        \left ( 
                \begin{array}{c}
                \bar{A}_{\rm var} \\
                \bar{N}_{\rm var} \\
                \bar{\omega}_{\rm var} \\
                V_{\rm var}  
                \end{array} 
        \right )  \nonumber \\
         && \hspace{10mm} =     
        \left ( 
                \begin{array}{cccc}
                G_1 & 0 & G_3 & G_4 \\
                H_1 & 0 & H_3 & 0 \\
                E_1 & E_2 & E_3 & E_4 \\
                F_1 & F_2 & F_3 & F_4 
                \end{array} 
        \right ) 
        \left ( 
                \begin{array}{c}
                \bar{A}_{\rm var} \\
                \bar{N}_{\rm var} \\
                \bar{\omega}_{\rm var} \\
                V_{\rm var}  
                \end{array} 
        \right )        
\end{eqnarray} 
with coefficients given in terms of self-similar solutions as 
\begin{eqnarray}
        A_s & \equiv &  1, \quad 
        B_s \equiv \frac{\gamma V}{1 - V^2}, \nonumber \\
        C_s & \equiv & (\gamma - 1) V, \quad 
        D_s \equiv \frac{\gamma}{1 - V^2},
\end{eqnarray} 
\begin{eqnarray}
        A_x & \equiv & 1 +  N V , \quad 
        B_x \equiv \frac{\gamma ( N+V)}{1 - V^2} , \nonumber \\
        C_x & \equiv & (\gamma - 1) (N + V), \quad 
        D_x \equiv \frac{\gamma ( 1 + N V)}{1 - V^2},
\end{eqnarray} 
\begin{mathletters}
\begin{eqnarray}
        E_1 & \equiv & - \frac{(\gamma + 2)}{2} A N V , \\
        E_2 & \equiv & \frac{6 - 3 \gamma}{2} N V 
                        - \frac{2 + \gamma}{2} A N V 
                        + (2 - \gamma) \omega N V \nonumber \\
                & &     - N V \bar{\omega}_{,x} 
                        - \frac{\gamma N V_{,x}}{1 - V^2} ,\\
        E_3 & \equiv & (2 - \gamma) \omega N V , \\
        E_4 & \equiv & \frac{6 - 3 \gamma}{2} N - \frac{2 + \gamma}{2} A N 
                        + (2 - \gamma) \omega N  \nonumber \\
                & &     - N  \bar{\omega}_{,x} 
                        - \frac{\gamma (1 + 2 N V + V^2) V_{,x}}{(1 - V^2)^2} , 
\end{eqnarray}
\end{mathletters}
\begin{mathletters}
\begin{eqnarray}
        F_1 & \equiv & \frac{2 - 3 \gamma}{2} A N , \\
        F_2 & \equiv & (2 - \gamma) (\gamma -1) \omega N 
                        + \frac{7 \gamma - 6}{2} N 
                        + \frac{2 - 3 \gamma}{2} A N  \nonumber \\
                & &     - (\gamma -1) N \bar{\omega}_{,x} 
                        - \frac{\gamma N V V_{,x}}{1 - V^2}, \\
        F_3 & \equiv & (2 - \gamma) (\gamma -1) \omega N , \\
        F_4 & \equiv &  
                        - (\gamma - 1) \bar{\omega}_{,x} 
                        - \frac{\gamma (N + 2 V + N V^2) V_{,x} }{( 1 - V^2)^2}, 
\end{eqnarray}
\end{mathletters}
\begin{eqnarray} 
        G_1 & \equiv & -A , \quad
        G_3 \equiv \frac{ 2 \{1 + (\gamma-1) V^2\} \omega}{1 - V^2} , 
        \nonumber \\
        G_4 & \equiv & \frac{4 \gamma \omega V}{(1 - V^2)^2} , 
\end{eqnarray} 
\begin{equation} 
        H_1 \equiv A, \quad
        H_3 \equiv (\gamma - 2) \omega .
\end{equation} 
We also have
\begin{eqnarray} 
        & & \partial_s \bar{A}_{\rm var} + \partial_x \bar{A}_{\rm var}
        = - \frac{2 \gamma N V \omega}{1 - V^2} 
        \left (\bar{N}_{\rm var} + \bar{\omega}_{\rm var} \right ) 
        \nonumber \\
        & & \hspace{10mm} 
        - \frac{2 \gamma N \omega (1 + V^2)}{(1 - V^2)^2} V_{\rm var}
\end{eqnarray} 

We note that there is a gauge freedom due to the coordinate
transformation of order $\epsilon$, namely
  (\ref{eq:CT:(s,x)}) with $f(s)=s+\epsilon f_1(s)$.
The transformation (\ref{eq:h}) now becomes [taking $O(\epsilon)$ terms; 
$' = \frac{d}{dx}, \; \dot{} = \frac{d}{ds}$]
\begin{equation}
\label{eq:CT.var}
        \tilde{h}_{\rm var}(s,x) = 
        \left \{
        \begin{array}{l} 
        h_{\rm var}(s,x) + f_1(s) h'_{\rm ss}(x) \\
                        \hspace{40mm}  (\bar{A}, \bar{\omega}, V),  \\ 
        h_{\rm var}(s,x) + f_1(s) h'_{\rm ss}(x) + \dot{f_1}(s) \\
                        \hspace{40mm}  (h = \bar{N})
        \end{array} 
        \right . 
\end{equation}
This in particular means that one can always require  
$\bar{N}_{\rm var} (s, x_0) \equiv 0$ for a fixed $x_0$.

%%%%%%%%%%%%%%%%%%%%%%%%%%%%%%%%%%%%%%%%%%%%%%%%%%%%%%%%%%%%%%%%%%%%%%%%
%%%%%%%%%%%%%%%%%%%%%%%%%%%%%%%%%%%%%%%%%%%%%%%%%%%%%%%%%%%%%%%%%%%%%%%%
\subsection{EOM for eigenmodes}

We consider 
eigenmodes of the form $h_{\rm var}(s,x) = h_{\rm p}(x)e^{\kappa s}$, 
with $\kappa \in {\Bbb C}$ being a constant.  
Substituting this form into (\ref{eq:EOM-var}) 
yields a set of linear, homogeneous first order 
ODE's for $(\bar{N}_{\rm p}, \bar{A}_{\rm p}, \bar{\omega}_{\rm p}, V_{\rm p})$. 
Explicitly, we have 
\begin{eqnarray}
\label{eq:EOM-eigenmodes}
        && \left [ \!
        \left ( 
                \begin{array}{cccc}
                1 & 0           & 0 & 0 \\
                0          & 1 & 0 & 0 \\
                0          & 0 & A_x & B_x \\
                0          & 0 & C_x & D_x  
                \end{array} 
        \right ) \!\! \frac{d}{dx}
        - 
        \left ( 
                \begin{array}{cccc}
                G_1 & 0 & G_3 & G_4 \\
                H_1 & 0 & H_3 & 0 \\
                E_1 & E_2 & E_3 & E_4 \\
                F_1 & F_2 & F_3 & F_4  
                \end{array} 
        \right ) \!
        \right ]
        \left ( \!
                \begin{array}{c}
                \bar{A}_{\rm p} \\
                \bar{N}_{\rm p} \\
                \bar{\omega}_{\rm p} \\
                V_{\rm p}  
                \end{array} 
                \!
        \right ) 
        \nonumber \\
        & & {} \qquad =       
        - \kappa \left ( 
                \begin{array}{cccc}
                0 & 0 & 0 & 0 \\
                0 & 0 & 0 & 0 \\
                0 & 0 & A_s & B_s \\
                0 & 0 & C_s & D_s 
                \end{array} 
        \right ) 
        \left ( 
                \begin{array}{c}
                \bar{A}_{\rm p} \\
                \bar{N}_{\rm p} \\
                \bar{\omega}_{\rm p} \\
                V_{\rm p}  
                \end{array} 
        \right )        
\end{eqnarray} 

and 
\begin{eqnarray}
\label{eq:EOM-eigenmodes.2}
        \kappa \bar{A}_{\rm p} + \partial_x \bar{A}_{\rm p}
        & = & - \frac{2 \gamma N V \omega}{1 - V^2} 
        \left (\bar{N}_{\rm p} + \bar{\omega}_{\rm p} \right ) 
        \nonumber \\
        & & \;\; - \frac{2 \gamma N \omega (1 + V^2)}{(1 - V^2)^2} V_{\rm p}
\end{eqnarray} 
Eq. (\ref{eq:EOM-eigenmodes.2}) and the first row 
of  (\ref{eq:EOM-eigenmodes}) provide an algebraic identity : 
\begin{eqnarray}
\label{eq:alg-PP}
        && (\kappa - A) \bar{A}_{\rm p} 
        + \left ( 
        \frac{2 \gamma N V \omega} { 1 - V^2} 
        \right ) 
        \bar{N}_{\rm p} \nonumber \\ 
        & & \quad 
        + 
        \left ( \frac{2 \omega \{1 + (\gamma-1) V^{2} + \gamma N V\} } 
        { 1 - V^2} \right ) 
        \bar\omega_{\rm p} \nonumber \\ 
        & & \quad 
        + \left ( \frac{2 \gamma \omega \{N ( 1 + V^2) + 2 V \} } 
                { ( 1 - V^2 ) ^2 } \right )   V_{\rm p}
        = 0 , 
\end{eqnarray}
just as in the case of self-similar solutions [cf. (\ref{eq:alg-ss})].  
As in our treatment of self-similar solutions, we here treat 
$\bar{A}_{\rm p}, \bar{N}_{\rm p}, \bar\omega_{\rm p}, V_{\rm p}$ as 
four unknown functions, and 
use (\ref{eq:alg-PP}) as a check at appropriate steps 
of computation. 

We require $\bar{N}_{\rm p}(s,0)=0$ to fix the coordinate freedom 
(\ref{eq:CT.var}).

%%%%%%%%%%%%%%%%%%%%%%%%%%%%%%%%%%%%%%%%%%%%%%%%%%%%%%%%%%%%%%%%%%%%%%%%
%%%%%%%%%%%%%%%%%%%%%%%%%%%%%%%%%%%%%%%%%%%%%%%%%%%%%%%%%%%%%%%%%%%%%%%%
\subsection{Conditions on perturbations}

As in the case of self-similar solutions, we require 
(i) that the perturbations are analytic for all $x \in {\Bbb R}$, and 
(ii) that the perturbed space-times are regular at the center 
($\bar{A}_{\rm p} = 0$ at $x = - \infty$).

%%%%%%%%%%%%%%%%%%%%%%%%%%%%%%%%%%%%%%%%%%%%%%%%%%%%%%%%%%%%%%%%%%%%%%%%
%%%%%%%%%%%%%%%%%%%%%%%%%%%%%%%%%%%%%%%%%%%%%%%%%%%%%%%%%%%%%%%%%%%%%%%%
\subsection{The sonic point} 

It can be seen  that  
apart {}from the overall multiplicative factor,
the perturbation solutions which satisfy the analyticity condition (i) 
at the sonic point 
are specified by one free parameter $\kappa$, as we now explain. 

(1) First we note that the sonic point is   a {\em regular singular 
point} for the perturbations. That is, the system of 
ODE's for perturbations (\ref{eq:EOM-eigenmodes}) 
is singular where the determinant of the matrix on the LHS of 
(\ref{eq:EOM-eigenmodes}) vanishes: $A_x D_x - B_x C_x = 0$; this is identical 
with the sonic point condition for self-similar solutions, (\ref{eq:det0}).

(2) Next, in order to have a smooth solution at the sonic point, 
the third and fourth row of  (\ref{eq:EOM-eigenmodes}) must be 
proportional to each other, like in the case of self-similar solution, 
 (\ref{eq:ss.rows}). 
This yields an algebraic relation between 
$\bar{A}_{\rm p}(0)$, $\bar{N}_{\rm p}(0)$, $\bar{\omega}_{\rm p}(0)$ 
and $V_{\rm p}(0)$ 
at the sonic point.

(3) Third, we have an algebraic identity  
between $\bar{A}_{\rm p}(x)$, $\bar{N}_{\rm p}(x)$, $\bar{\omega}_{\rm p}(x)$ 
and $V_{\rm p}(x)$, given by  (\ref{eq:alg-PP}). 

(4) The above two algebraic relations, together with our choice of gauge 
$\bar{N}_{\rm p}(0) = 0$ enables us to express, e.g. 
$\bar{\omega}_{\rm p}(0)$ 
and $V_{\rm p}(0)$ in terms of $\bar{A}_{\rm p}(0)$; and then higher order 
expansion coefficients of perturbations are given in terms of 
$\bar{A}_{\rm p}(0)$.  Because the system is linear and homogeneous, overall
normalization is irrelevant.  We thus  see that the solution which 
satisfies (i) is characterized by a single parameter $\kappa$.

This, together with the regularity condition (ii) at the center, 
in general allows only discrete values for $\kappa$ as now 
be explained.

%%%%%%%%%%%%%%%%%%%%%%%%%%%%%%%%%%%%%%%%%%%%%%%%%%%%%%%%%%%%%%%%%%%%%%%%
%%%%%%%%%%%%%%%%%%%%%%%%%%%%%%%%%%%%%%%%%%%%%%%%%%%%%%%%%%%%%%%%%%%%%%%%
\subsection{Asymptotic behaviour as $x \rightarrow \pm \infty$}

Searching numerically for eigenmodes of perturbations is 
in some sense easier than searching for self-similar solutions, 
because the ODE's are now {\em linear and homogeneous} in unknown 
functions. In particular, asymptotic behaviour of solutions as 
$x \rightarrow \pm \infty$ is easily obtained as follows.

%%%%%%%%%%%%%%%%%%%%%%%%%%%%%%%%%%%%%%%%%%%%%%%%%%%%%%%%%%%%%%
\subsubsection{Asymptotic behaviour as $x \rightarrow - \infty$}

Substituting the asymptotic behaviour, (\ref{eq:ss-asymp1}), 
of self-similar solutions as $x \rightarrow -\infty$,   
the system of ODE's for perturbations (\ref{eq:EOM-eigenmodes}) 
take on a simple form, given by 
\begin{equation}
\label{eq:PPasmp1}
        \partial_x \left ( \begin{array}{c} 
        \bar{A}_p \\  \bar{N}_p \\ \bar\omega_p \\ V_p 
        \end{array} \right ) 
        = 
        \left ( \begin{array}{cccc} 
        -1 & 0 & 0 & 0 \\  
        1  & 0 & 0 & 0 \\ 
        \frac{2 - 3 \gamma}{2(\gamma -1)} & 0 & 0 & 0 \\
        0 & 0 & 0 & -2 
        \end{array} \right )    \;\; 
        \left ( \begin{array}{c} 
        \bar{A}_p \\  \bar{N}_p \\ \bar\omega_p \\ V_p 
        \end{array} \right ).
\end{equation}
This homogeneous linear ODE with constant coefficients has four 
independent solutions, given by 
\begin{equation}
\label{eq:m-inf-ind-sol}
        \left ( \begin{array}{c} 
        0 \\  1 \\ 0 \\ 0 
        \end{array} \right ) , \quad 
        \left ( \begin{array}{c} 
        0 \\  0 \\ 1 \\ 0 
        \end{array} \right ) , \quad 
        \left ( \begin{array}{c} 
        0 \\  0 \\ 0 \\ 1
        \end{array} \right ) e^{-2x}, \quad 
        \left ( \begin{array}{c} 
        1 \\  -1 \\ \frac{2 - 3 \gamma}{2(\gamma -1)} \\ 0 
        \end{array} \right ) e^{-x},
\end{equation}
The last one should be excluded in view of the 
identity  (\ref{eq:alg-PP}), and thus asymptotic behaviour of 
arbitrary solutions of ODE's is described by suitable linear 
combinations of the first three independent solutions.  
Now, one of them (the third one) blows up as $x \rightarrow -\infty$. 
Therefore we see that any solution, which satisfies  our boundary condition,  
should thus be obtained by choosing $\kappa$ so as to eliminate the unwanted  
expanding mode.

%%%%%%%%%%%%%%%%%%%%%%%%%%%%%%%%%%%%%%%%%%%%%%%%%%%%%%%%%%%%%%%%%%%%%%%
\subsubsection{Asymptotic behaviour as $x \rightarrow \infty$} 

Similarly, the asymptotic behaviour as  $x \rightarrow \infty$ is 
obtained.  It is now seen that there are three independent solutions 
[after taking into account the identity  (\ref{eq:alg-PP})], 
whose $x$-dependence are 
\begin{equation}
\label{eq:p-inf-ind-sol}
        1, \quad  e^{-\kappa x},  \quad e^{-\kappa x}
\end{equation}
as $x \rightarrow \infty$. 
It is thus seen that every solution stays bounded 
as long as ${\rm Re}  \kappa > 0$, whereas for  ${\rm Re}  \kappa <0$  
it blows up as $x \rightarrow \infty$.

%%%%%%%%%%%%%%%%%%%%%%%%%%%%%%%%%%%%%%%%%%%%%%%%%%%%%%%%%%%%%%%%%%%%%%%%
%%%%%%%%%%%%%%%%%%%%%%%%%%%%%%%%%%%%%%%%%%%%%%%%%%%%%%%%%%%%%%%%%%%%%%%%
\subsection{The gauge mode}
\label{sec-pp.gauge-mode}
In searching for possible eigenmodes, one has to pay attention to a 
``gauge mode'' which emerges {}from a coordinate transformation 
applied to the self-similar solution, as we now explain.

Suppose one has a self-similar solution $h(s,x) = H_{\rm ss}(x)$.  
Seen {}from another coordinate 
[related by the coordinate transformation (\ref{eq:CT:(s,x)})], 
its perturbation is given by setting $h_{\rm p} \equiv 0$ in 
$\tilde{h}$ of   (\ref{eq:CT.var}): 
\begin{equation}
\label{eq:CT.no-p}
        \tilde{h}_{\rm var}(s,x) = \left \{ 
        \begin{array}{ll} 
        f_1(s) h_{\rm ss}'(x) & ( h = \bar{A}, \bar\omega, V) \\ 
        f_1(s) h_{\rm ss}'(x) + \dot{f}_1(s) & (h = \bar{N}) 
        \end {array} 
        \right . 
\end{equation}
For general $f_1$, this does not behave like an eigenmode.  
However,  with the choice
of  $f_1(s) \equiv e^{\bar\kappa s}$ ($\bar\kappa$ arbitrary), 
$\tilde{h}$ does behave as an eigenmode with eigenvalue $\bar\kappa$: 

\begin{eqnarray}
\label{eq:gauge-mode}
        && \hspace*{-5mm}
        \tilde{h}_{\rm gauge}(s,x; \bar\kappa) 
        = 
        e^{\bar\kappa s} 
        \left ( 
        \begin{array}{c} 
                h'_{\rm ss}(x) \\
                h'_{\rm ss}(x) + \bar \kappa    
        \end{array} 
        \right ) 
        \begin{array}{l}
        (h = \bar{A}, \bar\omega, V) \\
        (h = \bar{N}) 
        \end{array}
        . \nonumber \\
        & & 
\end{eqnarray}
The mode emerges via the coordinate
transformation of the self-similar solution $h_{\rm ss}$, and thus can 
be considered as a result of pure gauge transformation, like a ``zero 
mode'' in translation invariant systems. 
Because $\bar\kappa \in { \Bbb C}$ is {\em arbitrary}, 
this pure gauge mode forms a one-parameter family.

It can also be seen that other (non-gauge) modes are transformed under 
the gauge transformation as follows: 

Eq.(\ref{eq:CT.no-p}) for $h_{\rm var}(s,x) = h_{\rm p}(x) e^{\kappa s}$ 
becomes 
\begin{eqnarray}
\label{eq:CT.PP}
        \hspace*{-5mm}
        \tilde{h}_{\rm var}(s,x) 
        & = & 
        \left \{
        \begin{array}{l}                
        h_{\rm p}(x) e^{\kappa s} 
        + f_1(s) h'_{\rm ss}(x)  \\
        \hspace{35mm} (h = \bar{A}, \bar\omega, V) \\
        h_{\rm p}(x) e^{\kappa s} 
        + f_1(s) h'_{\rm ss}(x)
        + \dot{f}_1(s)   \\
        \hspace{35mm}   (h = \bar{N})
        \end{array} 
        \right . 
        \nonumber \\
        & & 
\end{eqnarray}

As long as $h_{\rm p} \not\equiv 0$, by taking  
$f_1(s) = e^{\kappa s}$, this means  transformed $\tilde{h}$ is also 
an eigenmode, with the same $\kappa$.  
Thus we have a one-parameter family of eigenmodes 
(with the same eigenvalue $\kappa$), which are mutually 
related by the gauge mode (\ref{eq:gauge-mode}).

%%%%%%%%%%%%%%%%%%%%%%%%%%%%%%%%%%%%%%%%%%%%%%%%%%%%%%%%%%%%%%%%%%%%%%%%
%%%%%%%%%%%%%%%%%%%%%%%%%%%%%%%%%%%%%%%%%%%%%%%%%%%%%%%%%%%%%%%%%%%%%%%%
\subsection{Numerical  Results} 
Based on the above observations, we searched for desired eigenmodes 
as follows.  We first fix a value of $\kappa$, and then starting {}from 
$x=0$ (the sonic point), integrate (\ref{eq:EOM-eigenmodes}) 
to $x = - \infty$.  When there appear nonzero components 
of the expanding modes (for sufficiently large $|x|$), we judge that this 
$\kappa$ does not give a desired eigenmode and stopped.   

In this way, we searched for eigenmodes in the range of 
${\rm Re}  \kappa \geq -1.5, \;\; |{\rm Im} \kappa | \leq 14$.  
We found several eigenmodes 
which are explained below. 

%%%%%%%%%%%%%%%%%%%%%%%%%%%%%%%%%%%%%%%%%%%%%%%%%%%%%%%%%%%%%%%%%%%%%
\subsubsection{The relevant mode for the Evans--Coleman solution}
\label{subsubsec-pp.ECmode}

\begin{figure}
\begin{center}
{\BoxedEPSF{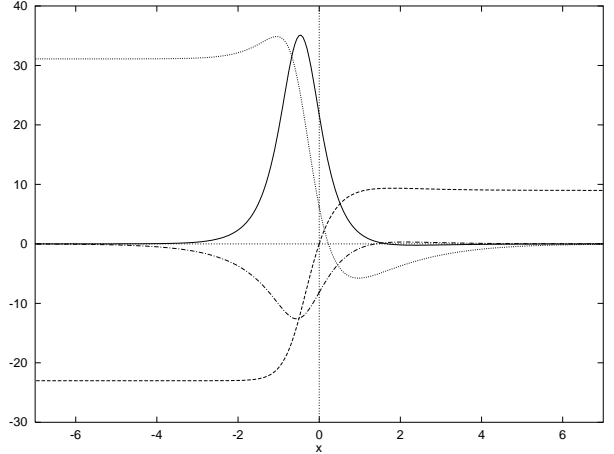 scaled 340} }
\end{center}
%{\bf // PLACE Fig.~\ref{fig-5.1} HERE. //} 
%\vspace*{60mm} 
\caption{Profile of the eigenmode with the largest eigenvalue.  
        Curves represent $\bar{A}_{\rm p}$ (solid 
        line), $\bar{N}_{\rm p}$ (dashed), 
        $\bar{\omega}_{\rm p}$ (dotted), 
        and $V_{\rm p}$ (dot-dashed).
}
\label{fig-5.1}
\end{figure}

The profile of the eigenmode with the largest ${\rm Re}  \kappa$ is shown 
in Fig.~\ref{fig-5.1}.  
It has the eigenvalue $\kappa \simeq 2.81055255$; which corresponds to 
the exponent value $\beta_{\rm BH} \simeq 0.35580192$, 
according to our scenario of section~\ref{sec-scenario}, 
and which we believe to be exact to the last digit.  
This is in good agreement with the value of \cite{EvCo94}.

%%%%%FIGFIGFIGFIGFIGFIGFIGFIGFIGFIGFIGFIGFIGFIGFIGFIGFIGFIGFIGFIG
%%%%%FIGFIGFIGFIGFIGFIGFIGFIGFIGFIGFIGFIGFIGFIGFIGFIGFIGFIGFIGFIG
%%%%%FIGFIGFIGFIGFIGFIGFIGFIGFIGFIGFIGFIGFIGFIGFIGFIGFIGFIGFIGFIG
%%%%%FIGFIGFIGFIGFIGFIGFIGFIGFIGFIGFIGFIGFIGFIGFIGFIGFIGFIGFIGFIG

%%%%%%%%%%%%%%%%%%%%%%%%%%%%%%%%%%%%%%%%%%%%%%%%%%%%%%%%%%%%%%%%%%%%%%
\subsubsection{The gauge mode}
\label{subsubsec-pp.gauge}

In our gauge, where  $\bar{N}_{\rm p}(s,0) \equiv 0$, 
we observe the gauge mode with 
$\bar\kappa = - \frac{d \bar{N}_{\rm ss}}{d x} (0) \simeq 0.35699$, 
as explained in section~\ref{sec-pp.gauge-mode}
\footnote{%
We take this opportunity to clarify the confusion expressed   
in ``Note Added'' of \cite{Mai96}, which states that the gauge mode 
reported in  \cite{KHA95} (same as reported here) ``seems to be erroneous.''  
The fact is that our report in \cite{KHA95} (and of course here) is correct. 
The confusion seems to be 
due to different gauges used in the analysis.  We here and in 
\cite{KHA95} use the gauge $\bar{N}_{p}(s,0) = 0$ at the sonic point, while 
Maison \cite{Mai96} and we in Sec.~\ref{sec-lyapunov} 
use the gauge $\bar{N}_{p}(s, -\infty) = 0$ 
at the origin.  The former gauge gives the gauge mode at 
$\kappa \simeq 0.35699$, while the latter gives at $\kappa = 1$; the 
difference in the values of $\kappa$ is well understood in view of  
(\ref{eq:gauge-mode}). 
}.

%%%%%%%%%%%%%%%%%%%%%%%%%%%%%%%%%%%%%%%%%%%%%%%%%%%%%%%%%%%%%%%%%%%%%%%
\subsubsection{Other modes for the Evans--Coleman solution}
\label{subsubsec.pp-others} 

To confirm our scenario, we 
performed a thorough search  of other eigenmodes in the region 
$0 \leq {\rm Re \,} \kappa \leq 15, \; | {\rm Im \,} \kappa | \leq 14$, 
and found {\em none}, except for the above mentioned 
relevant mode and the gauge mode. 
We also performed a less complete search in the region 
$-1.5 \leq {\rm Re}  \kappa < 0, | {\rm Im} \kappa | < 2$.  There is an 
eigenmode with ${\rm Re} \kappa \lesssim -1.4$, which is consistent 
with the results of the Lyapunov analysis in Sec.~\ref{sec-lyapunov}. 

It is not {\em a priori} obvious whether the eigenmodes form a complete 
set of basis functions.  
Due to the complicated structure of the equations of 
motion, we have not found a beautiful argument which can restrict 
possible eigenvalues (like that of \cite{Chan75} and
references therein).   

Our search of eigenmodes has a serious 
drawback that it is theoretically impossible to cover the 
whole values of $\kappa \in { \Bbb C}$ (unless, of course, one employs 
more sophisticated mathematical techniques).  To fill this gap, and 
to further confirm our scenario, we performed a Lyapunov analysis, 
as explained in Sec.~\ref{sec-lyapunov}.

%%%%%%%%%%%%%%%%%%%%%%%%%%%%%%%%%%%%%%%%%%%%%%%%%%%%%%%%%%%%%%%%%%%%%%%%
%%%%%%%%%%%%%%%%%%%%%%%%%%%%%%%%%%%%%%%%%%%%%%%%%%%%%%%%%%%%%%%%%%%%%%%%
\subsection{Modes for other self-similar solutions}
\label{subsec-pp.other-ss}
As has been stated in Sec.~\ref{sec-ss.results}, 
there exists a series of
self-similar solutions (specified by the number of zeros of $V$), 
in which the Evans--Coleman solution can be
considered as the first one (exactly one zero for $V$).
We have searched relevant eigenmodes for the first several
self-similar solutions and found {\em more than one} relevant modes 
(see Table~\ref{tab:other FP}).
This implies that the other self-similar solutions are irrelevant for
the generic critical behaviour.
However, we here admit that our analysis is less complete for these 
higher self-similar solutions, because due to a difficulty in coding, 
(1) we have not done the analysis for (possible) self-similar 
solutions with even number of zeros of $V$, 
(2) there can be more relevant 
modes than is reported in the Table. 

\begin{table}[htbp]
  \begin{center}
    \leavevmode
    \begin{tabular}{ld}
\hline
self-similar solution&
$\kappa$\\
\hline
\hline
1 (Evans--Coleman)
&2.8105525488\\
\hline
3
&8.456\\
&3.464\\
&1.665\\
&0.497\\
\hline
5
&15.80\\
&7.13\\
&3.22\\
&1.51\\
&0.500\\
\hline
7
&15.97\\
&6.92\\
&3.20\\
&1.50\\
\hline
    \end{tabular}
  \end{center}
  \caption{Relevant modes found for other self-similar solutions, 
  which are labeled by the number of zeros of $V$.  
}
  \label{tab:other FP}
\end{table}

%%%%%%%%%%%%%%%%%%%%%%%%%%%%%%%%%%%%%%%%%%%%%%%%%%%%%%%%%%%%%%%%%%%%
%%%%%%%%%%%%%%%%%%%%%%%%%%%%%%%%%%%%%%%%%%%%%%%%%%%%%%%%%%%%%%%%%%%%
%%%%%%%%%%%%%%%%%%%%%%%%%%%%%%%%%%%%%%%%%%%%%%%%%%%%%%%%%%%%%%%%%%%%
%%%%%%%%%%%%%%%%%%%%%%%%%%%%%%%%%%%%%%%%%%%%%%%%%%%%%%%%%%%%%%%%%%%%
\section{Lyapunov analysis}
\label{sec-lyapunov}

%%%%%%%%%%%%%%%%%%%%%%%%%%%%%%%%%%%%%%%%%%%%%%%%%%%%%%%%%%%%%%%%%%%%
%%%%%%%%%%%%%%%%%%%%%%%%%%%%%%%%%%%%%%%%%%%%%%%%%%%%%%%%%%%%%%%%%%%%
% \section{The Lyapunov analysis}
% \label{sub-lyapunov}

To further confirm the uniqueness of the relevant mode 
we performed a Lyapunov analysis around the critical solution.
A Lyapunov analysis and the shooting
method for an ordinary differential equation adopted in \cite{KHA95} 
and in the previous section 
are complementary to each other for the following reasons: (1) The
former extracts eigenmodes in the descending order of its real part,
whereas the latter affords information only of a finite region of
complex $\kappa$ plane.  (2) The Lyapunov analysis is useful even if
the (discrete) eigenmodes do not form a complete set.
(3) It is easier in the latter than in the
former to numerically obtain accurate values of the eigenvalues
$\kappa$, hence the critical exponent $\beta_{\rm BH}$.
We here present our method of analysis  and its results. 

%%%%%%%%%%%%%%%%%%%%%%%%%%%%%%%%%%%%%%%%%%%%%%%%%%%%%%%%%%%%%%%%%%%%
%%%%%%%%%%%%%%%%%%%%%%%%%%%%%%%%%%%%%%%%%%%%%%%%%%%%%%%%%%%%%%%%%%%%
\subsection{Numerical Methods}
\label{sub-lyp.num}

The Lyapunov analysis is a method of extracting eigenvalues in
descending order.
It involves time integration the linearized EOM (\ref{eq:EOM-var})
about the self-similar solution.
It takes advantage of the fact that the eigenmodes with large
${\rm Re} \kappa$ dominates at late times and that the volume, defined by
an (arbitrarily chosen) inner product on $\Gamma$, of the
parallelepiped spanned by the eigenmodes corresponding $\kappa_1$,
$\kappa_2$,..., $\kappa_n$ approaches  
$e^{\kappa_1\kappa_2...\kappa_n}$. 
During the integration, vectors in $\Gamma$ are orthonormalized
by the Gram--Schmidt procedure.  This is essential for avoiding
numerical overflows and underflows.
For details see Appendix~\ref{app-lyapunov}. 
We can also have information on the imaginary part of eigenvalues
by the period of oscillation in time $s$.

We have performed the above-described calculation by numerically
solving the PDE's for the perturbation, (\ref{eq:EOM-var}). 
We emphasize that solving the PDE's for perturbations,
(\ref{eq:EOM-var}), 
is much easier than solving the full EOM close to the critical 
point, because perturbations are well behaved in contrast with 
near-critical solutions whose derivatives are diverging in $(t,r)$ 
coordinate. 

We used the first order Lax scheme, and the second order Lax--Wendroff 
scheme for time evolution.  We show the results {}from the Lax--Wendroff 
scheme.
To simplify coding, we employed a new 
gauge, $\bar{N}_{\rm var} (s,-\infty) = 0$. In this gauge we observe the 
gauge mode at $\kappa = 1$. We have made appropriate 
coordinate transformations to stabilize the integration schemes. 

We imposed the free boundary condition at the sonic point $\xi=1$.
This can be done because no information can come in {}from outside the sonic
point (as is easily seen by studying the characteristic curves), 
for spherically symmetric metric perturbations contains no
gravitational waves and are determined by the matter degrees of
freedom.

%%%%%%%%%%%%%%%%%%%%%%%%%%%%%%%%%%%%%%%%%%%%%%%%%%%%%%%%%%%%%%%%%%%%
%%%%%%%%%%%%%%%%%%%%%%%%%%%%%%%%%%%%%%%%%%%%%%%%%%%%%%%%%%%%%%%%%%%%
\subsection{Numerical results}
\label{sub-lyp.res}

%%%%%FIGFIGFIGFIGFIGFIGFIGFIGFIGFIGFIGFIGFIGFIGFIGFIGFIGFIGFIGFIG
%%%%%FIGFIGFIGFIGFIGFIGFIGFIGFIGFIGFIGFIGFIGFIGFIGFIGFIGFIGFIGFIG
\begin{figure}
 \begin{center}
   {\BoxedEPSF{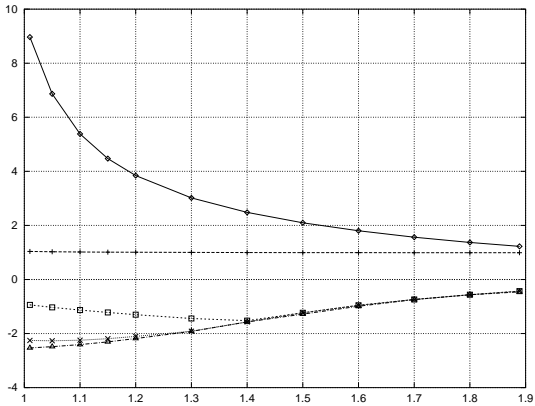 scaled 300} }
 \end{center}
%{\bf // PLACE Fig.~\ref{fig-kp-vs-gm} HERE. //} 
%  \vspace*{60mm}
\caption{Dependence of Lyapunov exponents ${\rm Re} \kappa$ on $\gamma$.
    The lines show the Lyapunov exponents.  The top line, labeled 
    ``relevant'' represents the relevant mode, 
        and the second line, labeled 
    ``gauge,'' represents the gauge mode, whose Lyapunov exponent 
    equals unity theoretically.  For $\gamma < 1.3$, the third and the
    fourth largest mode are real (labeled ``real'').
    For $\gamma=1.3$ and $\gamma=1.4$ the third is real and the fourth 
    and the fifth are complex conjugate.
    For $\gamma>1.4$ the third and the fourth complex conjugate
    (labeled ``cc'').
    We see that complex conjugate pair takes over the real eigenvalue
    twice in $1<\gamma\protect\lesssim 1.889$.
    The graph shows there is a {\em unique} relevant mode 
    for all values of $\gamma$ analyzed.
}
  \label{fig-kp-vs-gm}
\end{figure}
%%%%%FIGFIGFIGFIGFIGFIGFIGFIGFIGFIGFIGFIGFIGFIGFIGFIGFIGFIGFIGFIG
%%%%%FIGFIGFIGFIGFIGFIGFIGFIGFIGFIGFIGFIGFIGFIGFIGFIGFIGFIGFIGFIG

Fig.~\ref{fig-kp-vs-gm} shows the dependence of ${\rm Re}\kappa$ 
on $\gamma$.
The largest eigenvalue is real and agrees well with the first
eigenvalue (which is real) found by solving two-point boundary
problem of an ODE in Sec.~\ref{sec-pert}.  
The second largest eigenvalue is that of the gauge mode, which should 
be exactly equal to one. 
For $\gamma\le 1.2$ the third and the fourth largest eigenvalues are
real, and the fifth is imaginary.
For $\gamma= 1.3, 1.4$ the third is real and the fourth and
the fifth is imaginary, which are complex conjugate.
those for $\gamma\ge 1.4$ are complex. 
We can consider that there is a cross-over of Lyapunov exponents 
corresponding to the real eigenvalue and the complex pair between
$\gamma=1.2$ and $\gamma=1.3$.
For $\gamma\ge1.5$ the third and the fourth are a complex conjugate pair,
so that there is another cross-over of Lyapunov exponents 
corresponding to the real eigenvalue and the complex pair between
$\gamma=1.4$ and $\gamma=1.5$.
For large $\gamma$ it seems that many eigenmodes with close 
values of ${\rm Re}\kappa$ are present.  
Our preliminary computation of ten modes for $\gamma=1.889$ shows that
there at least nine modes in the range $-0.6\le{\rm Re}\kappa\le-0.4$, 
though it seems that not all of them are degenerate in ${\rm Re}\kappa$.

The figure establishes the 
uniqueness of the relevant mode in collapse of
perfect fluids with $p=(\gamma-1)\rho$, $1< \gamma \lesssim 1.889$.
{}{}from this we can conclude that the critical behaviour must be 
observed for all of these models  with $1 < \gamma \lesssim 1.889$.

%%%%%%%%%%%%%%%%%%%%%%%%%%%%%%%%%%%%%%%%%%%%%%%%%%%%%%%%%%%%%%%%%%%%
%%%%%%%%%%%%%%%%%%%%%%%%%%%%%%%%%%%%%%%%%%%%%%%%%%%%%%%%%%%%%%%%%%%%
%%%%%%%%%%%%%%%%%%%%%%%%%%%%%%%%%%%%%%%%%%%%%%%%%%%%%%%%%%%%%%%%%%%%
%%%%%%%%%%%%%%%%%%%%%%%%%%%%%%%%%%%%%%%%%%%%%%%%%%%%%%%%%%%%%%%%%%%%
%%%%%%%%%%%%%%%%%%%%%%%%%%%%%%%%%%%%%%%%%%%%%%%%%%%%%%%%%%%%%%%%%%%%
\section{Summary of results}
\label{sec-results}

We have performed the analysis presented in 
Secs.~\ref{sec-ss}--\ref{sec-lyapunov} for 
perfect fluid, with $1 < \gamma \leq 2$.  As has been noted in 
\cite{Bog77,OP90,FoHe93}, a 
self-similar solution of Evans--Coleman type (i.e. 
with one zero of $V$) ceases to exist for $\gamma \gtrsim 1.889$.  
For Evans--Coleman type self-similar solutions, the Lyapunov analysis 
of Sec.~\ref{sec-lyapunov} establishes that there is a unique 
relevant mode, and thus we can observe the critical behaviour 
(Fig.~\ref{fig-kp-vs-gm}).  Then, the precise values of the relevant 
eigenvalue $\kappa$ are obtained by the shooting method of 
Sec.~\ref{sec-pert}, and the critical exponent is given by 
$\beta_{\rm BH} = 1/ \kappa$.

The result is shown in Table~\ref{tab:kp vs gm}. 
This in particular shows that the relevant eigenmodes found by
Maison~\cite{Mai96} are actually unique, and thus are responsible
for the critical behaviour.
The value of the critical exponent $\beta_{\rm BH}$ depends strongly on
$\gamma$. Moreover, the limit $\gamma \rightarrow 1$ seems to be 
discontinuous. (For the dust, $\gamma =1$, we expect a trivial critical 
behaviour with $\beta_{\rm BH} = 1$.)  This may be because the domain 
of attraction of the Evans--Coleman type self-similar solution 
vanishes as $\gamma \rightarrow 1^+$.

\begin{table}[htbp]
\squeezetable
  \begin{center}
    \leavevmode
    \begin{tabular}{ldd}
\hline
$\gamma$ & $\kappa$ & $\beta_{\rm BH}$\\ 
\hline
1.00001 &  9.4629170  &   0.10567566 \\
1.0001  &  9.45592488 &   0.10575380 \\
1.001   &  9.38660322 &   0.10653481 \\
1.01    &  8.74868715 &   0.11430286 \\
1.03    &  7.61774326 &   0.13127247 \\
1.04    &  7.16334221 &   0.13959964 \\
1.05    &  6.76491004 &   0.14782163 \\
1.06    &  6.41269915 &   0.15594058 \\
1.08    &  5.81789124 &   0.17188358 \\
1.1     &  5.33435815 &   0.18746398 \\
1.12    &  4.93282886 &   0.20272343 \\
1.15    &  4.44235059 &   0.22510605 \\
1.18    &  4.0484584  &   0.2470076 \\
1.2     &  3.82545008 &   0.26140715 \\
1.22    &  3.62729455 &   0.27568756 \\
1.25    &  3.36750228 &   0.29695600 \\
1.28    &  3.14337431 &   0.31812947 \\
1.3     &  3.00990875 &   0.33223599 \\
1.32    &  2.88714829 &   0.34636253 \\
4/3     &  2.81055255 &   0.35580192 \\
1.36    &  2.66838221 &   0.37475891 \\
1.38    &  2.57025726 &   0.38906611 \\
1.4     &  2.47850858 &   0.40346844 \\
1.42    &  2.39245265 &   0.41798110 \\
1.44    &  2.31150728 &   0.43261815 \\
1.46    &  2.23517329 &   0.44739260 \\
1.48    &  2.16301995 &   0.46231659 \\
1.5     &  2.09467339 &   0.47740140 \\
1.52    &  2.02980720 &   0.49265763 \\
1.55    &  1.93841621 &   0.51588508 \\
1.58    &  1.85338883 &   0.53955219 \\
1.6     &  1.79989076 &   0.5555893 \\
%1.61  &  1.77402870 &   0.56368874 \\
1.62    &  1.74873002 &   0.5718436 \\
1.64    &  1.69974510 &   0.5883235 \\ 
1.66    &  1.65278973 &   0.6050376 \\
1.68    &  1.60773076 &   0.6219947 \\
1.7     &  1.56444628 &   0.6392038 \\
1.72    &  1.52282404 &   0.6566747 \\
1.74    &  1.48276003 &   0.6744180 \\
1.76    &  1.44415717 &   0.6924454 \\
1.78    &  1.40692422 &   0.7107703 \\
1.8     &  1.37097467 &   0.7294081 \\
% 1.82  &  
% 1.84  &  
% 1.86  &
1.88    &  1.2383842  &   0.8075039 \\
1.888   &  1.2259859  &   0.8156700 \\
1.889   &  1.2244458  &   0.8166960 \\
\hline
    \end{tabular}
  \end{center}
  \caption{Values of $\kappa$ and $\beta_{\rm BH}
  =1/\kappa$ for $1<\gamma \leq
    1.889$.  The last digit is rounded.}
  \label{tab:kp vs gm}
\end{table}

%%%%%%%%%%%%%%%%%%%%%%%%%%%%%%%%%%%%%%%%%%%%%%%%%%%%%%%%%%%%%%%%%%%%
%%%%%%%%%%%%%%%%%%%%%%%%%%%%%%%%%%%%%%%%%%%%%%%%%%%%%%%%%%%%%%%%%%%%
%%%%%%%%%%%%%%%%%%%%%%%%%%%%%%%%%%%%%%%%%%%%%%%%%%%%%%%%%%%%%%%%%%%%
%%%%%%%%%%%%%%%%%%%%%%%%%%%%%%%%%%%%%%%%%%%%%%%%%%%%%%%%%%%%%%%%%%%%
%%%%%%%%%%%%%%%%%%%%%%%%%%%%%%%%%%%%%%%%%%%%%%%%%%%%%%%%%%%%%%%%%%%%
\section{Conclusions and discussion}
\label{conc}

The aim of this paper was twofold: First, in Sec.~\ref{sec-RGscenario}  
we presented a mathematically rigorous framework, under which most of 
the critical behaviour observed in gravitational collapse are expected 
to be analyzed.  We have in particular shown rigorous sufficient 
conditions under which
(1) critical behaviour with a continuous self-similar solution is 
observed and the critical exponent is given {\em exactly} in terms of 
the unique relevant Lyapunov exponent $\kappa$ by 
$\beta_{\rm BH} = \beta / \kappa$ (Sec.~\ref{sub-rgflow}),    
(2) critical behaviour with a discrete 
self-similar solution is observed and the critical exponent is given 
by the same formula (Sec.~\ref{sec-discr.SS}), and 
(3) different models exhibit 
the same critical behaviour, i.e. {\em universality} in the sense of 
statistical mechanics (Sec.~\ref{sec-univ.class}).  
The sufficient conditions for (1) and (2) are essentially the {\em 
uniqueness} of the relevant mode, together with some regularity on 
the dynamics.  The condition for (3) is essentially that the 
difference of EOM is relatively small in the critical region.

% The key idea behind the analysis is the renormalization group.  The 
% renormalization group method provides rigorous control 
% of error terms in linear perturbation 
% analysis, and also provides nice framework to discuss models with 
% different EOM, both by clarifying  
% ``relevant'' terms for the phenomena we are interested in. 

The key idea behind the analysis is the renormalization group.  
It naturally casts a system described by partial
differential equations into a dynamical system on an infinite-
dimensional phase space,  where continuously and discretely
self-similar solutions of the original equations are described by
fixed points and limit cycles.  
This opened the way to the abstract analysis on dynamical systems 
described in Sec.~\ref{sec-RGscenario}.  With the help of techniques developed 
in statistical mechanics to study critical behaviour, we have been able to 
obtain detailed estimates on the time evolution of the
deviation from the critical solutions (or the stable manifold).  
We thus obtained general and precise results, 
including rigorous relations between the critical exponent and Lyapunov exponent.  

Another advantage of the renormalization group idea is that 
it enables us to treat systems with different EOM by naturally 
introducing a space of dynamical systems, 
as described in Sec.~\ref{sec-univ.class}. 
Each system is in general driven along a renormalization group flow in this 
space, and this enabled us to clarify
terms of EOM which are ``relevant'' for the phenomena we are interested in.  

These point of view  provide us a natural understanding on why 
critical behaviour are often observed in gravitational collapse. 

\medskip

We then proceeded to confirm the picture for a concrete model of 
perfect fluid in Secs.~\ref{sec-EOM}--\ref{sec-results}.  
We employed essentially two 
methods to show that the main assumption, uniqueness of the relevant 
mode, is satisfied: (1) usual shooting method, and (2) Lyapunov 
analysis.  Their results agree well, and strongly suggest that the 
relevant mode is unique.  We obtained values of the critical 
exponent $\beta_{\rm BH}$ in Sec.~\ref{sec-results}.

\bigskip

It should be emphasized that the general framework presented in this
paper is useful for both mathematical and numerical analysis on 
concrete models of critical behaviour in gravitational collapse,
because it tells that in which case one can rely on the intuitive
picture of the phase space shown in the schematic diagrams
Figs.~\ref{fig-flow0}, \ref{fig-poin}, \ref{fig-withV} 
and the argument presented in \cite{KHA95} 
which states that the linear perturbation is enough to
understand all aspects of the critical behaviour, 
apart {}from pathological behaviour mainly due to the
infinite dimensionality of the phase space.
For mathematical proof of the existence of the critical behaviour and
estimation of the critical exponent, the essential step is to find the
norm $\|\cdot\|$, which defines the functional space or the phase
space $\Gamma$, that satisfies the Assumptions in the
Sec. \ref{sub-rgflow}. 
This should be highly nontrivial and depends on each model considered.
For numerical analysis the nonlinear simulation is needed only to 
confirm Assumptions G1--3 in Sec. \ref{sub-rgflow}, especially 
Assumption G1.  This is qualitative one rather than quantitative
(because only the finiteness of the time before blowup is needed), 
and requires the time evolution of very small region of initial data,
essentially a set of initial data which is the fixed point
plus the small relevant mode. 
The calculation of this e.g. in $(s,x)$ coordinate should be much 
easier than adaptive mesh refinement calculation in $(t,r)$ coordinates
for wide region of the phase space 
and should allow a closer control of the numerical errors.
This together with close analysis of the spectrum of the linear operator
${\cal T}_\sigma$ around the fixed point may lead to a
computer--assisted proof of Theorem \ref{prop-final}.

\medskip

There remain several open questions to be answered. 

First, it should be emphasized that our analysis in 
Secs.~\ref{sec-EOM}--\ref{sec-results} confirms only the {\em local} 
behaviour of the flow, around the specific self-similar solutions.  
Although self-similar solutions with more than one zeros of $V$ seem 
to be irrelevant for generic critical behaviour (because they seem to 
have more than one relevant modes), it would be 
interesting to know the {\em global} behaviour of the flow, in 
particular, how these self-similar solutions are related.  

Second, now that various kinds of critical behaviour for 
different models have been observed, it would be interesting to ask 
what happens in a mixed system (e.g. perfect fluid $+$ scalar 
fields).  We performed a preliminary analysis in this direction.  
That is, we considered a ``mixed'' system of radiation fluid with a 
real scalar field, whose energy-momentum tensor is given by 
$T_{ab}=\rho u_a u_b+p(u_a u_b+g_{ab})
+\nabla_a\phi\nabla_b\phi
  -(1/2)g_{ab}\nabla^c\phi\nabla_c\phi$,
and studied the linear stability of the Evans--Coleman self-similar 
solution in this mixed system.  
(It is easily seen that eigenmodes decouple.) 
Lyapunov analysis in Sec.~\ref{sec-lyapunov}, applied to 
this mixed system,  shows that the Evans--Coleman self-similar 
solution has the unique relevant mode (given by the relevant mode of 
the radiation fluid), and all scalar eigenmodes are irrelevant.  
This means that the Evans--Coleman self-similar 
solution is stable under the scalar perturbation, and we will observe 
the same radiation fluid critical behaviour for the mixed system (at 
least when the scalar field is sufficiently weak).  It would be 
interesting to investigate the behaviour of the system without the 
restriction that the scalar field is weak; in particular, where does 
the cross-over between two types (fluid, scalar) of critical behaviour occur? 

As is suggested {}from our general framework in Sec.~\ref{sec-RGscenario}, 
a ``critical behaviour'' such as observed in 
gravitational collapse is not in fact limited to self-gravitating systems.  
Similar, but slightly simpler, 
phenomena can be observed in much simpler systems, such as 
nonlinear heat equation, where we can rigorously carry out the analysis of 
Sec.~\ref{sec-RGscenario}.  
However if a solution of EOM blows up in such simple systems, 
it usually means that the equation is not physically applicable in 
the blow-up region, and the blow-up is an artifact of bad approximation.  
General relativity provides rare examples where the 
blow-up of solutions does have a physical meaning --- formation of a 
black hole.

Our analysis in Secs.~\ref{sec-EOM}--\ref{sec-results} is not 
mathematically rigorous.  In view of the general framework given in 
Sec.~\ref{sec-RGscenario}, and in view of the above remark that the 
critical behaviour and blow-up of solutions are in fact physically 
meaningful, it is extremely desirable to develop a 
mathematically rigorous analysis of critical behaviour for a 
physically interesting model such as radiation fluid or scalar 
fields.  
We are planning to come back to this problem in the near future.

Our intuition on gravitational collapse still seems to be heavily
based on few exact solutions, especially the limiting case of
pressureless matter.
The critical behaviour may 
provide a different limiting case
that the final mass is small compared to the initial mass
for more realistic and wider range of matter contents.
It will be of great help to settle the problems in gravitational
collapse such as cosmic censorship conjecture.

%%%%%%%%%%%%%%%%%%%%%%%%%%%%%%%%%%%%%%%%%%%%%%%%%%%%%%%%%%%%%%%%%%%%%%%%%
%%%%%%%%%%%%%%%%%%%%%%%%%%%%%%%%%%%%%%%%%%%%%%%%%%%%%%%%%%%%%%%%%%%%%%%%%
%%%%%%%%%%%%%%%%%%%%%%%%%%%%%%%%%%%%%%%%%%%%%%%%%%%%%%%%%%%%%%%%%%%%%%%%%
%%%%%%%%%%%%%%%%%%%%%%%%%%%%%%%%%%%%%%%%%%%%%%%%%%%%%%%%%%%%%%%%%%%%%%%%%
\section*{Acknowledgments}
We thank Akio Hosoya, Hal Tasaki and Atsushi Inoue 
for stimulating discussions. 
This work is partially supported by the Japan Society for the 
Promotion of Science (T.K.) and the Ministry of Education, Science and
Culture (T.H. and T.K.).
All the computations were performed on several work stations 
of Tokyo Institute of Technology.

\appendix

%%%%%%%%%%%%%%%%%%%%%%%%%%%%%%%%%%%%%%%%%%%%%%%%%%%%%%%%%%%%%%%%%%%%%%%%%
%%%%%%%%%%%%%%%%%%%%%%%%%%%%%%%%%%%%%%%%%%%%%%%%%%%%%%%%%%%%%%%%%%%%%%%%%
%%%%%%%%%%%%%%%%%%%%%%%%%%%%%%%%%%%%%%%%%%%%%%%%%%%%%%%%%%%%%%%%%%%%%%%%%
%%%%%%%%%%%%%%%%%%%%%%%%%%%%%%%%%%%%%%%%%%%%%%%%%%%%%%%%%%%%%%%%%%%%%%%%%
\section{Sketch of the proof of Lemma~2} %%\protect\ref{lem-6}
\label{proof-lem-6}

The proof consists of several steps, and is somewhat complicated.  
Because the Lemma itself is of auxiliary importance, and because the 
proof uses techniques which have been explained in detail in the 
proofs of other lemmas and propositions already proved in 
Sec.~\ref{sub-rgflow},  we here sketch 
only the essence of each step.  

{\em Step~1. $\tilde{W}_{r}$ is a continuous curve, connecting $U_{0}$ 
and $U_{n_{1}+1}$, with $\| U_{0} - U^{*} \|_{\rm u} =r$, 
$\| U_{n_{1}+1} - U^{*} \|_{\rm u} > \epsilon_{2}$. } 

We write $U_{0} \equiv U^{*} + r \Frel$, and 
$U_{n} \equiv {\cal R}_{n\sigma} (U^{*} \! + \! r \Frel)$.  
Because $U_{0}$ satisfies the assumptions of Lemma~\ref{lem-Ftriangle}, 
\begin{itemize} 
\item $\| U_{n} - U^{*} \| _{\rm u}$ keeps growing until it 
finally exceeds $\delta_{1}$. This also shows 
$n_{1} (U_{0}; \epsilon_{2}) < \infty$.  
\item $\| U_{n} - U^{*} \| _{\rm s} < \| U_{n} - U^{*} \| _{\rm u}$ 
for $n \leq n_{1} (U_{0}; \epsilon_{2}) + 1$. 
\end{itemize}
Now $D_{r}$ is a straight line connecting $U_{0}$ and $U_{1}$; and 
$\tilde{W}_{r}$ is a union of images of $D_{r}$ under 
${\cal R}_{n\sigma}$ for $n \leq n_{1} (U_{0}; \epsilon_{2}) + 1$: 
 \begin{equation}
         \tilde{W}_{r} = \cup_{n=0}^{n_{1}} D_{r, n}, \qquad 
         D_{r, n} \equiv {\cal R}_{n\sigma} (D_{r}) . 
        \label{eq:WrDr}
 \end{equation}
Because ${\cal R}_{\sigma}$ is continuous and because $n_{1}$ is finite, 
${\cal R}_{n\sigma}$ is also continuous (for $n \leq n_{1}$).  
Thus $D_{r,n}$ is a 
continuous set.  Because endpoints of $D_{r,n}$ are overlapping, 
this proves the continuity of $\tilde{W}_{r}$.  
Bounds on the norm follow immediately {}from the 
definition of $n_{1}$. 

{\em Step~2. $\tilde{W}_{r}$ is ``one-dimensional'', and any 
$U, U' \in \tilde{W}_{r}$ satisfy 
$\| U - U' \|_{\rm s} \leq C_{4} \| U - U' \|_{\rm u}$.
} 

We first prove the claim for $U, U' \in D_{r,n}$ by induction in $n$, 
and then extend the result for general $U, U' \in \tilde{W}_{r}$.  
First, the claim is satisfied for $n=0$, because $D_{r,0} = D_{r}$ is 
a straight line connecting $U_{0}$ and $U_{1}$.  Now suppose the claim 
holds for $U, U' \in D_{r,n}$ and try to prove it for 
${\cal R}_{\sigma}(U), {\cal R}_{\sigma}(U') \in D_{r,n+1}$.  [Any 
$U \in D_{r,n+1}$ should have at least one inverse image in 
$D_{r,n}$, so this is sufficient.]
We compute as we did in deriving 
(\ref{eq:ap.ev1})--(\ref{eq:ap.ev11}) using (\ref{eq:RTrel2}), 
(write $F = U' - U$) 
\begin{eqnarray}
        {\cal R}_{\sigma} (U') & = & {\cal R}_{\sigma} (U+F) 
        = {\cal R}_{\sigma} (U) + {\cal T}_{\sigma, U^{*}} (F) 
        \nonumber \\
        && \hspace{-12pt}
        + \Oabs (6 K_{1} \| U - U^{*} \| \, \| F \|) 
        + \Oabs (K_{1} \| F \|^{2} ) .
        \label{eq:Wr.rec.1}
\end{eqnarray}
Now, by the inductive assumption, 
$\| F \|_{\rm s} < \| F \|_{\rm u}  = \| F \|$.  So we have 
\begin{eqnarray} 
        && \| {\cal R}_{\sigma} (U') - {\cal R}_{\sigma} (U) \|_{\rm s} 
        \leq   e^{-\bar\kappa \sigma} \| F^{\rm s} \| 
        \nonumber \\
        & & \qquad \qquad + K_{1} \bigl [ 
        6 \| U - U^{*} \| + \| F \|_{\rm u} 
        \bigr ] \, \| F \|_{\rm u} 
        \\ 
        && \| {\cal R}_{\sigma} (U') - {\cal R}_{\sigma} (U) \|_{\rm u} 
        \nonumber \\
        && \qquad
        \geq  
        \bigl [ e^{\kappa \sigma} 
        - 6 K_{1} \| U - U^{*} \| - K_{1} \| F \|_{\rm u} 
        \bigr ] \,
        \| F \|_{\rm u}
\end{eqnarray} 
Taking $\| U - U^{*} \| < \epsilon_{2}$ and $\| F \| < \epsilon_{2}$ 
into account, this leads to 

\begin{eqnarray}
        \frac{ \| {\cal R}_{\sigma} (U') - {\cal R}_{\sigma} (U) \|_{\rm s} } 
        {\| {\cal R}_{\sigma} (U') - {\cal R}_{\sigma} (U) \|_{\rm u} } 
        & \leq & 
        \frac{e^{-\bar\kappa \sigma} }
        {e^{\kappa\sigma} - 7 K_{1} \epsilon_{2}} 
        \frac{ \| U - U' \|_{\rm s} }{  \| U - U' \|_{\rm u} }
        \nonumber \\
        && {}
        + \frac{7 K_{1} \epsilon_{2} }{e^{\kappa\sigma} - 7 K_{1} \epsilon_{2}} 
        \label{eq:Wr.rec.4} .
\end{eqnarray}
It is now easily seen that this recursion preserves 
\begin{equation}
        \frac{ \| U - U' \|_{\rm s} }{  \| U - U' \|_{\rm u} } 
        \leq C_{4} , 
        \label{eq:Wr.rec.5}
\end{equation}
and $C_{4} <1$ follows from our choice of $\delta_{2}, \epsilon_{2}$, 
i.e. (\ref{eq:ep2del2def1}) and (\ref{eq:ep2del2def2}).  

All this proves the claim for $U, U' \in D_{r, n}$ for some $n$.  
Because $\| U_{n} - U^{*} \|$ is strictly increasing, we can combine 
this bound for each $D_{r, n}$, to conclude 
$\| U - U' \|_{\rm s} \leq C_{4} \| U - U' \|_{\rm u}$ for 
$U, U' \in \tilde{W}_{r}$ which are not necessarily in some $D_{r, n}$. 

{\em Step~{3}.  Contraction in $E^{\rm s}$-direction.} We now 
consider the distance between a flow ${\cal R}_{n\sigma}(U)$ 
and $\tilde{W}_{r}$, and show that their distance in $E^{\rm 
s}$-direction is decreasing.  To be more precise, 
we introduce a decomposition (see Fig.~\ref{fig-dcmp} (c)): 
\begin{equation}
        U^{(n\sigma)} = \tilde{U}_{n} + G^{\rm s}_{n}
        \label{eq:Lem6.dcmp1}
\end{equation}
where this time $\tilde{U}_{n} \in \tilde{W}_{r} (U^{*})$ 
and $F^{\rm s} \in E^{\rm s}(U^{*})$.  Because we have already shown 
that $\tilde{W}_{r} (U^{*})$ is one-dimensional and extends {}from 
$r$ to $\epsilon_{2}$ in $E^{\rm u}$-direction, we can always 
decompose this way, as long as 
$r \leq \| U^{(n\sigma)} - U^{*} \|_{\rm u} \leq \epsilon_{2}$. 

To derive a recursion for $\| G^{\rm s}_{n} \|$, we start as usual 
\begin{eqnarray}
        {\cal R}_{\sigma}(U^{(n\sigma)}) 
        & = & {\cal R}_{\sigma} (\tilde{U}_{n}) 
        + {\cal T}_{\sigma, \tilde{U}_{n}} (G^{\rm s}_{n}) 
        + \Oabs(K_{1}\| G^{\rm s}_{n} \|^{2}) 
        \nonumber \\
        & =  & {\cal R}_{\sigma} (\tilde{U}_{n}) 
        + {\cal T}_{\sigma, U^{*}} (G^{\rm s}_{n}) 
        + \Delta 
\end{eqnarray}
where 
\begin{equation}
        \Delta \equiv \Oabs(6 K_{1} \| \tilde{U}_{n} - U^{*} \| 
                \cdot \| G^{\rm s}_{n} \|) 
                + \Oabs(K_{1} \| G^{\rm s}_{n} \|^{2}) .
        \label{eq:Escon.Deldef}
\end{equation}
We have to define $\tilde{U}_{n+1} \in \tilde{W}_{r}$ so that 
$\Delta$ is decomposed as 
\begin{equation}
        \Delta = \tilde{U}_{n+1} -  {\cal R}_{\sigma} (\tilde{U}_{n})  
        + F^{\rm s}
        \qquad (F^{\rm s} \in E^{\rm s}(U^{*}))
        \label{eq:Escon.Deldcp1}
\end{equation}
so that the entire expression can 
be written as $\tilde{U}_{n+1} + G^{\rm s}_{n+1}$ with 
\begin{equation}
        G^{\rm s}_{n+1} \equiv {\cal T}_{\sigma, U^{*}} (G^{\rm s}_{n})  
        + F^{\rm s}_{n+1}.
        \label{eq:Escon.Deldcp2}
\end{equation}
Now (\ref{eq:Escon.Deldcp1}) implies 
\begin{equation}
        \| \Delta \|_{\rm u} = 
        \| \tilde{U}_{n+1} -  {\cal R}_{\sigma} (\tilde{U}_{n})  \|_{\rm u} 
        \label{eq:Escon.Deldcp3}
\end{equation}
and 
\begin{equation}
        \| F^{\rm s} \| = \| F^{\rm s} \|_{\rm s} \leq 
        \| \tilde{U}_{n+1} -  {\cal R}_{\sigma} (\tilde{U}_{n})  \|_{\rm s} 
        + \| \Delta \|_{\rm s} .
        \label{eq:Escon.Deldcp4}
\end{equation}
However, we have shown in Step~2 that $\tilde{W}_{r}$ is in some sense 
``parallel'' to $E^{\rm u}(U^{*})$, i.e. 
\begin{equation}
        \| \tilde{U}_{n+1} -  {\cal R}_{\sigma} (\tilde{U}_{n})  \|_{\rm s}
        \leq C_{4}
        \| \tilde{U}_{n+1} -  {\cal R}_{\sigma} (\tilde{U}_{n})  \|_{\rm u}
        \label{eq:Escon.Deldcp5}
\end{equation}
Combining (\ref{eq:Escon.Deldcp3})--(\ref{eq:Escon.Deldcp5}) implies 
\begin{equation}
        \| F^{\rm s} \| \leq (1 + C_{4}) \| \Delta \|
        \label{eq:Escon.Deldcp6}
\end{equation}
and we finally have 
% \begin{equation}
%         G' = {\cal T}_{\sigma, U^{*}} (G) 
%         + O(\| U^{\rm u} - U^{*} \| \cdot \| G \|) + O(\| G\|^{2}) , 
%         \label{eq:Lem6.Grec}
% \end{equation}
% which, after taking the norm of both sides and making use of 
% Assumption~L1, becomes 
\begin{eqnarray}
        \| G^{\rm s}_{n+1} 
        \| & \leq & e^{-\kappa' \sigma} \| G_{n} \| + 
        (1+C_4) \times \nonumber \\
        && \times 
        \left [ 
        \Oabs(6K_1\| \tilde{U}_{n} - U^{*} \| \cdot \| G_{n} \|) 
        + \Oabs(K_1\| G^{\rm s}_{n} \|^{2}) 
        \right ].  
        \nonumber \\ 
        && \label{eq:Lem6.Grec2}
\end{eqnarray}
By taking $\| \tilde{U}_{n} - U^{*} \|$ and $\| G_{n} \|$ sufficiently 
small as in (\ref{eq:ep2del2def1}) and (\ref{eq:ep2del2def2}), 
we can bound the right hand side by $e^{-\bar\kappa'' \sigma} \| G \|$ 
with some $\bar\kappa'' > 0$. 

This in particular shows that intersections of 
$\tilde{W}_{r}$ and the plane 
$U^{*} + E^{\rm s}(U^{*}) + a \Frel$ for various $r$'s are Cauchy 
sequences, and thus the limit $r \searrow 0$ exists: existence of 
$W^{\rm u}_{\rm loc}(U^{*})$. 

Finally, repeating the above argument replacing 
$\tilde{W}_{r}$ by $W^{\rm u}_{\rm loc}(U^{*})$, we get the 
contraction around $W^{\rm u}_{\rm loc}(U^{*})$ as well. 
\hfill$\Box$

%%%%%%%%%%%%%%%%%%%%%%%%%%%%%%%%%%%%%%%%%%%%%%%%%%%%%%%%%%%%%%%%%%%%%%%%%
%%%%%%%%%%%%%%%%%%%%%%%%%%%%%%%%%%%%%%%%%%%%%%%%%%%%%%%%%%%%%%%%%%%%%%%%%
%%%%%%%%%%%%%%%%%%%%%%%%%%%%%%%%%%%%%%%%%%%%%%%%%%%%%%%%%%%%%%%%%%%%%%%%%
%%%%%%%%%%%%%%%%%%%%%%%%%%%%%%%%%%%%%%%%%%%%%%%%%%%%%%%%%%%%%%%%%%%%%%%%%
\section{Solving recursive inequalities}
\label{proof-lem-5}
We here provide a lemma, which effectively solves recursive inequalities 
encountered in Sec.~\ref{sec-RGscenario}, such as 
(\ref{eq:ap.ev33}),  (\ref{eq:ap.ev34}), 
(\ref{eq:rec-ineq-V1}),  (\ref{eq:rec-ineq-V2}).

%\bigskip\noindent  {\bf Lemma~5.}
%  {\em 
\begin{lemma}
\label{lem-5}
   Let $\Lambda$, $\bar{\Lambda}$, $C$, $\delta$ and $\epsilon$ 
   be positive constants satisfying
   $1 < \bar{\Lambda}^{-1} \le\Lambda$, 
   $C \geq 1$, and
   \begin{equation}
     \delta +  \epsilon < \frac{\Lambda -1}{2 C} . 
     \label{eq:del2cond}
   \end{equation}
   Let $b_{n}, f_{n}$ be nonnegative sequences satisfying
   \begin{eqnarray}
     &&0 < b_{0} < \epsilon,\quad f_{0} =0,
     \\
     &&b_{n+1} 
     \grless 
     \Lambda  b_{n} 
     \pm  C \left [ 
       g_{n} ( f_{n} \vee b_{n}) +  ( f_{n} \vee b_{n})^{2} 
     \right ]
     \label{eq:ap.ev42} \\ 
     &&f_{n+1}   \leq  
     \bar{\Lambda} f_{n} + C   \left [ 
       g_{n} ( f_{n} \vee b_{n}) +  ( f_{n} \vee b_{n})^{2} 
     \right ], 
     \label{eq:ap.ev41} 
   \end{eqnarray}
   where $g_{n}$ ($n \geq 0$) satisfies 
   \begin{equation}
          0 \leq g_{n} \leq \delta.
   \end{equation}
   Define $n_{2} \equiv \max \bigl \{ n \in {\Bbb Z} \bigl |   \;  
   b_{k} \leq \epsilon \, \mbox{ for } \, k \leq n \bigr \}$.
   \\ \noindent
   (i) $n_{2}$ is finite and the following hold for $n \leq n_{2}$:
   \begin{eqnarray}
     & &  0 \leq f_{n} \leq b_{n} , 
     \label{eq:f-b-bound} \\
     && b_{n} \leq \epsilon
     \left ( \frac{\Lambda +1}{2} \right ) ^{-(n_{2}-n)}.
     \label{eq:detlem.best} 
   \end{eqnarray}
   (ii) Suppose in addition that $g_n$ satisfies
   \begin{equation}
     \sum_{n=0}^{\infty} g_{n} < \infty. 
     \label{eq:del2cond2}
   \end{equation}
   Then the following hold for $n \leq n_{2}$:
   \begin{eqnarray}
         &&  \frac{2 \epsilon}{3 \Lambda -1} 
         \leq b_{n_{2}} \leq \epsilon, 
         \label{eq:detlem.bn2rgh} \\ 
         &&
         \frac{\epsilon}{C'}  
         \leq b_{0} \Lambda^{n_{2}} 
         \leq C' \epsilon,
         \label{eq:detailedlemma} 
   \end{eqnarray}  
   where $C'$ is some positive constant.
\end{lemma}

The proof is an elementary exercise in calculus, 
which is done in several steps in a bootstrapping manner.  
We only consider $n \leq n_{2}$, which in particular means 
$0 \leq b_{n} \leq \epsilon$.

(i) {\em Step 1.  Rough estimates and $f_{n} \leq b_{n}$.}\/    
We first prove by induction $f_{n} \leq b_{n}$ 
(i.e. (\ref{eq:f-b-bound})), and 
\begin{equation}
        \frac{\Lambda +1}{2} b_{n} \leq b_{n+1} 
        \leq \frac{3 \Lambda -1}{2} b_{n} 
        \label{eq:roughbd3}. 
\end{equation}
These are satisfied for $n=0$, because $f_{0} = 0$, and because of 
our choice of $\epsilon$.  Now suppose they hold for $n$.  
Our inductive assumption in particular includes $f_{n} \leq b_{n}$, which 
reduces the recursion (\ref{eq:ap.ev42}) to  
\begin{eqnarray}
        b_{n+1} & \; \grless \; & 
        \Lambda b_{n} \pm C b_{n} (g_{n} + b_{n}) 
        \label{eq:bn-rec-red} \\ 
        & \; \grless  \; & 
        b_{n} \left [ \Lambda \pm C (\delta + \epsilon) 
        \right ] 
\end{eqnarray} 
where we used uniform bounds on $g_{n}$ and $b_{n}$.  
Eq.(\ref{eq:roughbd3}) immediately follows {}from our choice of 
$\delta$ and $\epsilon$.  

On the other hand, $f_{n} \leq b_{n}$ reduces the recursion 
(\ref{eq:ap.ev41}) to 
\begin{eqnarray}
        f_{n+1} & \leq   & 
        \bar{\Lambda} f_{n} + C b_{n} (g_{n} + b_{n}) 
        \nonumber \\ 
        & \leq   & 
        b_{n} \left [ \bar{\Lambda} + C (\delta + \epsilon) 
        \right ] 
        \nonumber \\ 
        & \leq & \left [ \bar{\Lambda} + C (\delta + \epsilon) 
        \right ] \frac{2}{\Lambda+1} b_{n+1} 
\end{eqnarray} 
where in the last step we used (\ref{eq:roughbd3}) just proved.  
The factor before $b_{n+1}$ is easily seen to be less than one, and 
(\ref{eq:f-b-bound}) is proved.  

{\em Step 2.  Rough estimate on $b_{n}$.}\/  
First note that (\ref{eq:roughbd3}) for $n = n_{2}$ and the 
definition of $n_{2}$ implies 
\begin{equation}
        \epsilon \leq b_{n_{2}+1} \leq \frac{3 \Lambda -1 }{2} 
        b_{n_{2}}, 
        \label{eq:bnibd3} 
\end{equation}
i.e. (\ref{eq:detlem.bn2rgh}).  Also solving (\ref{eq:roughbd3}) 
backwards {}from $n_{2}$ to $n$, we have 
\begin{equation}
        \frac{b_{n}}{b_{n_{2}}} \leq 
        \left ( \frac{\Lambda+1}{2} \right )^{n- n_{2}}
        \label{eq:bnrough4}, 
\end{equation}
which, in view of $b_{n_{2}} \leq \epsilon$,  
is nothing but (\ref{eq:detlem.best}). 

{\em Step 3. Refined estimate on $b_{n}$.}\/  
For later use, we note the following refined estimate, which is valid 
without the additional assumption (\ref{eq:del2cond2}).  
First note that $g_{n} \leq \delta$, 
$b_{n} \leq \epsilon$,  and $f_{n} \leq b_{n}$ guarantee 
$\Lambda - C (g_{k} + b_{k} ) \geq (\Lambda +1)/2 > 1$.  
So the recursion (\ref{eq:bn-rec-red}) implies 
($n \leq n_{2}$) 
\begin{eqnarray}
        \frac{b_{n}}{b_{0}} 
        & \; \grless \; & 
        \prod_{k=0}^{n-1} \left [ \Lambda \pm 
        C (g_{k}+ b_{k}) 
        \right ]  
        \nonumber \\ 
        & \; \grless \; & 
        \Lambda^{n} \prod_{k=0}^{n-1} \left [ 1 \pm 
        \frac{C (g_{k}+ b_{k})}{\Lambda} 
        \right ]  
        \nonumber \\ 
        & \; \grless \;  & 
        \Lambda^{n} \, 
        \exp         
        \left [ 
        \pm \frac{(2 \ln 2) C}{\Lambda} 
        \sum_{k=0}^{n-1}  (g_{k}+ b_{k})  
        \right ] .
        \label{eq:refinedb.2}
\end{eqnarray}
In the second step above we used $| \ln(1+x) | \leq (2 \ln 2) |x|$, which 
is valid for $|x| \leq  1/2$.  [Note that 
$C (g_{k} + b_{k} ) / \Lambda \leq (\Lambda -1) / (2 \Lambda) < 
1/2$.]

(ii) 
Given the above estimate (\ref{eq:refinedb.2}), the rest is easy. 
Making use of the 
assumption (\ref{eq:del2cond2})  on $g_{n}$,  and 
the estimate (\ref{eq:detlem.best}) 
on $b_{k}$ proven so far, it can be easily seen that the 
sum in the exponent of (\ref{eq:refinedb.2}) 
is bounded by a constant depending only on 
$C, \Lambda, \bar{\Lambda}$ and 
$\sum_{k=0}^\infty g_k$,  uniformly in $n$ and  $n_{2}$ as 
long as $n \leq n_{2}$.  
Taking (\ref{eq:detlem.bn2rgh}) into 
account,  (\ref{eq:refinedb.2}) with $n = n_{2}$ 
gives (\ref{eq:detailedlemma}). 
\hfill$\Box$

%%%%%%%%%%%%%%%%%%%%%%%%%%%%%%%%%%%%%%%%%%%%%%%%%%%%%%%%%%%%%%%%%%%%%%%%%
%%%%%%%%%%%%%%%%%%%%%%%%%%%%%%%%%%%%%%%%%%%%%%%%%%%%%%%%%%%%%%%%%%%%%%%%%
%%%%%%%%%%%%%%%%%%%%%%%%%%%%%%%%%%%%%%%%%%%%%%%%%%%%%%%%%%%%%%%%%%%%%%%%%
%%%%%%%%%%%%%%%%%%%%%%%%%%%%%%%%%%%%%%%%%%%%%%%%%%%%%%%%%%%%%%%%%%%%%%%%%
\section{Spherically symmetric, self-similar space-time}
\label{sec-ssss}

Defriese and Carter (see \cite{Ear74}) showed 
that the line element of any self-similar,
spherically symmetric space-time is written as
\begin{equation}
   ds^2=e^{-2\eta}(-G_1(y)d\eta^2 \! + \! G_2(y)dy^2 
   \! + \!e^{2y}(d\theta^2 \! + \! \sin\theta d \phi^2)). 
     \label{eq:DC}
\end{equation}
Let $t=-e^{-\eta}Y(y)$ and $r=e^{y-\eta}$, where 
\begin{equation}
  Y^2(y)=\int_{\rm constant}^y dy{G_1e^{2y}\over 2G_2}.
\end{equation}
One has 
\begin{equation}
  ds^2=-\alpha^2dt^2+a^2dr^2+r^2(d\theta^2+\sin\theta d\phi^2),
  \label{eq:app:met}
\end{equation}
where $\alpha$ and $a$ are functions of $y$.
Since $r/(-t)$ is a function of $y$ only, $\alpha$ and $a$ are
functions of $r/(-t)$.

%%%%%%%%%%%%%%%%%%%%%%%%%%%%%%%%%%%%%%%%%%%%%%%%%%%%%%%%%%%%%%%%%%%%%%%%%
%%%%%%%%%%%%%%%%%%%%%%%%%%%%%%%%%%%%%%%%%%%%%%%%%%%%%%%%%%%%%%%%%%%%%%%%%
%%%%%%%%%%%%%%%%%%%%%%%%%%%%%%%%%%%%%%%%%%%%%%%%%%%%%%%%%%%%%%%%%%%%%%%%%
%%%%%%%%%%%%%%%%%%%%%%%%%%%%%%%%%%%%%%%%%%%%%%%%%%%%%%%%%%%%%%%%%%%%%%%%%

\section{The Lyapunov analysis}
\label{app-lyapunov}

%%%%%%%%%%%%%%%%%%%%%%%%%%%%%%%%%%%%%%%%%%%%%%%%%%%%%%%%%%%%%%%%%%%%%%%%%
%%%%%%%%%%%%%%%%%%%%%%%%%%%%%%%%%%%%%%%%%%%%%%%%%%%%%%%%%%%%%%%%%%%%%%%%%
The Lyapunov method has been extensively used in nonlinear 
analysis \cite{Osel68}.  To make our presentation self-contained and 
transparent, we present a concise mathematical description of the 
method, restricted to the situation we are interested in. 

Let us consider the linearization 
of a flow on a real Hilbert space (or a real Hilbert manifold)
around an orbit. 
It is determined by an equation 
\begin{equation}
  \frac{df}{ds}(s) = A(s) f(s),
  \label{eq:Lyp-PDE}
\end{equation}
where each $f(s)$ is an element of real Hilbert space $V$,
a complete vector space with inner product, and $A(s)$ is a real
function determined by the (background) orbit which we are
considering. 
Let  $\mbox{$(\cdot, \cdot)$}$ denote the inner product and 
$\mbox{$\| \cdot \|$} \equiv \sqrt{(\cdot, \cdot)}$
denote the norm defined by the inner product. 
We have $A(s)=A$ in (\ref{eq:Lyp-PDE}) 
when the orbit is a fixed point, and
$A(s+\Delta)=A(s)$ when it is an periodic orbit with periodicity
$\Delta$. 
We concentrate on these cases below.

Let us define time evolution operator $T_s:V\rightarrow V$ by
\begin{equation}
  f(0)=F,\quad f(s)=T_s F,
\end{equation}
where $f$ is a solution of (\ref{eq:Lyp-PDE}).
We wish to find eigenvalues 
$\kappa$ and eigenvectors $E^{\rm c}$ of $T_s$ satisfying
\begin{equation}
  \label{eq:Lyp-diff'ce eq.}
  T_s E^{\rm c} = e^{\kappa s} E^{\rm c},
  \label{eq:Lyp-eigen}
\end{equation}
in particular those with large ${\rm Re} \kappa$.
Here $\kappa$ is a complex number and $E^{\rm c}$ is a complex vector with
${\rm Re} E^{\rm c}, {\rm Im} E^{\rm c}\in V$.
In (\ref{eq:Lyp-eigen}) and below,
$s$ (and $\sigma$) can be any number in the case of
a fixed point, 
whereas it is restricted to be integer multiple of 
$\Delta$ in the case of a periodic orbit. 

We assume that the eigenvalue 
problem (\ref{eq:Lyp-eigen}) has a complete set of discrete
eigenvectors, denoted by ${\cal E}^{\rm c} \equiv 
\{E_i^{\rm c}\}_{i = 1, 2, ...}$,
with the corresponding eigenvalues
$\{\kappa_i\}_{i = 1, 2, ...}$, 
${\rm Re} \kappa_1 \geq {\rm Re} \kappa_2 \geq 
{\rm Re} \kappa_3 \geq ...$. 
Each ${\rm Re}\kappa_i$ is called the $i$th Lyapunov exponent.
(Here we implicitly assume that we can always find $\kappa_i$ with 
the largest real part after we have defined $\kappa_1$,...,
$\kappa_{i-1}$.)
We observe that if $\kappa_i$ is not real there is an integer $i^* \ne i$ 
with $\kappa_{i^*}=\kappa_i{}^*$ and $E^{\rm c}_{i^*}=E^{\rm c}_i{}^*$, 
where asterisks (except for those on $i$ or $j$) 
denote complex conjugate in this appendix.  
We define ${\cal E}=\{ E_i \}_{i = 1, 2, ...}$ by $E_i=E^{\rm c}_i$ if
$\kappa_i$ is real, and $E_i={\rm Re} E^{\rm c}_i$, 
$E_{i^*}={\rm Im} E^{\rm c}_i$ if
$\kappa_i$ is not real, $i<i^*$ and $E^{\rm c}_{i^*} = E^{\rm c}_i{}^*$.
Any real vector $F$ can be expanded by ${\cal E}$ or ${\cal E}^{\rm c}$ as 
\begin{equation}
  F = \sum_{i=1}^\infty f^iE_i
  = {\rm Re}\sum_{i=1}^\infty f_{\rm c}^i E^{\rm c}_i,
  \label{eq:expand}
\end{equation}
where $f_{\rm c}^i=f_{\rm c}^{i^*}{}^*=(1/2)(f^i+\sqrt{-1}\, f^{i^*})$. 
Without loss of generality, we assume all eigenvectors are 
normalized so that $\| {\rm Re} E^{\rm c}_i \|^2 
+ \| {\rm Im} E^{\rm c}_i \|^2 =1$,
although $E_i$'s are not necessarily  
orthogonal with respect to the inner product
$(\cdot,\cdot)$.  

Let ${\cal F}=\{F_i\}_{1\le i \le n}$ denote an $n$-frame of
(linearly independent, real) vectors and let 
$\mbox{\rm span}{\cal F}$ be the subspace of $V$ 
spanned by ${\cal F}$.
Below we often omit $1\le i \le n$ in 
${\cal F} = \{F_i\}_{1\le i \le n}$ if no 
confusion occurs.

We define several operations on frames.  First given an 
operator $X:V\mapsto V$ we define ${\cal X}$, an operator on frames 
induced {}from $X$, by ${\cal X} {\cal F} \equiv \{  X(F_i) \}_{i}$.  
We in the following consider the cases 
where $X$ is a linear operator (such as $T_{\sigma}$), or the 
normalization operator $N$: $N F \equiv F / \| F \| $. 
Second, we define the orthogonalization operator 
${\cal O} {\cal F} \mapsto {\cal F}' = \{F'_{i}\}_{1\le i\le n}$, 
where ${\cal F}'$ is defined by 
\begin{eqnarray}
  && F'_{1} \equiv F_1; \nonumber\\
  && F'_i\equiv F_i-\sum_{j=1}^{i-1}
  \frac {( F'_j,F_i )} {\|  F'_j \|^2}  F'_j,
  \qquad 2\le i \le n.
\end{eqnarray}
We adopt the convention that 
${\cal X} {\cal F}_i \equiv ( {\cal XF})_{i}$ denote
the $i$th vector of $n$-frame ${\cal X} {\cal F}$, when ${\cal X}$ 
is one of the operations on a frame defined above.  
Distinct {}from ${\cal X}$ induced {}from $X: V \rightarrow V$, 
${\cal X} {\cal F}_{1\le i\le n}$ in general depends on $F_{j}$ with
$j\ne i$. 
Note that $\mbox{\rm span} {\cal O}{\cal F} 
= \mbox{\rm span} {\cal F}$.

\bigskip\noindent 
 {\bf Proposition.}
  {\em For almost every ${\cal F}$, i.e.,  except for ``measure zero''
    cases whose precise meaning is given by (\ref{eq:generic-frame}), 
    we have
    \begin{eqnarray}
      {\rm  Re} \kappa_i
      & = & \lim_{m \rightarrow \infty} 
      \frac 1{m\sigma}
      \ln \frac{\| {\cal O} {\cal T}_{m\sigma} {\cal F}_{i} \|} {F_{i}} 
      \label{eq:Rekappa0} \\ 
      & = & 
      \lim_{m \rightarrow \infty} 
      \frac 1{m\sigma}
      \sum_{l=1}^m {\lambda_i(l,\sigma)},
      \label{eq:Rekappa}
    \end{eqnarray}
    for $1 \leq i \leq n$, where
    \begin{eqnarray}
      \lambda_i(\ell,\sigma) &\equiv &
      \ln  \|  {\cal O} {\cal T}_{\sigma} 
      ({\cal N}{\cal O} {\cal T}_{\sigma})^{l-1} {\cal F}_i \|.
      \label{eq:lambda}
    \end{eqnarray}
  }

\medskip

\noindent {\bf Remark.}
The first expression (\ref{eq:Rekappa0}) of the Proposition is 
mathematically straightforward, but is not suitable for numerical 
calculations, due to serious overflow and underflow problems 
in large time integration. To overcome this difficulty, we 
employ the second expression of the Proposition, which is performed 
in the following manner in practice. 
\begin{enumerate}
\item [(0)] Prepare ${\cal F}$ and let $\Lambda_i=0$.
\item [(1)] Evolve ${\cal F}$ in time by $T_{\sigma}$, and define 
        ${\cal F}' = T_{\sigma}{\cal F}$.
\item [(2)] Find ${\cal F}'' = {\cal O} T_{\sigma}{\cal F}$ and 
  ${\cal F}''' = {\cal N}{\cal O} T_{\sigma}{\cal F}$ by the 
  Gram--Schmidt procedure:
  \begin{eqnarray}
    &&\qquad
    F''_1 = F'_1,
    \quad
    F'''_1 = F''_1 / \| F''_1 \|;
    \nonumber\\
    &&
    F''_i = F'_i-\sum_{j=1}^{i-1}
    {( F'''_j,F'_i)}  F'''_j,  
    \quad
    F'''_i = F''_i / \| F''_i \|,
    \nonumber\\
    &&
    \hspace*{5.5cm} i\ge2.
  \end{eqnarray}
\item[(3)] Add $\lambda_i=\ln \| F''_i \|$ to $\Lambda_i$.
\item[(4)] Define new ${\cal F}={\cal F}'''$ and go back to (1).
\end{enumerate}
$\Lambda_i/m \sigma$, where $m$ is the number of iteration, gives the
Lyapunov exponent ${\rm Re}\kappa_i$ according to the Proposition, because 
${\cal N}{\cal O}$ is equivalent to the Gram--Schmidt procedure.

Before proceeding to the proof of the Proposition we list several 
properties of operations on frames. 

\bigskip 
\noindent  
  {\bf Lemma.} {\em Let ${\cal X, Y}$ be operations on frames. 
    \begin{enumerate}
    \item[(i)] 
      If ${\cal X}$ is induced {}from a {\em linear} operator 
        $X:V\rightarrow V$, 
      then as operators on frames, 
      ${\cal O} {\cal X} {\cal O} = {\cal O} {\cal X}$.  In 
      particular, 
        \begin{equation}
                {\cal O}^2 = {\cal O}, \quad 
                {\cal O} {\cal T}_s {\cal O} = {\cal O} {\cal T}_s.
                \label{eq:relation:O}
        \end{equation}
    \item[(ii)] If ${\cal X, Y}$ are induced {}from operators $X, Y$ on 
    $V$, we have 
    ${\cal XYF} = \left\{ X\left(Y(F_{i})\right)\right\}_{1\le i\le n}$.  
    In particular, 
    \begin{eqnarray}
    &&  {\cal N}^{2} = {\cal N}, \quad
        {\cal N}{\cal T_{\sigma}}{\cal N} = {\cal N}{\cal T}_{\sigma}. 
                \label{eq:relation:N} \\
        &&   {\cal T}_s {\cal N}{\cal F} 
        =\{ \| F_i \|^{-1} T_s (F_i) \}_{1\le i\le n}.
        \label{eq:relation:TN}
        \end{eqnarray}
    \item[(iii)] For $c_{i}\neq 0$  ($1 \leq i \leq n$) we have
        \begin{equation}
                {\cal O} ( \{ c_{i} F_{i} \}_{1\le i\le n} ) = 
                \{ c_{i} {\cal OF}_{i} \}_{1\le i\le n}
                \label{eq:relation:OO}
        \end{equation}
        and 
        \begin{equation}
                {\cal NON}= {\cal NO}. 
                \label{eq:relation:N0}
        \end{equation}
    \end{enumerate}
  }

\medskip

\noindent 
{\em Proof.}\/
(i) 
By the definition of ${\cal O}$, any frame ${\cal F}$ 
is written as 
${\cal F}=
\left\{  
  {\cal O} F_i+\sum_{j=1}^{i-1}a^j{}_i {\cal O} F_j 
\right\}_{1\le i\le n}$ 
with suitable scalars $a^j{}_i$'s. 
Since $X$ is linear we have
        ${\cal X} {\cal F}=\left\{  {\cal XO}{\cal F}_{i} 
        +\sum_{j=1}^{i-1}a^j{}_i {\cal XO}{\cal F}_j \right\}_{1\le i\le n}$.  
However, for scalars $a^j{}_i$'s,
\begin{equation}
  {\cal O} \left ( 
  \Bigl\{ F_i+\sum_{j=1}^{i-1}a^j{}_iF_j \Bigr\}_{1\le i\le n} \right ) 
  = {\cal O} \left (\{ F_i\}_{1\le i\le n} \right ), 
  \label{eq:equiv.class-of-O}
\end{equation}
which can be shown by induction in $i$.
So we have $ {\cal OX}{\cal F}= {\cal OXO}{\cal F}$.

(ii) is just a definition. 

(iii) can be shown by induction in $n$, noting that ${\cal OF}_{i}$ 
depends only on $F_{j}$ ($j \leq i$).  \hfill$\Box$

{\em Proof of the Proposition.} 
We prove the proposition in two steps.  First, we prove 
(\ref{eq:Rekappa0}), and then  prove the equivalence of expressions 
in (\ref{eq:Rekappa0}) and (\ref{eq:Rekappa}).

We assume ${\cal F}$ to be {\em generic} in the sense that
\begin{eqnarray}
  \mbox{\rm span}\{ F_j \}_{j\le i} \, &\cap& \,
  \left (
    \mbox{\rm span} \{ E_j \}_{j\ge i} \backslash 
    \mbox{\rm span} \{ E_j \}_{j\ge i+1} \neq \{ 0 \} 
  \right ),
  \nonumber\\
  && \hspace*{2cm} 1\le i \le n.
  \label{eq:generic-frame}
\end{eqnarray}
The set of all nongeneric ${\cal F}$'s is of measure zero in the space of
$n$-frames, because each nongeneric ${\cal F}$ contains
at least one such $i$ that $F_i$ lies in a subspace of $V$ 
of codimension one,  $\mbox{\rm span} \{ E_j \}_{j\ne i}$.
{}{}from (\ref{eq:generic-frame}) we can define 
$\tilde{\cal F} = \{ \tilde F_{i} \}_{i}$ such that 
(for $1 \leq i \leq n$) 
\begin{eqnarray}
  \tilde F_i \,&&
  \in 
  \mbox{\rm span} \{ E_j \}_{j\ge i} \backslash 
  \mbox{\rm span} \{ E_j \}_{j\ge i+1},
  \nonumber\\
  F_i-\tilde F_i \,  &&
  \in \mbox{\rm span}\{ F_j \}_{j\le i-1}. 
\end{eqnarray}
Note that we have
\begin{equation}
  O\tilde{\cal F}=O{\cal F}.
  \label{eq:OFF}
\end{equation}

Let us expand $\tilde{\cal F}$ in terms of ${\cal E}$ as
\begin{equation}
  \tilde F_i=\sum_{j=i}^{\infty} f^j{}_i E_j
  = {\rm Re}\sum_{j=i}^{\infty} f_{\rm c}^j{}_i E^{\rm c}_j,
\end{equation}
where $f_{\rm c}^i{}_j$'s are defined similarly as 
$f_{\rm c}^i$'s in (\ref{eq:expand}).
Condition (\ref{eq:generic-frame}) implies that
$f^i{}_i\ne0$ for all $i\le n$, and that no $f^j{}_i$'s 
appear for $j <i$.  
Time  evolution of $\tilde F_i$ is given by
\begin{eqnarray}
  T_s\tilde F_i
  & = & {\rm Re}\sum_{j=i}^{\infty} e^{\kappa s}f_{\rm c}^j{}_i E^{\rm c}_j
  = e^{{\rm Re}\kappa_i s} 
  \left( \tilde E_i(s) + O(e^{-\delta s}) \right), 
  \nonumber \\ 
  & & 
\end{eqnarray}
where
\begin{equation}
  \tilde E_i(s) 
  \equiv {\rm Re}\sum_{j: \ j\ge i, {\rm Re}\kappa_j = {\rm Re}\kappa_i}
  e^{i({\rm Im}\kappa_j) s}f_{\rm c}^j{}_i E^{\rm c}_j,
  \label{eq:EEtilde}
\end{equation}
and $\delta$ is some positive constant.
Since $f^i{}_i\ne0$, $\| \tilde E_i(s) \|$ is bounded and nonzero 
uniformly in $s$. 
Operating ${\cal O}$ on both sides of (\ref{eq:EEtilde}), we thus 
have by (\ref{eq:relation:OO}), 
\begin{eqnarray}
  {\cal O}{\cal T}_s\tilde {\cal F}_i
  = e^{{\rm Re}\kappa_i s} 
  \left( {\cal O} \tilde {\cal E}_i(s)+O(e^{-\delta s}) \right)
  \label{eq:OTFFtilde}
\end{eqnarray}
for sufficiently large $s$.
The left hand side can be replaced by ${\cal OT}_s {\cal F}_i$ 
leading to  
\begin{eqnarray}
  {\cal OT}_s {\cal F}_i
  = e^{{\rm Re}\kappa_i s} 
  \left( {\cal O} \tilde {\cal E}_i(s)+O(e^{-\delta s}) \right),
  \label{eq:OTFF}
\end{eqnarray}
because we have {}from (\ref{eq:relation:O}) and (\ref{eq:OFF})
that ${\cal OT}_s{\cal F}= {\cal OT}_s {\cal OF} = {\cal OT}_s {\cal 
O} \tilde {\cal F} =  {\cal OT}_s\tilde {\cal F}$ 
for any $s$.
Thus we have 
\begin{equation}
  C_1 \le \frac {\| {\cal OT}_s {\cal F}_i \|} {e^{{\rm Re}\kappa_i s}} 
  \le C_2
  \label{eq:|OTFF|}
\end{equation}
for sufficiently large $s$, 
where $C_1$, $C_2$ are positive constants.
Eqs. (\ref{eq:Lambda:simple}) and (\ref{eq:|OTFF|}) prove 
(\ref{eq:Rekappa0}). 

Now we turn to the proof of equivalence of (\ref{eq:Rekappa0}) and 
(\ref{eq:Rekappa}).  
Using  (\ref{eq:relation:N}) and (\ref{eq:relation:N0}) we have 
\[
        {\cal NOTN } = {\cal NONTN} = {\cal NONT} = {\cal NOT} 
\] 
and thus 
\[
        ( {\cal NOT} )^{\ell} = {\cal N}({\cal OT})^{\ell}. 
\]
So by the definition of $\lambda_i(l,\sigma)$, we have 
\begin{eqnarray}
  \lambda_i(l,\sigma)
  &=&
  \ln  \| {\cal OT}_{\sigma} {\cal N} ({\cal OT}_{\sigma})^{l-1} 
        {\cal F}_i \|
  \nonumber\\
  &=&
  \ln  
  \frac {\| ({\cal OT}_{\sigma})^l {\cal F}_i \|}
  {\| ({\cal OT}_{\sigma})^{l-1} {\cal F}_i \|}, 
\end{eqnarray}
where in the last step we used  (\ref{eq:relation:TN}) and 
(\ref{eq:relation:OO}). 
We thus have
\begin{eqnarray}
  &&
  \sum_{l=1}^m\lambda_i(l,\sigma) 
  = 
  \ln \frac {\| {\cal O}{\cal T}_{m \sigma} {\cal F}_i \|}
  {\| {\cal F}_i \|}.
  \label{eq:Lambda:simple}
\end{eqnarray}
and the proposition is proved. 
\hfill$\Box$

%%%%%%%%%%%%%%%%%%%%%%%%%%%%%%%%%%%%%%%%%%%%%%%%%%%%%%%%%%%%%%%%%%%%%%%%%
%%%%%%%%%%%%%%%%%%%%%%%%%%%%%%%%%%%%%%%%%%%%%%%%%%%%%%%%%%%%%%%%%%%%%%%%%
%%%%%%%%%%%%%%%%%%%%%%%%%%%%%%%%%%%%%%%%%%%%%%%%%%%%%%%%%%%%%%%%%%%%%%%%
%%%%%%%%%%%%%%%%%%%%%%%%%%%%%%%%%%%%%%%%%%%%%%%%%%%%%%%%%%%%%%%%%%%%%%%%
%\bibliographystyle{prsty}
%\bibliography{rf_full} 

\end{document}